\documentclass[graybox]{svmult}

\usepackage{type1cm}        
\usepackage{makeidx}         
\usepackage{graphicx}        
\usepackage{multicol}        
\usepackage[bottom]{footmisc}

\usepackage{newtxtext}       %
\usepackage[varvw]{newtxmath}       
\usepackage{cite}
\usepackage{orcidlink}
\newcommand{\orcid}[1]{\orcidlink{#1}}

\makeindex             

\usepackage{etoolbox}
\usepackage{xcolor}

\newcommand{\chapterpreprintnotice}{%
  \noindent{\color{gray}\bfseries Chapter 14}\par
  \vspace{2mm}
  \noindent{\color{blue}\small
    This is a preprint of the following chapter:
    M.~Chambonneau, M.~Garcia-Lechuga, D.~Grojo, and S.~Nolte,
    ``Scaling the Wavelength to Optimize Ultrafast Laser
    Processing of Bandgap Materials,''
    published in \emph{Scaling of Laser Processing -- Making Light Matter},
    edited by J.~Bonse and A.F.~Lasagni, 2026,
    Springer Nature Switzerland AG;
    reproduced with permission of Springer Nature Switzerland AG.
    The final authenticated version is available online at:\\
    \url{https://doi.org/10.1007/978-3-032-17133-7_14}.
    \par
  }
  \vspace{2mm}
}

\makeatletter
\patchcmd{\@maketitle}
  {\newpage}
  {\newpage\chapterpreprintnotice}
  {}
  {\PackageError{chapter}{Could not insert preprint notice}{}}
\makeatother

\makeatletter
\providecommand*{\toclevel@title}{0}
\providecommand*{\toclevel@author}{0}
\makeatother

\begin{document}

\title*{Scaling the Wavelength to Optimize Ultrafast Laser Processing of Bandgap Materials}
\author{Maxime Chambonneau \orcid{0000-0003-1910-2296} \and Mario Garcia-Lechuga \orcid{0000-0002-9642-2982} \and David Grojo \orcid{0000-0002-8189-3781} \and Stefan Nolte \orcid{0000-0002-2919-2662}}
\authorrunning{Scaling the Wavelength to Optimize Ultrafast Laser Processing of Bandgap Materials}
\institute{Maxime Chambonneau \at Friedrich Schiller University Jena, Institute of Applied Physics, Abbe Center of Photonics, Albert-Einstein-Straße 15, 07745 Jena, Germany \email{maxime.chambonneau@hotmail.fr}
\and Mario Garcia-Lechuga \at Laser Processing Group, Instituto de Óptica ``Daza de Valdés'', Consejo Superior de Investigaciones Científicas (CSIC), C. Serrano, 121, 28006 Madrid, Spain \email{mario.garcia.lechuga@csic.es} \and David Grojo \at Aix Marseille Université, Centre national de la recherche scientifique (CNRS), Lasers, Plasmas et Procédés Photoniques (LP3), unité mixte de recherche (UMR) 7341, 163 Avenue de Luminy, 13009 Marseille, France \email{david.grojo@univ-amu.fr} \and Stefan Nolte \at Friedrich Schiller University Jena, Institute of Applied Physics, Abbe Center of Photonics, Albert-Einstein-Straße 15, 07745 Jena, Germany \at Fraunhofer Institute for Applied Optics and Precision Engineering IOF, Center of Excellence in Photonics, Albert-Einstein-Straße 7, 07745 Jena, Germany \email{stefan.nolte@uni-jena.de}}

%
\tocauthor{Maxime Chambonneau, Mario Garcia-Lechuga, David Grojo, and Stefan Nolte}
\maketitle
 
\abstract{The spectral characteristics of ultrashort laser pulses offer unique possibilities to control interactions in space and time as well as the subsequent material responses. In particular, selecting the wavelength predefines the fundamental mechanisms governing laser--matter interaction, and thus, how energy is deposited. This Chapter provides an overview on how the essential physical effects in ultrafast laser--matter interaction scale with the wavelength, and the consequences for bandgap material processing applications. First, we recall the main spectral properties of bandgap materials and discuss the important question of the expected processing resolutions depending on wavelength. Then, recent studies on the impact of the wavelength in ultrafast laser--dielectric interaction are presented. Finally, we present how the latest developments in infrared ultrafast laser sources can be advantageously utilized to internally process narrow-gap materials despite challenges related to increased nonlinearities.}

\pagebreak


\renewcommand{\theequation}{14.\arabic{equation}}
\renewcommand{\thefigure}{14.\arabic{figure}}
\renewcommand{\thetable}{14.\arabic{table}}

\section{Introduction}
\label{sec:intro}

The efficiency of ultrafast laser processing of bandgap materials (i.e., non-metallic materials) is directly related to energy deposition. The latter is determined by linear and nonlinear mechanisms which in turn depend on how concentrated is the delivered photon packet in space and time. Most common ways to vary energy deposition experimentally consist of tuning independently the pulse energy, as well as the spatial and temporal pulse properties. As the overwhelming majority of mechanisms dictating laser--matter interaction are wavelength-dependent, another method to optimize the irradiation and achieve the desired material transformations is to adjust the wavelength. When aiming for ultrafast laser processing applications, it is therefore paramount to understand how laser--matter interaction scales with the wavelength so that the most appropriate laser system can be selected with respect to the expected outcomes.\\

In this Chapter, we first recall how experimental conditions vary with the wavelength, including the refractive index of the irradiated materials, the three-dimensional (3D) focus characteristics of the laser beams used for processing, and the driving light absorption mechanisms. From these considerations, we introduce the compromises existing between the minimum focal spot size and the ideally needed nonlinearities for high processing resolution on the surface and in the bulk of transparent bandgap materials. Then, the impact of the wavelength on laser--dielectric interaction is detailed. This includes recent and non-intuitive results indicating a modest wavelength dependence of the performance of ultrafast laser ablation in dielectrics from the ultraviolet to the mid-infrared domain. However, the wavelength is a unique control parameter on the period of surface and in-volume nanostructures, as well as on the performance of laser-written waveguides. Combining ultrashort pulses at different wavelengths can also represent a smart approach to individually control different mechanisms and improve the process efficiency as investigated in recent works on machining and drilling applications. Finally, we make a special emphasis on emerging near- and mid-infrared laser technologies which hold promises to transfer ultrafast laser 3D writing from wide- to narrow-gap materials, i.e., semiconductors. This is strongly supported by the advent of a new generation of stable and powerful ultrashort pulse sources which emit in the transparency range of semiconductors. In contrast to dielectrics, enhanced nonlinearities are found when focusing intense infrared laser pulses in semiconductors. In practice, this strongly limits the space-time energy localization, and thus, the energy deposition in these materials. We review the different strategies which have been recently developed to solve these limitations. This has led to experimental demonstrations of new laser processing capabilities and 3D solutions to meet important challenges in the semiconductor industry. 

\section{Wavelength-Dependence of Optical Properties}
\label{sec:properties}

\subsection{Refractive Index}
\label{subsec:index}

For all media, the first and fundamental parameter to exhibit wavelength dependence is the linear refractive index $n_{0}$, defined as the ratio between the phase velocity of light in the medium, and the speed of light in vacuum $c$. The empirical relationship between $n_{0}$ and $\lambda$ was established in the 19th century by Sellmeier \cite{Sellmeier1871} on the basis of the works of Cauchy. The Sellmeier equation is
\begin{equation}
\label{eq:Sellmeier}
    n_{0}^{2}(\lambda)=1+\sum_{k} \frac{B_{k}\lambda^{2}}{\lambda^{2}-C_{k}} ,
\end{equation}
where $B_{k}$ and $C_{k}$ are coefficients experimentally determined. A noticeable consequence of Eq.~(\ref{eq:Sellmeier}) is that $n_{0}$ is a decreasing function of $\lambda$, which implies that ``red'' light propagates faster than ``blue'' light. In this case, the medium exhibits ``normal'' dispersion ($\frac{dn_{0}}{d\lambda}<0$). Technically, taking the first three terms of the sum on the right-hand side of Eq.~(\ref{eq:Sellmeier}) gives already an excellent fit of $n_{0}$ over a broad range of wavelengths $\lambda$. However, because Sellmeier's model assumes optical resonances in the medium which are calculated with Kramers-Kronig relations, Eq.~(\ref{eq:Sellmeier}) is not valid close to absorption edges, where the medium exhibits ``anomalous'' dispersion ($\frac{dn_{0}}{d\lambda}>0$).\\

When aiming at internal ultrafast laser material processing, dispersion may strongly affect the pulse duration. Indeed, assuming a bandwidth-limited ultrashort laser pulse, i.e., all spectral components are in phase, group velocity dispersion (GVD) during propagation through the material causes a delay between the different wavelengths---the pulse duration thus increases. The increased pulse duration $\tau_{\text{out}}$ after propagation to a distance $z$ in the material reads
\begin{equation}
\label{eq:GVD}
    \tau_{\text{out}}=\tau_{\text{in}}\sqrt{1+\left( 4\ln{(2)}\text{GVD} \frac{z}{\tau_{\text{in}}^{2}} \right)^{2}},
\end{equation}
where $\tau_{\text{in}}$ is the pulse duration at the entrance of the material (the pulse is assumed to be Gaussian and bandwidth-limited), and $\text{GVD}$ is the group velocity dispersion, typically expressed in fs$^{2}$/mm. The temporal broadening described by Eq.~(\ref{eq:GVD}) causes an intensity decrease, in turn impacting on nonlinear absorption phenomena (see Sect.~\ref{subsec:absorption}).\\

Considering a laser pulse with a typical duration of 100~fs, Eq.~(\ref{eq:GVD}) implies that the pulse duration will increase by $10\%$ after tens of millimeters of propagation in fused silica ($\text{GVD} \approx 20$--100~fs$^{2}$/mm for $\lambda=0.4$--1~$\mu$m \cite{Malitson1965}) which is one of the most studied dielectric material (see Sect.~\ref{sec:dielectrics}). This suggests that group velocity dispersion can be neglected when working with $\tau_{\text{in}} \ge 100$~fs and focusing depths of a few millimeters, corresponding to the thickness of the fused silica sample. In contrast, group velocity dispersion may represent a critical issue in extremely dispersive media such as narrow-gap materials (see Sect.~\ref{sec:semiconductors}), as can be exemplified with silicon ($\text{GVD} \approx 1120$~fs$^{2}$/mm at $\lambda=1.55$~$\mu$m \cite{Salzberg1957}). Considering the same laser pulse with $\tau_{\text{in}}=100$~fs, a temporal broadening of $10\%$ is obtained after propagation of 1.5~mm in silicon. Therefore, group velocity dispersion cannot be neglected when working with ultrashort pulses in silicon.

\subsection{Spatial Focus Properties}
\label{subsec:focus}

As the wavelength $\lambda$ plays a major role in diffraction, all spatial properties directly depend on this parameter. In laser material microprocessing where Gaussian pulses are focused to the sample, the beam radius at 1/e$^{2}$  (i.e., the waist $w_{0}$) and the Rayleigh length $z_{R}$ (where the intensity is half of the maximum at the focus) can be used to describe the off- and on-axis spatial extension of the focal zone. When the clear aperture of the focusing optics is filled, the minimum $w_{0}$ value under paraxial approximation reads
\begin{equation}
\label{eq:w0}
    w_{0}=\frac{\lambda}{\pi \text{NA}},
\end{equation}
where $\text{NA}=n_{0}\sin{\theta}$ is the numerical aperture of the focusing optics, and $\theta$ is the maximum half-angle of the cone of light. The Rayleigh length $z_{R}$ reads
\begin{equation}
\label{eq:zR}
    z_{R}=\frac{\pi n_{0} w_{0}^{2}}{\lambda}=\frac{n_{0}\lambda}{\pi \text{NA}^{2}}.
\end{equation}
From Eqs.~(\ref{eq:w0}) and (\ref{eq:zR}), one can conclude that longer wavelengths cause a broadening of the off- and on-axis focus dimensions.\\

The above relations implicitly assume the absence of any aberrations. The goal of the present Chapter is not to provide an extensive description of optical aberrations. Nevertheless, let us discuss two aberrations where wavelength plays an important role in the context of material processing with ultrashort laser pulses which intrinsically exhibit a large spectral width.
\begin{itemize}
    \item The first is chromatic aberration which originates from the wavelength-dependent focal distance for focusing optics working in transmission, as a direct consequence of the wavelength dependence of $n_{0}$ [Eq.~(\ref{eq:Sellmeier})]. Chromatic aberration delocalizes energy deposition due to on-axis spatial extension of the focal zone. Also, since all spectral components do not spatially overlap at the focus, the pulse duration is increased---and thus, the intensity is decreased. Chromatic aberration can be limited by using achromatic transmission optics (e.g., doublets, wavelength-corrected objectives), or reflective optics (e.g., parabolic mirrors, Schwarzschild objectives).
    \item The second type of aberration which must be taken into account for internal ultrafast laser material processing is spherical aberration. This type of aberration can be caused by (1) the different refraction of rays at and far from the center of the focusing optics, and (2) in the context of internal processing, by the refractive index mismatch at the air--sample interface. In the latter case, spherical aberration is more pronounced when working deep inside the material. High-NA objective lenses equipped with correction collar can be technically used to precompensate for this aberration, and obtain the smallest possible focus dimensions. Nevertheless, it has been recently shown that, for a given material and focusing depth, the collar position to produce the smallest possible modifications is not necessarily the same as the collar position where spherical aberrations are precompensated \cite{Lapointe2024}. This has been explained by nonlinear refraction, which tends to elongate the focal zone. Also, the wavelength dependence of spherical aberration called ``spherochromatism'' \cite{Kingslake2010,Berner2018} implies that particular attention should be paid to the selected laser wavelength, which should match with the correction wavelength of the focusing optics. In addition, spherical aberrations along the whole pathway have to be considered. As shown in Ref.~\cite{Chambonneau2021b} where an aspherical lens designed at $\lambda=9.5$~$\mu$m, and ultrashort laser pulses at $\lambda=1.95$~$\mu$m have been used, negative spherical aberration caused by the optics (1) is produced. In contrast, when the focus is positioned in the bulk of silicon, positive spherical aberration caused by the refractive index mismatch (2) is obtained. As the prominence of the latter strongly depends on the focus position, the existence of a particular depth in silicon for which spherical aberrations (1) and (2) counterbalance has been identified. The higher intensity at this depth has led to transverse inscription in this material (see Sect.~\ref{sec:semiconductors}).\\
\end{itemize}

One must note that all the above-mentioned spatial aspects are valid in the linear propagation regime. Nonlinear absorption and refraction may change the spatial beam properties.

\subsection{Light Absorption}
\label{subsec:absorption}

For photon energies $hc/\lambda$ (where $h$ is the Planck constant, and $c$ is the speed of light in vacuum) higher than the bandgap $E_{g}$ of the material, single-photon ionization is the dominant process. The disadvantage of this regime is that energy is mostly deposited on the surface, as the intensity in the bulk exponentially decays with the depth in the material according to Beer--Lambert law. In contrast, for moderate intensities (i.e., not orders of magnitude above the modification threshold of the material), lower photon energies can advantageously be utilized to trigger nonlinear absorption for ultrafast laser processing with an excellent space selectivity of the energy deposition, either on the surface or in the bulk of the considered bandgap material. In this intensity range, the three main nonlinear absorption mechanisms which may govern the interaction are multi-photon, tunnel and avalanche ionization.\\

Multi-photon ionization (MPI) consists of simultaneous absorption of multiple photons by a valence electron, so that the bandgap is crossed and the electron ends in the conduction band---thus becoming free. The absorbed energy is proportional to $I^{K}$, where $I$ is the laser intensity and $K$ is the multi-photon ionization order. Another free-electron production mechanism is tunnel ionization (TI), which consists of a distortion of the potential barrier of an atom under the action of intense electric field, so that a bound electron can tunnel through this barrier. Keldysh ionization theory allows one to discriminate which photoionization mechanism (between MPI or TI) dominates \cite{Keldysh1965,Boroumand2022}. To examine this, the Keldysh parameter $\gamma$ can be estimated as
\begin{equation}
\label{eq:Keldysh}
    \gamma=\frac{2 \pi c}{\lambda e \mathcal{E}}\sqrt{m^{*} E_{g}},
\end{equation}
where $e$ is the elementary charge, $m^{*}$ is the reduced mass of the electron-hole pair, and $E_{g}$ is the bandgap. The maximum value for the magnitude of the electric field $\mathcal{E}$ can be deduced from experimentally measurable laser parameters as
\begin{equation}
\label{eq:Efield}
    \mathcal{E}^{2}=\frac{2I}{c n_{0} \varepsilon_{0}}=\frac{4E}{\pi c n_{0} \varepsilon_{0} w_{0}^2 \tau},
\end{equation}
where $I$ is the maximum intensity of the Gaussian laser pulse, $E$ is the laser energy, $n_{0}$ is the refractive index of the medium, $\varepsilon_{0}$ is the vacuum permittivity, $w_{0}$ is the beam waist defined in Eq.~(\ref{eq:w0}), and $\tau$ is the pulse duration. While for surface processing Eq.~(\ref{eq:Efield}) can generally be directly used to estimate $\mathcal{E}$, in the case of internal processing, the intensity inside the material has to be considered by accounting for potential nonlinear propagation effects and a multiplication factor $T_{F}=1-(1-n_{0})^{2}/(1+n_{0})^{2}$ corresponding to Fresnel losses at the air--material interface. Typical values for fused silica [$n_{0}(\text{SiO}_{2}) \approx 1.45$] and silicon [$n_{0}(\text{Si}) \approx 3.50$] are $T_{F} \approx 0.97$ and $\approx 0.69$, respectively. For Keldysh parameters much larger than unity ($\gamma \gg 1$) multi-photon ionization dominates, i.e., for shorter laser wavelengths $\lambda$, moderate intensities $I$, and large bandgaps $E_{g}$. It was shown that the efficiency of this mechanism is higher for linear polarization than for circular polarization \cite{Temnov2006}. Moreover, recent studies have demonstrated that, at moderate intensities ($I \le 10^{15}$~W/cm$^{2}$) this difference in ionization rate between linear and circular polarizations is more pronounced for high multi-photon ionization orders---and thus, high $\lambda$ values for a given material \cite{Nuter2023,Nuter2025}. In contrast, when $\gamma \ll 1$, i.e., at longer $\lambda$, higher $I$, and small $E_{g}$, tunnel ionization is expected to dominate.\\

In parallel to these processes, avalanche ionization takes place. This collision-based mechanism relies on a synergy between free-carrier absorption (i.e., inverse \textit{Bremsstrahlung} absorption) and impact ionization. Let us consider an electron at the bottom of the conduction band, either promoted by the aforementioned mechanisms or because of thermal population. This electron linearly gains energy under the action of free-carrier absorption. Once the electron energy is $E_{g}$ with respect to the bottom of the conduction band, it can collide with an electron in the valence band and transfer half of its energy, which produces two electrons at the bottom of the conduction band by the so-called impact ionization mechanism. Avalanche ionization takes its name from the repetition of this process as long as the laser intensity is sufficiently high, leading to an exponential growth of free-carrier population during the pulse. One must note that avalanche ionization is more efficient at long wavelengths, as, in a simple approach, free-carrier absorption scales as $\lambda^{2}$ according to the Drude model \cite{Ashcroft1976}.\\

The rate of increase in the electron density $\rho$ during the laser pulse can be expressed in its most general form as
\begin{equation}
\label{eq:rateeq}
    \frac{\partial \rho}{\partial t}=W_{\text{MPI}}+W_{\text{TI}}+W_{\text{avalanche}}-W_{\text{recombination}},
\end{equation}
where the terms $W_{\text{MPI}}$, $W_{\text{TI}}$, and $W_{\text{avalanche}}$ are the time-dependent contributions from multi-photon, tunnel, and avalanche ionization, respectively. The term $W_{\text{recombination}}$ corresponds to electron-hole recombination, which contains a characteristic recombination time which can be time-dependent (e.g., Auger recombination) or time-independent (e.g., Shockley--Read--Hall recombination).\\

The produced free electrons during the laser pulse can then transfer their energy to the lattice. The typical time scale for this energy transfer is 1--10~ps, corresponding to electron--phonon scattering \cite{Rousse2001,Sokolowski-Tinten2003}. While this transfer will occur independently of the pulse duration, the main benefit of femtosecond laser pulses is that the energy deposition and the material response do not overlap temporally. This prevents for thermal runaway during the pulse, thus limiting the heat-affected zone, in turn improving the degree of control on the material modifications. Once energy is transferred to the lattice, thermomechanical events such as heat diffusion, mechanical waves, and resolidification take place \cite{Sundaram2002c,Gattass2008}. This leads to permanent material modification which can take various forms (see hereafter).

\subsection{Processing Resolution}
\label{subsec:resolution}

\noindent \textbf{Optical far-field limit.} In laser processing, resolution refers to the level of spatial precision with which a material can be modified. It is a critical aspect for the most demanding applications. Several factors define the resolution but the laser spot size is a primary aspect. By the same way one can refer to the Abbe diffraction limit to describe the minimum structure size which can be observed with an optical microscope. The minimum laser spot diameter which can be obtained when focusing the beam with a diffraction-limited system scales as $\lambda/\text{NA}$ [see Eq.~(\ref{eq:w0})], which means that shorter wavelengths [e.g., ultraviolet (UV) lasers] lead to smaller spot sizes, resulting in potentially higher resolution. However precision and resolution do not only depend on the delivered energy profile on the workpieces but also on the physical material response. The latter depends on the wavelength but also on various other parameters including the required laser power and pulse duration, as well as the type of processed materials. For instance, the employed laser power can affect the produced features at small scales. High power can be desirable for high throughput but can also cause excessive heat, leading to so-called heat-affected zones (HAZ) around the processed spots, while a limited power (close to threshold conditions) can help to achieve finer, more controlled ablation or marking. In that respect the use of ultrashort laser pulses is also decisive as these are capable of higher precision and resolution compared to longer interaction regimes (nanosecond pulses to continuous wave).\\

\begin{figure}[t]
\centering
\includegraphics[scale=1]{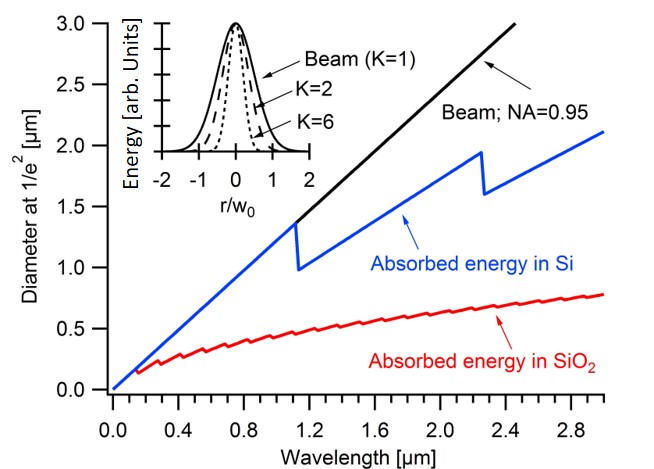}
\caption{Wavelength dependence of the resolution which can be achieved in energy deposition in silicon and fused silica for a beam focused with $\text{NA}=0.95$. The inset illustrates the steepening of a normalized beam profile when only the multi-photon absorption order $K$ is varied. In practice, as $K=\lceil E_{g}/\hbar\omega \rceil$, where $\hbar$ is the reduced Planck constant and $\omega=2\pi c/\lambda$ is the laser angular frequency, the size of the region where energy is deposited grows in step-like function. \emph{© D.~Grojo}}
\label{fig:resolution}    
\end{figure}

\noindent \textbf{Nonlinear energy deposition.} It is usually admitted that a process based on multi-photon absorption leads to superior spatial resolutions in comparison to single-photon absorption (linear absorption). This relies on the comparison done between the laser spot size and the narrower distribution of absorbed energy which is proportional to $I^{K}$, where $K$ is the necessary number of photons to cross the bandgap, i.e., the multi-photon absorption order (see Sect.~\ref{subsec:absorption}). The inset in Fig.~\ref{fig:resolution} illustrates this aspect with a Gaussian laser energy distribution corresponding to the absorbed energy distribution for single-photon absorption and the narrowing of the energy distribution for $K=2$ (valid for Si irradiated at $\lambda=1100$--2200~nm) and $K=6$ (valid for SiO$_2$ at $\lambda=700$--800~nm). However, it is important to highlight here that there is in practice no direct way to take benefit of this characteristic for a given material to process. This is because the control parameter on the multi-photon absorption order is the wavelength which simultaneously changes the spot size (see Sect.~\ref{subsec:focus}). Figure~\ref{fig:resolution} shows the linear dependence to the wavelength of a beam diameter $d$ at its resolution limit with $\text{NA}=0.95$. It compares on the same graph the corresponding diameter of the energy deposition as a function of the wavelength for silicon and fused silica. According to the diameter of energy deposition that scales as  $d'=d/\sqrt{K(\lambda)}$, one can observe  that the energy is deposited in regions smaller than the beam size but the diameter increases with the wavelength and the multi-photon absorption order. Then, shorter wavelengths associated with smaller multi-photon absorption orders is expected to be always beneficial for resolution on the basis of these simple considerations. In Sect.~\ref{sec:dielectrics}, we describe research results that examine in more details these questions in practical conditions. This is of importance as multi-photon absorption is obviously not the only nonlinear absorption mechanism depending on processing intensities and pulse durations. Beside this question of potential lateral resolution improvements, one should not forget a direct benefit of nonlinear energy deposition as multi-photon absorption order $K \ge 2$ is required to perform internal local modifications which is the basis for 3D writing technologies in bandgap materials.\\

\noindent \textbf{Deterministic response.} In practice, the highest femtosecond laser machining precisions have been mostly demonstrated in dielectrics at near-infrared wavelengths \cite{joglekar2004optics}. As will be discussed in more details in Sect.~\ref{sec:dielectrics}, these results were obtained based on a remarkably well-defined threshold for laser-induced material damage associated with the femtosecond regime. This enables the properties of materials to be modified with nanometer precision when only the central part of the Gaussian focus exceeds the threshold. In principle the threshold response provides no limit to the resolution but careful control and stabilization of laser energy is required (typical energy fluctuations in most laser systems are $\approx 1 \%$). From experiments with ultrashort laser pulses\cite{joglekar2004optics,joglekar2003study} and pulse durations tested down to few optical cycles \cite{sanner2010toward}, this deterministic character of the response has been associated with the importance of tunnel ionization with respect to multi-photon ionization \cite{joglekar2004optics,joglekar2003study,sanner2010toward}. According to Keldysh theory (see Sect.~\ref{subsec:absorption}), ultrashort pulse durations and long wavelengths \cite{grojo2013long} should be highly favorable in this prospect. However, more recent works have demonstrated the capacity to rely on this deterministic threshold-based response with conventional femtosecond laser pulses for various wavelengths from UV to mid-IR applied to dielectrics \cite{Garcia-Lechuga2020}. Similar conclusions have been also extended to various materials: dielectrics, semiconductors and metals \cite{Zhou2021}, raising questions on the initial interpretations. 

\begin{figure}[!ht]
\centering
\includegraphics[scale=0.58]{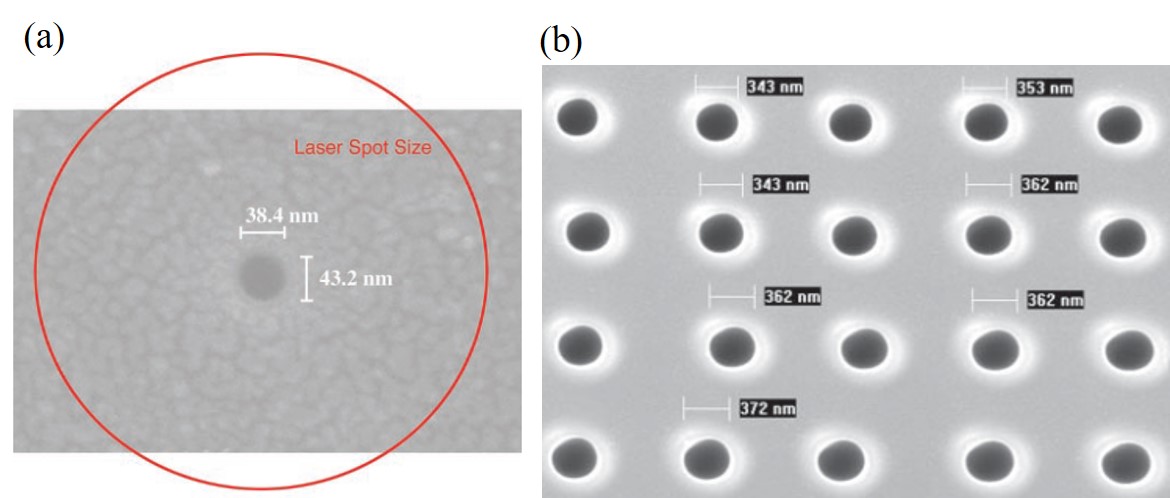}
\caption{(a) Scanning electron microscopy (SEM) image of a nanometer-scale crater produced by laser ablation on glass. The red circle indicates the focus spot size at $1/e^{2}$ according to the processing femtosecond laser beam at $\lambda=1053$~nm with a high-numerical-aperture immersion objective. (b) Array of nano-craters in glass repeatedly produced at a glass--water interface. One can note the absence of features or heat-affected zones surrounding the craters. \emph{Copyright (2004) National Academy of Sciences, U.S.A. Adapted with permission from A.~P.~Joglekar et al. ``Optics at critical intensity: Applications to nanomorphing,'' Proc. Natl. Acad. Sci. U.S.A. \textbf{101}(16), 5856 (2004) \cite{joglekar2004optics}}}
\label{fig:joglekar}
\end{figure}

\section{Impact of the Wavelength on Laser-Dielectric Interaction}
\label{sec:dielectrics}

In this Section, we discuss the role of laser wavelength in the modifications produced in dielectric materials, where the use of ultrashort lasers is essential to trigger nonlinear laser--matter interactions, ultimately leading to material modifications. We review aspects related to material ablation, specifically exploring the impact of laser wavelength on spatial resolution and precision in depth control. Additionally, we examine the proposed spectral shaping strategies to control and understand the process of laser energy deposition. Finally, we discuss the role of laser wavelength in other strategies for transforming dielectric materials that impact material functionality, such as the use of direct laser interference patterning (DLIP), the formation of laser-induced periodic surface structures (LIPSS) on material surfaces, as well as the internal inscription of waveguides, nanogratings, and photopolymerization for 3D printing.\\

\subsection{Scaling Spatial Resolution in Ultrashort Laser Ablation}
\label{subsec:ablationresolution}

As already addressed in Sect.~\ref{subsec:resolution}, the concept of nonlinear resolution, associated with an energy deposition profile confined to a region smaller than the beam size (and thus, the diffraction limit, potentially), has often been invoked in laser processing, influenced by a similar concept widely utilized in multi-photon microscopy applied for bio-imaging \cite{Zipfel2003nonlinearmicroscopy}. In the specific context of surface modification of dielectrics, it is common to find references attributing the ablation features to the size of the nonlinear absorption profile rather than the laser beam itself \cite{Lenzner1999,Korte2003,Sugioka2014}.\\

Somehow, this attributed idea has served as a motivation to primarily keep using near-infrared radiation, which is the standard wavelength of modern ultrashort laser technology ($\lambda=800$~nm and $\lambda=1030$~nm for Ti:sapphire and Yb-doped crystal lasers, respectively). This spectral range, when applied to processing dielectrics, ensures a high degree of nonlinearity, as the photon energy is $hc/\lambda \approx 1.5$~eV at $\lambda=800$~nm, and 1.2~eV at $\lambda=1030$~nm, respectively, while dielectric materials of interest (e.g., sapphire or fused silica) exhibit a bandgap $E_{g}>6$~eV corresponding to a multi-photon absorption order $K>4$.\\

Given access nowadays to ultrashort laser pulses in other spectral ranges [harmonic generation for visible light (VIS) or near-UV, optical parametric amplification for near-infrared (NIR) and mid-infrared (MIR), high-power fiber lasers emitting at increasingly longer wavelengths (see Sect.~\ref{subsec:sources})] with sufficient power to modify dielectrics, the question arises on how to exploit them to further enhance resolution. Should one take advantage of reduced optical diffraction limits with the use of shorter wavelengths (VIS or UV), or rely on highly nonlinear interactions turning to the NIR or MIR domains?\\

This question has been explored in a recent work \cite{Garcia-Lechuga2020}. In this study, the goal was to determine whether the ablation features in different dielectrics follow the shape and size of the laser beam profile itself or appear smaller due to the concept of nonlinear resolution. This was investigated in different nonlinear regimes with ultrashort irradiation of dielectrics such as sapphire ($E_{g}=8.8$~eV) and soda-lime glass ($E_{g}=3.8$~eV) at $\lambda=1550$~nm (photon energy $hc/\lambda=0.8$~eV) and other wavelengths.\\

\begin{figure}
    \centering
    \includegraphics[width=1\linewidth]{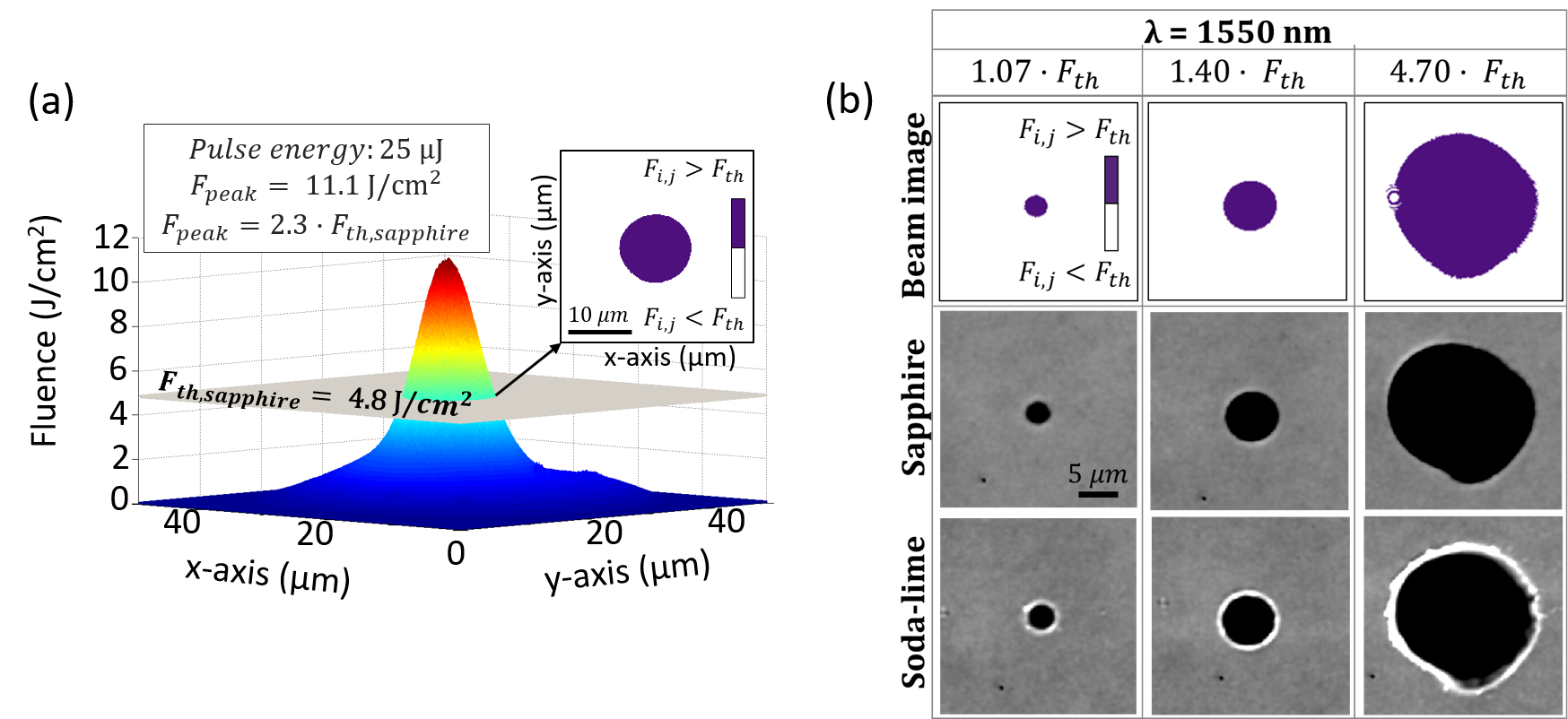}
    \caption{(a) Fluence distribution at the sample position for a laser beam at $\lambda=1550$~nm as obtained using a calibrated infrared microscopy system (pulse duration: 200~fs). The inset displays the beam portion exhibiting values above the ablation fluence threshold of the considered material. (b) (first row) Beam portions above the ablation fluence threshold for different excitation levels above ($F_{\text{peak}}/F_{\text{th}}$). (second and third rows) Confocal microscopy images of surface modifications on sapphire and soda lime induced at the indicated peak fluence above the ablation fluence threshold. Surface depressions and elevations are represented with darker and brighter colors, respectively. \textit{Adapted with permission from Ref.~\cite{Garcia-Lechuga2020}. © Optical Society of America}}
    \label{fig:beamspot-and-crater}
\end{figure}

The experimental diagnostic is based on ``thresholding'' the laser beam profile, as sketched in Fig.~\ref{fig:beamspot-and-crater}(a). The first step is to characterize the laser beam profile at the focus using an imaging system consisting of a microscope objective and an IR camera. The second step involves scaling the recorded image to the corresponding pulse energy under consideration. In this case, it is scaled to a pulse energy of 25~$\mu$J, which, given the spatial distribution of this profile, corresponds to a peak fluence of 11.1~J/cm$^2$. The third and final step is to extract the portion of the beam showing fluence values above the ablation fluence threshold. In the example of Fig.~\ref{fig:beamspot-and-crater}(a), the peak fluence value corresponds to $2.3 \times$ the fluence threshold for sapphire, with the corresponding portion of the beam in the inset.\\

In Fig.~\ref{fig:beamspot-and-crater}(b) (first row), three different beam portions extracted using peak fluence values slightly above ($1.07\times$ and $1.40\times$) and significantly above ($4.70\times$) the ablation fluence threshold are shown. As expected, for increased peak fluence, the portion of the beam with values above the fluence threshold becomes larger. When performing irradiations at these same excitation levels in sapphire ($K=11$) and soda-lime glass ($K=4$), the resulting ablation features are shown in Fig.~\ref{fig:beamspot-and-crater}(b) (second and third row). The comparison with the top-row images in Fig.~\ref{fig:beamspot-and-crater}(b) shows an excellent correlation between the shapes of the crater contours and the spatial beam distribution. This observation is a clear evidence that the ablated areas consistently correspond to precise imprints of the measured beam profile at a specific threshold. This demonstrates the absence of nonlinear resolution in femtosecond laser ablation, resolving the contradiction of associating nonlinearity and intensity-threshold responses with resolution enhancement. This same observation and conclusion were further confirmed through experiments in dielectrics, semiconductors, and metals, along with numerical deductions \cite{Zhou2021}.\\

\begin{figure}
    \centering
    \includegraphics[width=1\linewidth]{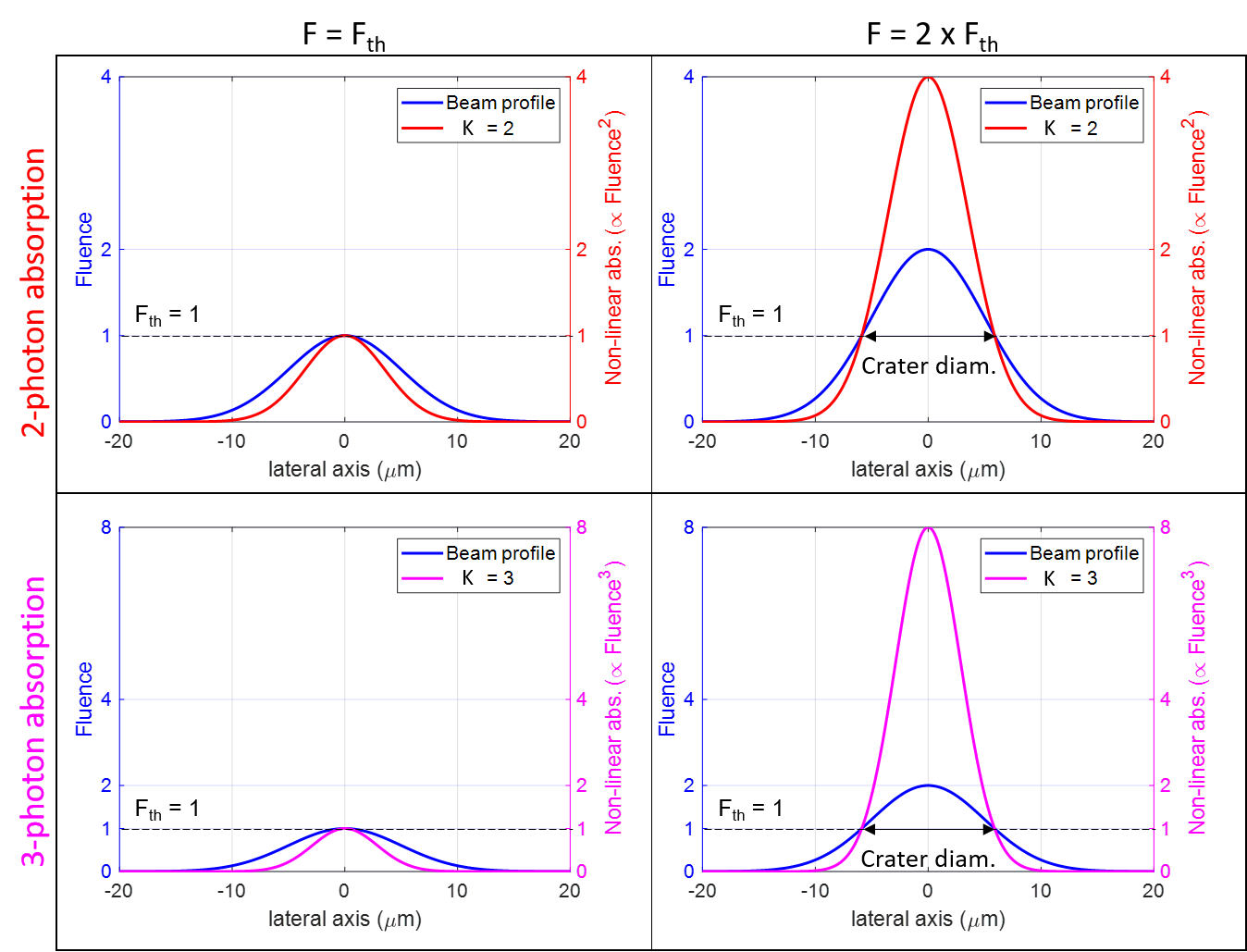}
    \caption{Correlation between the laser beam profile and the nonlinear absorption profile, representing two nonlinear conditions (2- and 3-photon absorption) and two levels of excitation (equal to and twice the fluence threshold for ablation). \emph{© M.~Garcia-Lechuga}}
    \label{fig:non_linearity_correct}
\end{figure}

This misconception has persisted partly because nonlinear energy deposition is often compared to excitation in multi-photon microscopy. However, the key difference is in the response. In microscopy, fluorescence gradually follows nonlinear absorption. In material processing, the response is threshold-based, meaning that the result depends only on whether the fluence is above or below the threshold, leading to no resolution enhancement from nonlinear absorption. This concept is illustrated in Fig.~\ref{fig:non_linearity_correct} for the specific cases of 2- and 3-photon nonlinear absorption ($K=2$ and $K=3$), in which, as previously shown in the inset of Fig.~\ref{fig:resolution}, the nonlinear absorption profiles are indeed narrower than the beam profile by a factor of $1/\sqrt{K}$. In the case of  $ F = F_{\text{th}}$, this narrowing of the absorption profile has no impact on the material transformation (at least under single-shot irradiation), since threshold (represented with a dashed line) is not surpassed. For the case of  $F = 2 \times F_{\text{th}}$, it can be observed that nonlinear absorption profile  scales quadratically for $K=2$ (absorption peak proportional to 4) and cubically for $K=3$ (absorption peak proportional to 8). It is under this representation that one can observe that the size of the crater diameter is predicted either by ``thresholding'' the beam profile or the absorption profile, without any difference in size due to the narrower absorption profile resulting from nonlinear absorption. Nonetheless, this steeper nonlinear absorption should impact the modification profile (e.g., crater depth, marking contrast), which will be discussed in Sect.~\ref{subsec:ablationdepth}.\\

To conclude this Section and clearly answer the question raised in the beginning, following a trend similar as extreme ultraviolet lithography for integrated circuit manufacturing, the use of tightly focused UV ultrashort pulses appears to remain the most favorable option to scale down the ablation resolution. While we believe this aspect should not be neglected for future technology developments, it is important also to mention here that these conclusions are based on single-shot ablation studies. More work needs to be performed to study how this translates for multi-shot experiments accounting for incubation effects.  

\subsection{Precision Depth Control in Ultrashort Laser Ablation}
\label{subsec:ablationdepth}

Studies on the role of wavelength in ultrashort laser ablation typically focus on determining the fluence threshold for ablation, as this experimental parameter indirectly indicates how efficiently energy is absorbed. On this topic, it should be highlighted the works reporting the fluence ablation threshold in ZnS, ZnSe, LiF, CaF$_{2}$, BaF$_{2}$ and MgF$_{2}$ under ultrashort pulse irradiation in the visible and in the mid-infrared range \cite{Simanovskii2003}, in fused silica and CaF$_{2}$ for $\lambda = 267$--2200~nm \cite{Jia2006}, in various bandgap materials with $E_{g} =1$--10~eV for $\lambda = 310$--1030~nm \cite{Gallais2015a}, and in fused silica, sapphire and soda-lime glass for $\lambda=258$--3500~nm \cite{Garcia-Lechuga2023}. In addition to these studies on ablation, ionization mechanisms at sub-ablation threshold conditions have been explored in SiO$_2$ and MgF$_2$ as a function of wavelength, ranging from $\lambda = 620$ to 4780~nm \cite{Migal2020plasmathrehold}.\\

Even if a unique picture cannot be drawn due to the different pulse durations and irradiation conditions (some involving multiple pulses), it can still be inferred—at least for large bandgap dielectrics such as fused silica, as studied in \cite{Jia2006, Gallais2015a, Garcia-Lechuga2023}—that the ablation threshold increases from the ultraviolet to the near-infrared range, later exhibiting a wavelength-independent behavior up to the mid-infrared. Most of these studies include numerical modeling based on the Keldysh formalism (strong-field ionization, see Sect.~\ref{subsec:absorption}) and the Drude model (free-carrier absorption \cite{Ashcroft1976}). These models help deepen the understanding of the relative contributions of different ionization processes depending on the irradiation wavelength. For instance, the fluence threshold behavior of fused silica described above is interpreted as a change in the dominant photoionization mechanism---from multi-photon to tunnel ionization---when shifting toward the infrared range. Additionally, modeling of the experimental data reveals that both multi-photon and tunnel ionization are necessary to seed the impact ionization mechanism, which becomes extremely dominant across the entire spectral range, except in the ultraviolet \cite{Garcia-Lechuga2023}.\\

\begin{figure}
    \centering
    \includegraphics[width=1\linewidth]{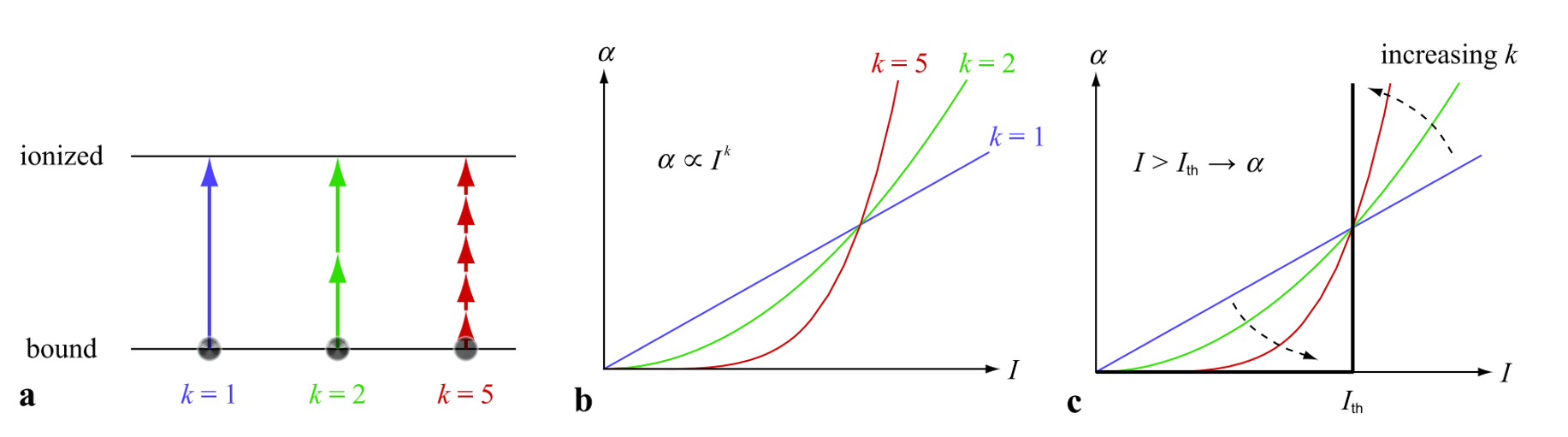}
    \caption{(a) Linear and multi-photon ionization. (b) Absorption $\alpha$ at a given wavelength is proportional to the $k$th power of the intensity, where $k$ is the number of photons absorbed simultaneously. (c) The intensity dependence of the nonlinear absorption approaches a threshold condition as $k$ increases. \emph{Reproduced with permission from Springer Nature from S.~H.~Chung and E.~Mazur, ``Femtosecond laser ablation of neurons in C. elegans for behavioral studies,'' Appl. Phys. A \textbf{96}, 335 (2009) \cite{Chung2009}}}
    \label{fig:Chung-and-Mazur}
\end{figure}

From an engineering perspective, the goal of triggering absorption differently by tuning the laser wavelength is to enhance the performance of the ablation process, not only in terms of ablated volume, but also in the controllability of producing a desired crater depth. Since laser-matter interaction depends on the dynamic evolution of the dielectric function of the excited material, which in turn affects the propagation characteristics of the incoming laser beam, predicting the distribution of deposited energy is of utmost complexity \cite{Christensen2009}. As a result, aprioristic estimation of crater depth variations with wavelength and irradiation fluence represents a considerable physical challenge. However, as highlighted by Chung and Mazur \cite{Chung2009} and shown in Fig.~\ref{fig:Chung-and-Mazur} where linear or multi-photon absorption regimes are considered ($k$ is the number of photons), exploring different spectral ranges is motivated by the potential to induce different absorption profiles (represented as $\alpha$). As depicted in Fig.~\ref{fig:Chung-and-Mazur}(c), knowing that ablation is produced when exceeding a given absorption value (corresponding to the intensity threshold $I_{\text{th}}$), the multi-photon absorption process has a dual impact on the crater formation. First, the energy deposited in the crater periphery (close to but below $I_{\text{th}}$) is reduced compared to linear absorption. Second, the energy deposited when exceeding the threshold becomes more abrupt, which, in turn, affects the final crater shape and maximum depth.\\

Nonetheless, experimental investigations into the potential impact of nonlinearity on the crater depth or shape remain scarce. Most studies have primarily focused on the influence of the pulse duration, which is another factor that alters absorption mechanisms. The effect of pulse duration using different wavelengths on crater depth and ablation efficiency has been explored at $\lambda=800$~nm laser wavelength (pulse duration: 7--300~fs) \cite{Uteza2011fusedsilica}, and at $\lambda=400$~nm (pulse duration: 100 and 600~fs) \cite{Terasawa2020fusedsilica400nm}.\\

\begin{figure}
    \centering
    \includegraphics[width=0.70\linewidth]{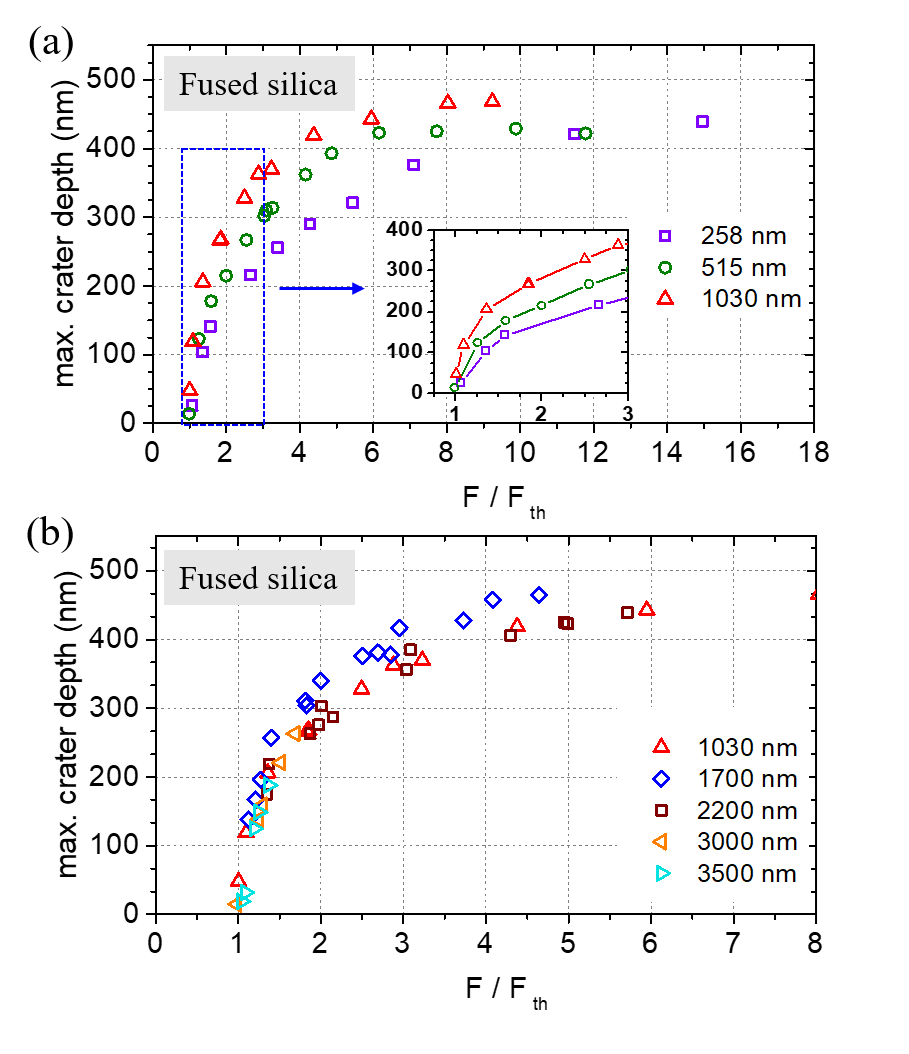}
    \caption{Maximum crater depth produced in fused silica under a single-shot irradiation at different excitation levels (ratio $F/F_{\text{th}}$ between the peak fluence and the fluence threshold). The inset in (a) shows data close to the fluence threshold (pulse duration: 160--200~fs). \emph{Reprinted figure with permission from M.~Garcia-Lechuga et al. ``Wavelength-Independent Performance of Femtosecond Laser Dielectric Ablation Spanning Over Three Octaves,'' Phys. Rev. Applied \textbf{19}, 044047 (2023) \cite{Garcia-Lechuga2023}. Copyright (2023) by the American Physical Society}}
    \label{fig:PRA-mario-2023}
\end{figure}

For a comprehensive study to date on the impact of wavelength on crater shape and depth, one can refer to a recent study \cite{Garcia-Lechuga2023}. In this work, laser ablation in dielectrics is investigated using single-shot irradiation with 200~fs pulses across a wide spectral range, from deep ultraviolet (258~nm) to mid-infrared (3.5~$\mu$m). As shown in Fig.~\ref{fig:PRA-mario-2023} for fused silica, the study demonstrates that despite the different nonlinearities influencing light absorption, the effect on the maximum crater depth as a function of excitation level does not show significant changes across different wavelengths. This is particularly evident in two aspects. First, regardless of the wavelength, a ``saturation'' on the maximum crater depth is reached at $\lambda=450$~nm. Second, for NIR and MIR irradiations [Fig.~\ref{fig:PRA-mario-2023}(b)], the results are nearly identical. This second observation can only be explained if the absorption dynamics are similar, regardless of the wavelength. This can be attributed to the role of tunnel ionization (not only multi-photon ionization) as the initial stage of the absorption process, combined with a predominant contribution from free-carriers absorption in that spectral range \cite{Garcia-Lechuga2023}. Nonetheless, despite small changes, the results shown in Fig.~\ref{fig:PRA-mario-2023}(a) for $\lambda=258$, 515, and 1030~nm (near-UV to near-IR) indicate that the increase in crater depth as a function of irradiation fluence becomes steeper with increasing nonlinearity [see, for instance, the inset in Fig.~\ref{fig:PRA-mario-2023}(a)]. For example, reaching a crater depth of 200~nm requires approximately $ 1.3\times F_{\text{th}}$ at 1030~nm, $2\times F_{\text{th}}$ at 515~nm, and $3\times F_{\text{th}}$ at 258~nm. This observation tends to confirm the predictions in Fig.~\ref{fig:Chung-and-Mazur}. The simplest physical explanation for this behavior could be that, at shorter wavelengths, absorption is initiated by multi-photon ionization, which influences the absorption profile and, consequently, the crater profile.

\subsection{Material Functionalization and Applications}
\label{subsec:functionalization}

In the previous Sections, we have discussed the role of wavelength for laser ablation of dielectrics, impacting both resolution and crater depth. In this subsection, we explore how different wavelengths enable targeted functionalization, enhancing material applications both on the surface and within the volume.\\

Regarding surface transformations, aside from ablation studies on common dielectric materials (e.g., SiO$_2$), the bio-related work on corneal stroma ablation \cite{Olivie2008}, covering a wavelength range from $\lambda = 800$--1400~nm, should be mentioned. Additionally, with the advent of ultrafast mid-infrared laser sources, ablation studies governed by nonlinear absorption in semiconductors (narrow-gap materials) are investigated, trying to understand the differences and similarities compared to dielectrics for linear ($hc/\lambda \geq E_{g} $) or nonlinear absorption ($hc/\lambda < E_{g} $). In Ref.~\cite{Borowiec2005InP}, the ablation threshold of indium phosphide ($E_{g} = 1.34$~eV) has been characterized for $\lambda=400$--2050~nm. The reported values, even though they increase with increasing wavelength, exhibit larger changes in the linear regime than in the nonlinear regime. For germanium ($E_{g} = 0.66$~eV), the ablation fluence threshold for wavelengths ranging from $\lambda=780$--3600~nm is given in Ref.~\cite{Austin2018germanium}. In this work, the fluence values are nearly constant. For silicon ($E_{g} = 1.12$~eV), the ablation fluence threshold is discussed in Ref.~\cite{GarciaLechuga2024LPR} across wavelengths from $\lambda=258$--4000~nm. Similarly to the work on indium phosphide, a moderate increase is observed in the nonlinear regime compared to the linear regime. Finally, the linear regime of interaction (at $\lambda=1030$, 515 and 343~nm) is further explored in Ref.~\cite{Prada-Rodrigo2024} under multi-pulse irradiation. The ablation threshold decreases with the wavelength in silicon. The ablation threshold of germanium at $\lambda=515$~nm is lower than at 1030~nm, it is similar at $\lambda=515$ and 343~nm.\\

\noindent \textbf{Amorphization.} Besides ablation, another transformation of interest is laser-induced amorphization of crystalline materials. On this topic, apart from the range of fluences leading to amorphization of silicon (limited by recrystallization or ablation) as characterized in Ref.~\cite{GarciaLechuga2024LPR} where UV radiation exhibits its highest effectiveness, it is also of interest to determine how to achieve the maximum amorphization depth. On this topic, it has been demonstrated that irradiation at $\lambda=3000$~nm tends to offer proper conditions, together with certain sample characteristics (such as crystal orientation and the thickness of the top SiO$_{2}$ layer), to maximize the amorphized thickness up to 128~nm \cite{GarciaLechuga2021AdvOptMat}. The potential application is that by inducing sufficiently thick amorphization in silicon, it would become possible to create surface waveguides to guide NIR radiation, as already demonstrated in Ref.~\cite{Nejadmalayeri2005}. Other examples are reported in Sect.~\ref{sec:semiconductors}, offering a method for modifying the optical properties of silicon in localized regions for photonic applications.\\

\noindent \textbf{Laser-induced periodic surface structures (LIPSS).} So far, all the mentioned examples are related to direct laser writing (DLW) strategies. However, wavelength also plays a crucial role in other laser surface structuring strategies or techniques, such as direct laser interference patterning (DLIP) and laser-induced periodic surface structures (LIPSS). First, DLIP is based on imprinting the intensity of the electric field formed by the interference pattern created through the overlap of laser beams onto the material surface \cite{Ihlemann2023chapterDLIP, MulkoSolderaLasagni2022}. Complete information and applications of DLIP technique can be found in Chap.~10 (Ancona~\textit{et al.}). The periodicity of the interference pattern ($\Lambda_{\text{DLIP}}$) formed by two laser beams scales with the wavelength as
\begin{equation}
\Lambda_{\text{DLIP}} = \frac{\lambda}{2 \sin(\phi/2)},
\end{equation}
where $\phi$ is the angle between the interfering beams. Therefore, the spatial resolution and dimensions of the resulting micro- and nanostructures highly depend on the experimental configuration (e.g., numerical aperture) \cite{Oliveira2013DLIP, Blumenstein2023DLIP} and the wavelength \cite{Solana2024}, making DLIP an adaptable and precise technique for surface material structuring. \\

\begin{figure}
    \centering
    \includegraphics[width=0.8\linewidth]{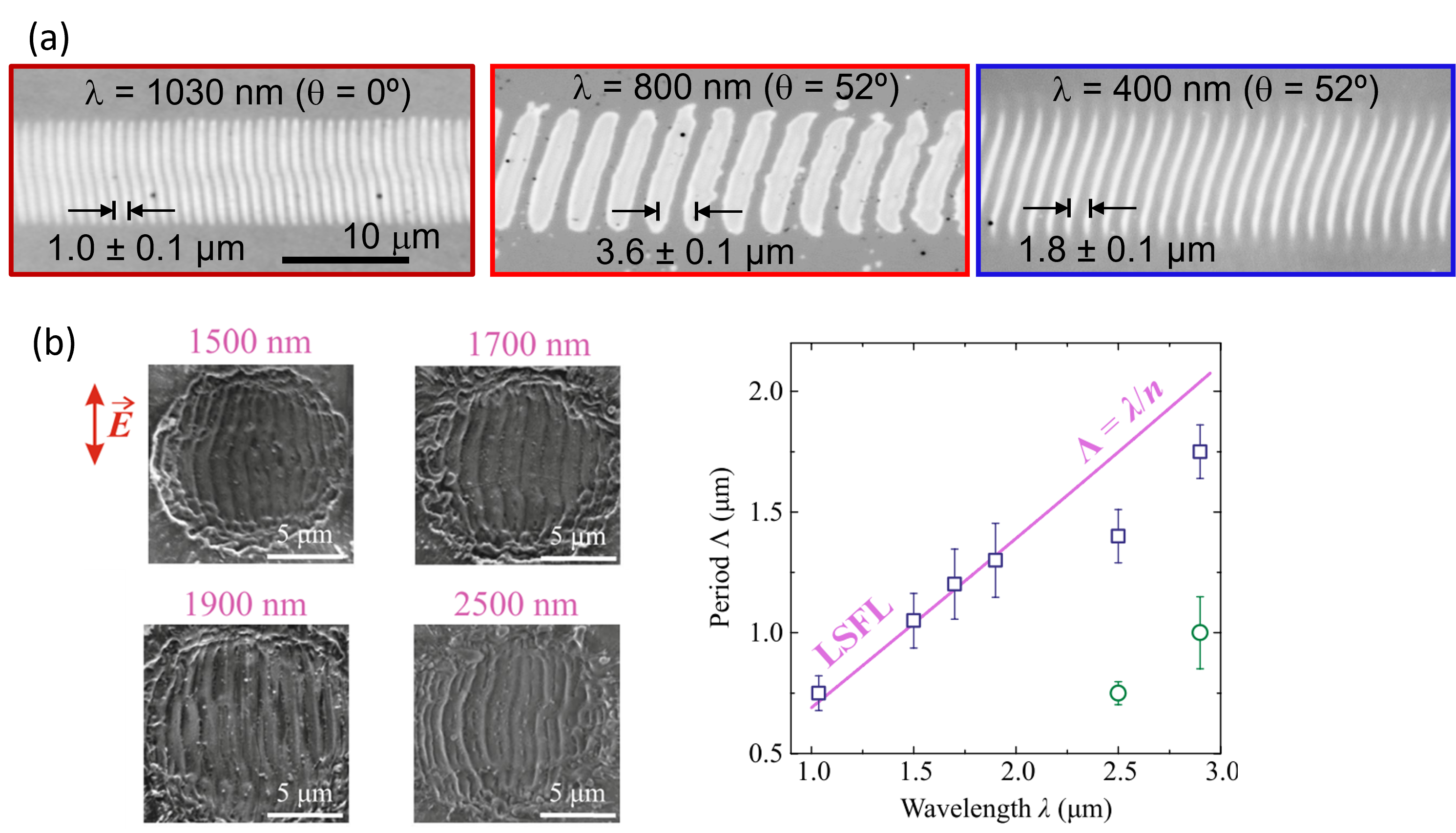}
    \caption{(a) Optical microscopy images of alternate amorphous and crystalline LSFL in silicon produced at different laser wavelengths and angles of incidence. In all three cases the beam is horizontally polarized (pulse duration: 120~fs). \textit{Adapted from D.~Puerto et al. ``Femtosecond laser-controlled self-assembly of amorphous-crystalline nanogratings in silicon,'' Nanotechnology \textbf{27}(26), 265602 (2016) \cite{Puerto2016LIPSS}. © IOP Publishing. Reproduced with permission. All rights reserved.} (b) SEM images and periodicity analysis of LSFL in fused silica produced at different wavelengths (pulse duration: 200~fs). \emph{Adapted from M.~Navickas et al. ``Low spatial frequency laser-induced periodic surface structures in fused silica inscribed by widely tunable femtosecond laser pulses,'' Sci. Rep. \textbf{12}, 20231 (2022) \cite{Navickas2022LIPSSfusedsilica}, Copyright 2022 under CC BY 4.0 (https://creativecommons.org/licenses/by/4.0/)}}
    \label{fig:LIPSSwavelength}
\end{figure}

Second, LIPSS are periodic structures that originate from the interference between the incident laser beam and surface-scattered waves \cite{Bonse2012LIPPSreview}, enabling the production of sub-wavelength structuring without the need for strong focusing conditions. Complete information and applications of LIPSS can be found in Chap.~10. Depending on their periodicity, LIPSS can be categorized into two main types which are low spatial frequency LIPSS (LSFL), and high spatial frequency LIPSS (HSFL). The typical periodicity $\Lambda_{\text{LSFL}}$ of LSFL is close to the laser wavelength. Considering a pure scattering model \cite{Sipe1983}, the periodicity follows
\begin{equation}
\Lambda_{\text{LSFL}} = \frac{\lambda}{1-  \sin(\theta)},
\end{equation}
where $\theta$ is the angle of incidence of the laser beam with respect the sample surface. The influence of these experimental parameters is demonstrated when fabricating amorphous-crystalline LIPSS in silicon \cite{Puerto2016LIPSS,Garcia-Lechuga2017thesis}, as displayed in Fig.~\ref{fig:LIPSSwavelength}(a), where three different wavelengths and two different angles of incidence were used. An example of tuning the periodicity in transparent materials (fused silica) by changing the laser wavelength is reported in Ref.~\cite{Navickas2022LIPSSfusedsilica}, where high nonlinear excitation conditions are explored to produce LIPSS using wavelengths ranging from 1030--2900~nm. Figure~\ref{fig:LIPSSwavelength}(b) shows transformations induced by static irradiations in fused silica at four different wavelengths. The periodicity follows a linear trend ($\Lambda_{\text{LSFL}}=\lambda/n$) up to $\lambda=2000$~nm, beyond which two distinct periodicity regimes start to appear. A proposed explanation for such two-period features is that linear absorption in fused silica is not negligible for $\lambda\approx2500$~nm due to vibrational overtones and combination bands associated with the Si--O bonds in the glass network.\\

Typically, HSFL exhibit periodicities much smaller than the irradiation wavelength. The formation of HSFL is influenced by complex interactions involving higher-order multi-photon absorption processes and localized field enhancements. In dielectrics, the periodicity $\Lambda_{\text{HSFL}}$ for HSFL is often given by
\begin{equation}
\Lambda_{\text{HSFL}} \approx \frac{\lambda}{2n_{0}}.
\label{eq:HSFL}
\end{equation}
As an example, the formation of HSFL under laser irradiation at $\lambda = 800$~nm on TiO$_2$ and Al$_{2}$O$_{3}$ with periodicities of $\Lambda_{\text{HSFL}} =172$~nm and 260~nm, respectively, was shown \cite{Buividas2010HSFL}. These results are consistent with the larger refractive index of TiO$_2$ [$n_{0}(\text{TiO}_{2}) = 2.49$] compared to that of Al$_{2}$O$_{3}$ [$n_{0}(\text{Al}_{2}\text{O}_{3}) = 1.76$]. The universality of this behavior is also confirmed for fused silica [$n_{0}(\text{SiO}_{2}) = 1.45$], with reported periods ranging from 200--400~nm \cite{Rohloff2011}.\\

\begin{figure}
    \centering
    \includegraphics[width=1\linewidth]{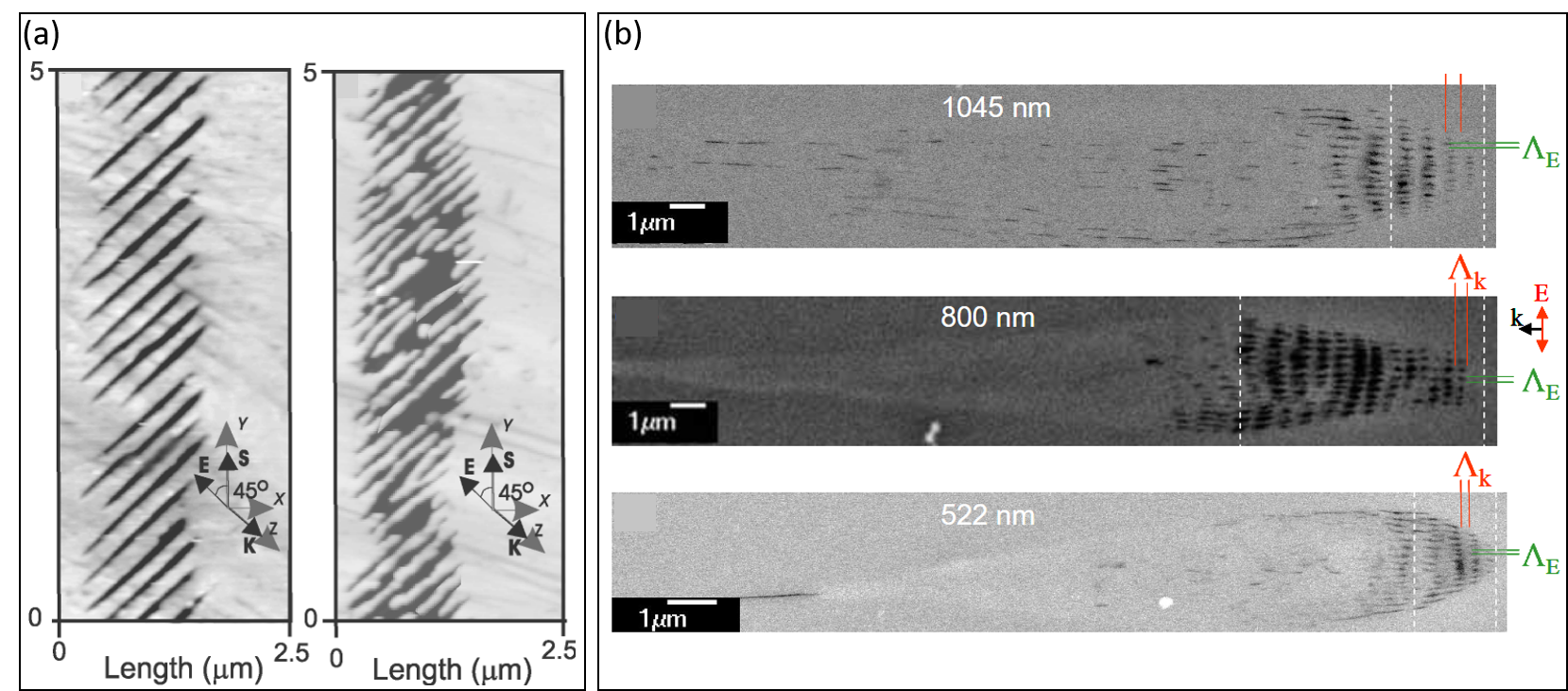}
    \caption{(a) Atomic force microscopy (AFM) images of chemically etched modified regions with laser irradiations at  $\lambda=800$~nm (left) and at $\lambda=400$~nm (right) (pulse duration: 50~fs). \textit{Reprinted figure with permission from V.~R.~Bhardwaj et al. ``Optically Produced Arrays of Planar Nanostructures inside Fused Silica,'' Phys. Rev. Lett. \textbf{96}, 057404 (2006) \cite{Bhardwaj2006periodNG}. Copyright (2006) by the American Physical Society.} (b) SEM images of nanogratings formed at three different wavelengths. Two distinct periodicities are identified ($\Lambda_k$ and $\Lambda_E$). $E$: electric field of the writing laser, $k$: wave vector of the writing laser beam. (pulse duration: 520~fs, 150~fs and 490~fs, respectively). \emph{Adapted with permission from Ref.~\cite{Yang2006NGswavelength}. © Optical Society of America}}
    \label{fig:nanogratings}
\end{figure}

\noindent \textbf{Nanogratings.} Analogous to surface nanostructuring, in-volume sub-wavelength structures can be written in the form of nanogratings \cite{Shimotsuma2003nanograting, Lancry2024reviewNGs}. Nanogratings are important for tailoring optical properties such as birefringence, and add new functionalities of devices such as waveplates and polarization-sensitive components \cite{Beresna2011PolSensitive,Chavilkkadan2025}, as well as 5D optical data storage systems \cite{Zhang2014}. The laser wavelength impacts on the periodicity $\Lambda_{\text{NG}}$ of nanogratings, as demonstrated in Ref.~\cite{Bhardwaj2006periodNG} with femtosecond laser pulses at $\lambda=800$~nm and 400~nm focused inside fused silica [Fig.~\ref{fig:nanogratings}(a)]. The proposed empirical relation between the periodicity of nanogratings and the laser wavelength is  
\begin{equation}
    \Lambda_{\text{NG}} \approx \frac{\lambda}{2n_{0}}.
\label{eq:NGs}
\end{equation}
Two other important works on this topic are Ref.~\cite{Yang2006NGswavelength}, where two periodicities ($\Lambda_k$ and $\Lambda_E$) tunable with the wavelength [Fig.~\ref{fig:nanogratings}(b)] have been identified, and Ref.~\cite{Richter2012NGs}, where it has been observed that, in addition to wavelength, an increase in the number of applied pulses also reduces the periodicity of nanogratings. As a last remark, one may note that Eq.~(\ref{eq:NGs}) closely resembles the periodicity obtained at the surface during HSFL formation [Eq.~(\ref{eq:HSFL})]. A discussion of this common behavior has been reviewed \cite{Buividas2014HSFLandNG}, and a theoretical demonstration of the same electromagnetic behavior is presented in Ref.~\cite{Rudenko2017}. Both phenomena originate from the growth of nanoplasmas seeded by initial material inhomogeneities such as defects, nanopores, or voids. The characteristic periodicity arises from the coherent superposition of near-field scattered waves emitted by these nanoplasmas interfering with the incident laser field, resulting in a spatially modulated energy deposition pattern. The ripple orientation arises because the nanoplasmas are aligned perpendicular to the laser polarization due to local field enhancement.\\

\begin{figure}
    \centering
    \includegraphics[width=0.7\linewidth]{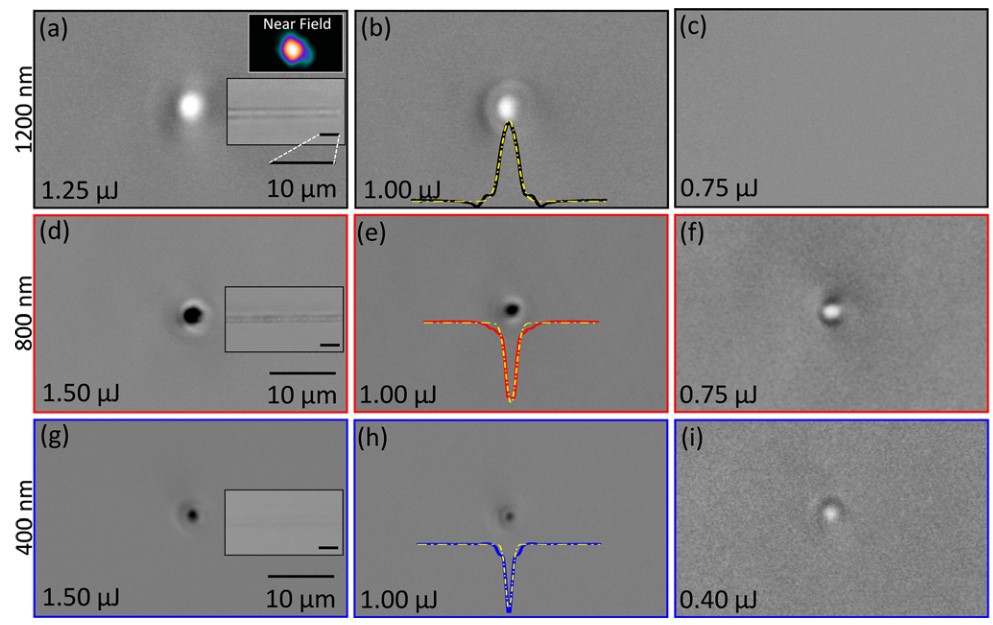}
    \caption{Optical micrographs of the cross section of waveguides inscribed inside fused silica at various wavelengths and pulse energies. The images feature three processing regimes: (c) unmodified material, (a), (b), (f), and (i) smooth optical changes, and (d), (e), (g) and (h) damaged material (pulse duration: 200~fs). \emph{Reprinted from J.~Hernandez-Rueda et al. ``The influence of femtosecond laser wavelength on waveguide fabrication inside fused silica,'' Appl. Phys. Lett. \textbf{110}, 161109 (2017) \cite{Hernandez-Rueda2017waveguideLambda}, with the permission of AIP Publishing}}
    \label{fig:waveguideDielectric}
\end{figure}

\noindent \textbf{Waveguide writing.} Another well-established application of internal modification of dielectrics with femtosecond lasers, since its first demonstration in 1996 \cite{Davis1996waveguide}, is the fabrication of optical waveguides. In this direct laser writing technique, the beam is focused inside the sample, creating a confined region where nonlinear absorption occurs, in turn leading to localized refractive index changes of various natures and origins (e.g., voids, thermal accumulation, ion migration) \cite{Rodenas2009RIchange,Eaton2005,Fernandez2015ionmigration}. Multiple laser pulses are applied while the sample is moved, effectively scanning the laser across the material to create a continuous waveguide. Extensive research has optimized this multi-pulse process by exploring different pulse durations, repetition rates and material compositions \cite{chen2014waveguide}. Obviously, the laser wavelength plays a crucial role in the determination of the excited region, and therefore, the finally formed waveguide. Nonetheless, only few studies have directly explored this topic. Among them, in Ref.~\cite{Hernandez-Rueda2017waveguideLambda}, waveguides in fused silica have been written using laser wavelengths of $\lambda=400$~nm, 800~nm, 1100~nm, and 1200~nm (Fig.~\ref{fig:waveguideDielectric}). The waveguide cross section, material modification threshold, and structural changes strongly depend on the laser wavelength which impacts on the ionization dynamics. Also, in Ref.~\cite{michele2019near}, waveguides in silica glasses have been inscribed at $\lambda=343$~nm and 1030~nm. It is found that UV pulses create waveguides with stronger structural network compaction and higher refractive index changes than IR pulses. Thermal treatments show that the observed differences rely on the generation of distinct laser-induced defects and densification.\\

\noindent \textbf{Photopolymerization.} Finally, another significant 3D modification approach is based on polymerization in a photosensitive material, commonly referred to as two-photon polymerization (TPP) or multi-photon lithography (MPL). As these names suggest, the selected photosensitive materials are transparent at common wavelengths for ultrafast laser sources in the NIR, requiring at least two-photon absorption to initiate polymerization. This technique enables fine-scale printing, exemplified by a bull micro-sculpture in the study by Kawata~\textit{et al.} in 2001 \cite{Kawata2001}. Since that work, numerous 3D-printed structures and applications have emerged, along with extensive research on photosensitive materials and resolution limits \cite{GordonANDFarsari2024}. Regarding the role of laser wavelength in this later aspect, it has often been suggested that increasing the absorption nonlinearity can be beneficial for improving resolution, as it relies on the confinement of the nonlinear absorption profile. However, this question was only recently addressed experimentally, using ultrashort laser pulses for $\lambda=500$--1200~nm \cite{Skliutas2023xphoton}. Analyses of the horizontal and vertical dimensions of the cross-sectional photopolymerized lines lead to the conclusion that, in contrast to the literature, no specific wavelength leading to higher structuring resolution could be identified. This observation can, in fact, serve as another indicator of the absence of nonlinear resolution in intensity-threshold based material responses, as described in Sect.~\ref{subsec:ablationresolution} for ablation \cite{Garcia-Lechuga2020}, and applied here to the photopolymerization process.  

\subsection{Spectrally Shaped Irradiation}
\label{subsec:spectralshaping}

Looking for solutions to improve process performances and, in particular, to solve the contradiction between spatial resolution and energy deposition which is inherent to single-pulse interaction, several groups have developed two-pulse strategies. The idea relies on the conclusion that even for extremely short laser pulses ($<50$~fs), multi-photon ionization serves as initial step, but it is the subsequent interaction with the generated carriers which defines energy deposition \cite{Stuart1996,grojo2010exciton}. Even in these cases where avalanche ionization has no time to develop, free-carrier absorption is expected to represent a major contribution to energy deposition after plasma ignition \cite{shen1989laser}. Therefore, higher control would be obtained if ignition and energy deposition are produced by two separate pulses, i.e., a seed pulse and a driving pulse, respectively.\\

Depending on the pulse-to-pulse delay, seeding can be obtained by free-carriers or any transient intraband states which will make ionization easier compared to ionization starting from the valence band. For instance, in SiO$_2$, it is possible to take advantage of self-trapped excitons (STEs) that form with the decay of free-carriers and exhibit a lifetime exceeding tens of picoseconds at room temperature. For example, a density of $\approx 10^{19}$~cm$^{-3}$ characterizes the initial population of STEs that needs to be prepared to trigger avalanche with 50-fs pulses at $\lambda=800$~nm \cite{grojo2010exciton}. However, with tighter synchronization (delays $<150$~fs) seeding can be ensured by free carriers themselves at much lower densities \cite{peng2013control}. Apart from these experiments concentrating on SiO$_{2}$, other studies have been conducted on doped silicon with various initial free-carrier concentration without the need of a pre-pulse. In such cases, electron densities $>10^{18}$~cm$^{-3}$ are required to measure significant linear absorption by free carriers with a driving pulse of $\approx~100$~fs at $\lambda=1300$~nm \cite{leyder2013non}.\\

By employing two pulses at the same wavelength with different durations, direct control of energy deposition in a volume defined by the seeding pulse interaction was shown \cite{peng2013control}. In particular increased deposited energy that can be achieved in SiO$_{2}$ was demonstrated by increasing the pulse duration of the driving pulse but adjusting its energy so that it is always below the intensity level for significant in-volume absorption alone. In comparison to the energy deposited by a single pulse, an increased energy density of one order of magnitude is demonstrated without sacrificing resolution.\\

\begin{figure}
    \centering
    \includegraphics[width=0.6\linewidth]{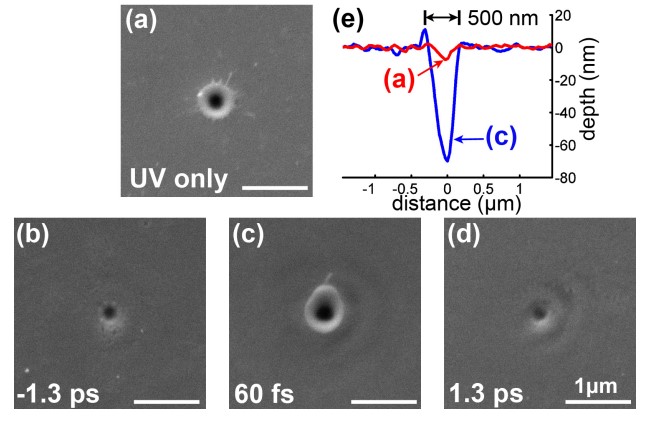}
    \caption{(a--d) SEM images of the damage spots with UV femtosecond laser pulses only ($\lambda=266$~nm), UV-IR double-pulse sequences with $-1.3$~ps (UV first), 60~fs, and $+1.3$~ps pulse-to-pulse delay, respectively ($\lambda=266$ and 800~nm). (e) Crater profiles in (a) and (c) measured with atomic force microscopy (AFM). \emph{Reprinted with permission from Ref.~\cite{yu2014fabricating}. © Optical Society of America}}
    \label{fig:twocolor}
\end{figure}

\noindent \textbf{Multi-wavelength irradiation.} Obviously, when replacing single-pulse processing by an irradiation with two individual components, it is then also natural to examine configurations with two different wavelengths. For instance, by employing two laser pulses, the first in the ultraviolet and the second in the infrared, the processing resolution and efficiency can be optimized as the diffraction limit which defines the smallest achievable interaction volume applies to the seeding pulse and free-carrier absorption is much more efficient in the infrared. In practice, the method does not require complex experimental arrangements as UV and IR pulses can be generated from the same laser (for instance, the fundamental beam and its third harmonic). One can find several demonstrations for surface ablation experiments performed with these two-color schemes \cite{yu2013near, yu2014fabricating, mouskeftaras2013mechanisms, zoppel2005enhancement}. To illustrate the potential benefits from the multi-wavelength approach, Fig.~\ref{fig:twocolor} shows results combining a very large (20~$\mu$m) IR beam at $\lambda=800$~nm and a UV spot size of $\approx 500$~nm \cite{yu2014fabricating}. The applied UV pulse energy ($\lambda=266$~nm, third harmonic) in this experiment was on a sub-nanojoule level revealing that pre-ionization can be accomplished far-below the material modification threshold. By varying the pulse-to-pulse delay, increased precision and change of the ablation rate were noted, as shown in Fig.~\ref{fig:twocolor}. Ultimately, one can extrapolate from this response that irradiation with soft x-rays from high-harmonic generation might be utilized for free-carrier injection to modify surfaces at the nanometer level \cite{grojo2010exciton}.\\

Beside the precision controllability, other works have explored configurations for improving ablation efficiency and throughput. A recent work examined energy deposition in SiO$_2$ and ablation using a combination of two pulses exhibiting different durations and wavelengths (500~fs at $\lambda=515$~nm, and 1~ps or 10~ps at $\lambda=1030$~nm) with various fluences and delays \cite{gaudfrin2022fused}. It is worth also noting a similar study performed on ablation of silicon with the same wavelength combination \cite{bulgakov2024dual}. With experimental results supported by simulations, the work confirmed a best configuration for energy deposition and efficient material removal taking place when the green pulse first generates free electrons, followed by energy deposition in the pre-excited material by the infrared pulse. In the study in SiO$_2$ \cite{gaudfrin2022fused} an optimal cooperative effect was found for a 1-ps delay so that the IR pulse interacts with the material in a state where maximum free-carrier density is generated by the green pulse. This is consistent with a previous study on electron excitation using a femtosecond pump-probe interferometry technique, in which a sequence of two excitation pulses in different dielectrics has been applied \cite{mouskeftaras2013mechanisms}. A first short pulse at $\lambda=400$~nm was followed by a second pulse at $\lambda=800$~nm with variable pulse duration, from 50~fs to 10~ps. In Al$_{2}$O$_{3}$, they show that the total free-carrier density never exceeds the sum of the densities produced by each individual pulse. This means that higher energy is deposited in the material by the second pulse by heating the previously excited carriers, and that no electron multiplication occurs. Interestingly, in SiO$_{2}$, it was possible, under specific conditions, to reveal an increased carrier density due to impact ionization, but not at a level supporting a major role played by electron avalanche to reach material breakdown conditions. For the study on silicon, experiments and simulations based on a two-temperature model confirmed that the short-wavelength pulse is beneficial for plasma generation while the delayed IR pulse is efficiently absorbed by the plasma enhancing energy coupling to the target \cite{bulgakov2024dual}. In addition, it has been interestingly found that at long delays of 30--100~ps, dual-wavelength ablation can be particularly strong with formation of deep smooth craters. This was attributed in this case to a drastic decrease in the surface reflectivity with the expansion of a hot liquid layer produced by the first pulse.\\

All these results surely provide insight into the processes of dual-wavelength laser ablation offering new degrees of control for energy deposition. However, the question of a decisive advantage achievable with two-color schemes remains a matter of debate as the results are not systematically compared to single-pulse irradiation with optimal characteristics (wavelength, duration) depending the target goal (e.g., resolution, ablation efficiency). On this aspect, it is interesting to highlight the comparison between the dual-color sequence and the short-wavelength single-pulse case in Ref.~\cite{gaudfrin2022fused}, leading to the conclusion of no net benefit of the dual-color approach for efficient surface ablation in SiO$_{2}$.\\

\noindent \textbf{Chirped pulses.} Besides irradiation with two pulses in extremely different spectral ranges, similar synergy between different wavelengths may also exist for single-pulse irradiation. Indeed, for systems based on chirped pulse amplification (CPA) which deliver bandwidth-limited pulses, detuning dispersive elements (e.g., gratings) in the optical stretcher or compressor results in a time difference between the shortest and the longest spectral components of the pulse (hereafter referred to as the ``blue'' and the ``red'', respectively). A direct consequence of playing with the chronology of spectral components is an increase of the pulse duration (see Sect.~\ref{subsec:index}). Thus, for a given pulse duration $\tau$, one must distinguish three types of pulses, which are (1) bandwidth-limited pulses (i.e., all spectral components in phase), (2) up-chirped pulses (or positive-chirp pulses, i.e., the red arrives before the blue), and (3) down-chirped pulses (or negative-chirp pulses, i.e., the blue arrives before the red).\\

In strong-field physics (e.g., laser-plasma particle acceleration) with laser intensities $I \ge 10^{16}$~W/cm$^{2}$, the chirp may profoundly impact on the result of the interaction \cite{Chen2008}. This is due to the fact that accounting only for the smooth envelope of the intensity profile is not sufficient, and the modulated magnitude of the electric field which dictates the acceleration and deceleration phases of carriers must be taken into consideration. In contrast, studies on the influence of the chirp in the laser processing regime are rarer. Nevertheless, it has been shown that down-chirped pulses lead to lower surface damage thresholds on the entrance surface of fused silica and MgF$_{2}$ compared to up-chirped pulses \cite{Louzon2005}. These results were obtained with pulse durations in the 60~fs--1~ps range at $\lambda=800$~nm, and were explained by the fact that the blue and red part of the spectrum lead to more efficient multi-photon and avalanche ionization, respectively \cite{Louzon2005,Duchateau2014}. Free electrons are more efficiently produced by the blue part of the spectrum through multi-photon ionization. These free electrons seed the avalanche ionization process which is more efficient for the red part of the spectrum. Interestingly, for narrow-gap materials such as silicon ($E_{g}=1.11$~eV) and gallium arsenide ($E_{g}=1.43$~eV) irradiated at $\lambda=800$~nm (i.e., linear absorption regime), the chirp was found to have only a marginal effect on the production of LIPSS \cite{Sun2022}, or no effect at all \cite{Louzon2005}. In contrast, in the 2-photon absorption regime ($\lambda=1960$~nm), it has been recently shown that the maximum fluence reached in the bulk of silicon was $2.4 \times$ higher for down-chirped pulses than for up-chirped pulses \cite{Chambonneau2026}.\\

\begin{figure}[!ht]
\centering
\includegraphics[scale=0.5]{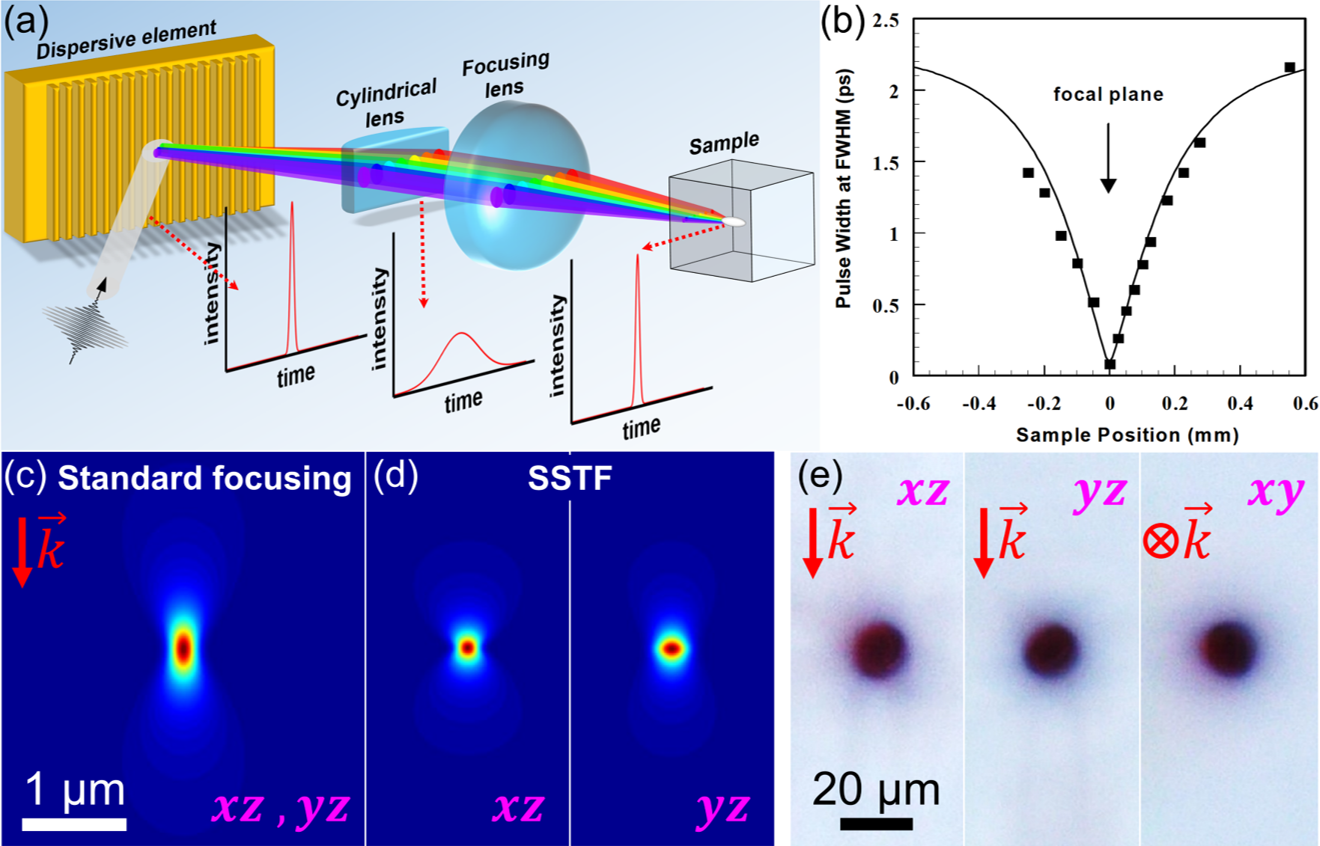}
\caption{(a) Schematic principle of simultaneous spatial and temporal focusing (SSTF). \emph{© M.~Chambonneau.} (b) Evolution of the pulse duration as a function of the sample position. The focal plane corresponds to a sample position of zero. \emph{Adapted with permission from Ref.~\cite{Zhu2005}. © Optical Society of America.} (c), (d) Calculations of the intensity distribution in different planes near the focus for (c) standard focusing, and (d) SSTF. The vector $\vec{k}$ indicates the direction of laser propagation along the $z$ axis, and the spatial scale is identical for (c) and (d). \emph{Adapted from F.~He et al. ``Independent control of aspect ratios in the axial and lateral cross sections of a focal spot for three-dimensional femtosecond laser micromachining,'' New J. Phys. \textbf{13}, 083014 (2011) \cite{He2011}. © Deutsche Physikalische Gesellschaft.  Reproduced by permission of IOP Publishing. CC BY-NC-SA.} (e) Cross sections of lines written in different directions inside glass with the SSTF technique. \emph{Adapted from Y.~Tan et al. ``Three-dimensional isotropic microfabrication in glass using spatiotemporal focusing of high-repetition-rate femtosecond laser pulses,'' Opto-Electron. Adv. \textbf{6}, 230066 (2023) \cite{Tan2023}, Copyright 2023 under CC BY 4.0 (https://creativecommons.org/licenses/by/4.0/)}}
\label{fig:FigSSTF}    
\end{figure}

\noindent \textbf{Simultaneous spatial and temporal focusing.} Spectral structuring can advantageously be utilized for simultaneous spatial and temporal focusing (SSTF). While focusing in general refers to the spatial concentration of light, this concept can also be applied to the time domain, as proposed in the late 1980s \cite{Kolner1989}. Decades later, SSTF was experimentally demonstrated for microscopy applications \cite{Zhu2005}. As illustrated in Fig.~\ref{fig:FigSSTF}(a), the principle of SSTF is to separate spatially all spectral components of an ultrashort laser pulse with a dispersive element (e.g., grating, prism) and recollimate the beam. The spatially chirped pulse is then focused, so that all spectral components spatially coincide solely in the focal region. Temporal focusing is consequently produced concurrently to spatial focusing [Fig.~\ref{fig:FigSSTF}(b)], and the intensity in the prefocal region is much lower than for conventional space-only focusing [Fig.~\ref{fig:FigSSTF}(c) and (d)]. Capitalizing on this feature, it was shown that plasmas produced in water are much more confined using SSTF, even at very high pulse energies where nonlinear propagation effects govern the interaction when space-only focusing is employed \cite{Kammel2014}. To obtain extreme localization of the energy deposition, this technique was also applied to in-volume material processing. It has been shown that SSTF improves the achievable degree of control to produce permanent modifications in glass for applications in microfluidics \cite{He2010,He2011,Tan2023}. This is illustrated in Fig.~\ref{fig:FigSSTF}(e) with the circular cross section of lines written in different directions inside glass with the SSTF technique \cite{Tan2023}. One must highlight the remarkably constant line diameter for all writing directions, in contrast to what is expected from standard Gaussian focusing [Fig.~\ref{fig:FigSSTF}(c)], where transversely written lines are generally elliptical rather than circular.

\section{Internal Processing of Narrow-Gap Materials}
\label{sec:semiconductors}

Besides wide-gap materials such as dielectrics, ultrafast laser processing has been recently extended to the bulk of narrow-gap materials such as semiconductors, which typically exhibit extreme nonlinear optical coefficients. Recent achievements in this emerging field \cite{Chambonneau2021a,Grojo2023} owe their success to the challenging development of stable high-energy ultrafast laser sources emitting at increasingly longer infrared wavelengths. While nonlinear propagation phenomena prevent sufficient energy deposition in narrow-gap materials, multiple techniques have been developed to bypass these limitations and demonstrate the applicability of ultrafast laser processing in semiconductors.

\subsection{The Challenging Development of ``Infraredder'' Laser Sources}
\label{subsec:sources}

A prerequisite for internal ultrafast laser processing is to work at a wavelength in the transparency range of the material. As a first and raw approximation, the material can be considered as transparent when linear absorption is negligible. In other words, nonlinear ionization---especially, multi-photon ionization---should be the dominant initial absorption mechanism. This condition is fulfilled when the bandgap $E_{g}$ exceeds the photon energy, which implies
\begin{equation}
\label{eq:cutofflambda}
  \lambda > \frac{hc}{E_{g}},
\end{equation}
where $h$ is the Planck constant, and $c$ is the speed of light in vacuum. Equation~(\ref{eq:cutofflambda}) highlights that materials exhibiting a bandgap $E_{g} \le 1.2$~eV cannot be processed internally at $\lambda \approx 1$~$\mu$m standardly emitted by Yb- and Nd-based lasers. As a consequence, ``infraredder'' (i.e., longer) wavelengths must be utilized to process narrow-gap materials such as semiconductors (e.g., silicon) or transparent conductive oxides (e.g., indium tin oxide).\\

A considerable challenge lies in the apparent contradiction in terms between ``infraredder'' and ``ultrashort'', which represents a hurdle for the development of laser sources emitting ultrashort pulses in this spectral range. Indeed, the time-bandwidth product reads
\begin{equation}
\label{eq:TBP1}
  \Delta t \Delta \nu \ge P,
\end{equation}
where $\Delta t$ is the pulse duration [full width at half maximum (FWHM)], $\Delta \nu$ is the bandwidth (FWHM), and $P$ is a constant which depends on the temporal pulse shape (e.g., $P=2\ln{2}/\pi \approx 0.441$ and $P \approx 0.315$ for Gaussian and sech$^{2}$ pulses, respectively). Noticing that $\Delta \nu = c\Delta \lambda / \lambda^{2}$, where $\Delta \lambda$ is the spectral width experimentally measured with a spectrometer, and $\lambda$ is the center wavelength, Eq.~(\ref{eq:TBP1}) can be rewritten as
\begin{equation}
\label{eq:TBP2}
  \Delta t \ge \frac{\lambda^{2}}{c \Delta \lambda} P.
\end{equation}

For bandwidth-limited pulses (i.e., all spectral components are in phase) where the left- and right-hand sides of Eq.~(\ref{eq:TBP2}) equalize, the shortest achievable pulse duration $\Delta t$ scales as $\lambda^{2}$. This implies that, in contrast with pulses in the UV and visible, the spectral width $\Delta \lambda$ must be larger to achieve short durations with infraredder pulses. As an example, Gaussian pulses with $\Delta t=100$~fs require a bandwidth of at least $\Delta \lambda \approx 15$~nm at $\lambda=1$~$\mu$m, $\Delta \lambda \approx 59$~nm at $\lambda=2$~$\mu$m, and $\Delta \lambda \approx 368$~nm at $\lambda=5$~$\mu$m. In the latter case, such a large bandwidth may result in strong dispersion during the propagation through the different optics of the setup (see Sect.~\ref{subsec:index}), which in turn changes the pulse duration at the focus. Moreover, the wavelength dependence of absorption (e.g., of water molecules in air) may change the spectral shape along the beam path, which also modifies the temporal pulse shape.\\

Traditionally, infraredder wavelengths are reached with nonlinear optical systems such as optical parametric amplifiers (OPAs), which show advantages including wavelength tunability and high temporal contrast. However, the intrinsic nonlinear nature of frequency conversion in those systems implies that pulse-to-pulse energy stability is more challenging to achieve, and the complexity of the system can represent a major hurdle for industrial applications. For this reason, it is interesting to note that various femtosecond laser systems emitting directly in the short-wavelength infrared (SWIR) or mid-infrared (MIR) have been recently developed. Examples of active elements and the associated emitted wavelength are given in Table~\ref{tab:IRlasers}.\\

\begin{table}[!ht]
\caption{Selected active elements for SWIR and MIR ultrafast laser sources, and the corresponding emitted wavelength. For each laser type, typical pulse durations obtained in different studies are indicated.}
\label{tab:IRlasers}
\begin{tabular}{p{3cm}p{3cm}p{3cm}p{2cm}}

\hline\noalign{\smallskip}

    Active element & Wavelength ($\mu$m) & Pulse duration (fs) & Reference\\
    
\noalign{\smallskip}\svhline\noalign{\smallskip}

    Cr & 1.3 & 14 & \cite{Chudoba2001}\\
    Er & 1.55 & 60 & \cite{Nakazawa1990}\\
    Tm & 1.95 & 10.2 & \cite{Wang2023}\\
    Ho & 2.1 & 88 & \cite{Suzuki2024}\\
    Cr & 2.4 & 69 & \cite{Tolstik2013}\\
    Er & 2.8--3.6 & 160 & \cite{Duval2016}\\
\noalign{\smallskip}\hline\noalign{\smallskip}
\end{tabular}
\end{table}

\subsection{Limitations in Ultrafast Energy Deposition}
\label{subsec:limitations}

\begin{figure}[!ht]
\centering
\includegraphics[scale=0.35]{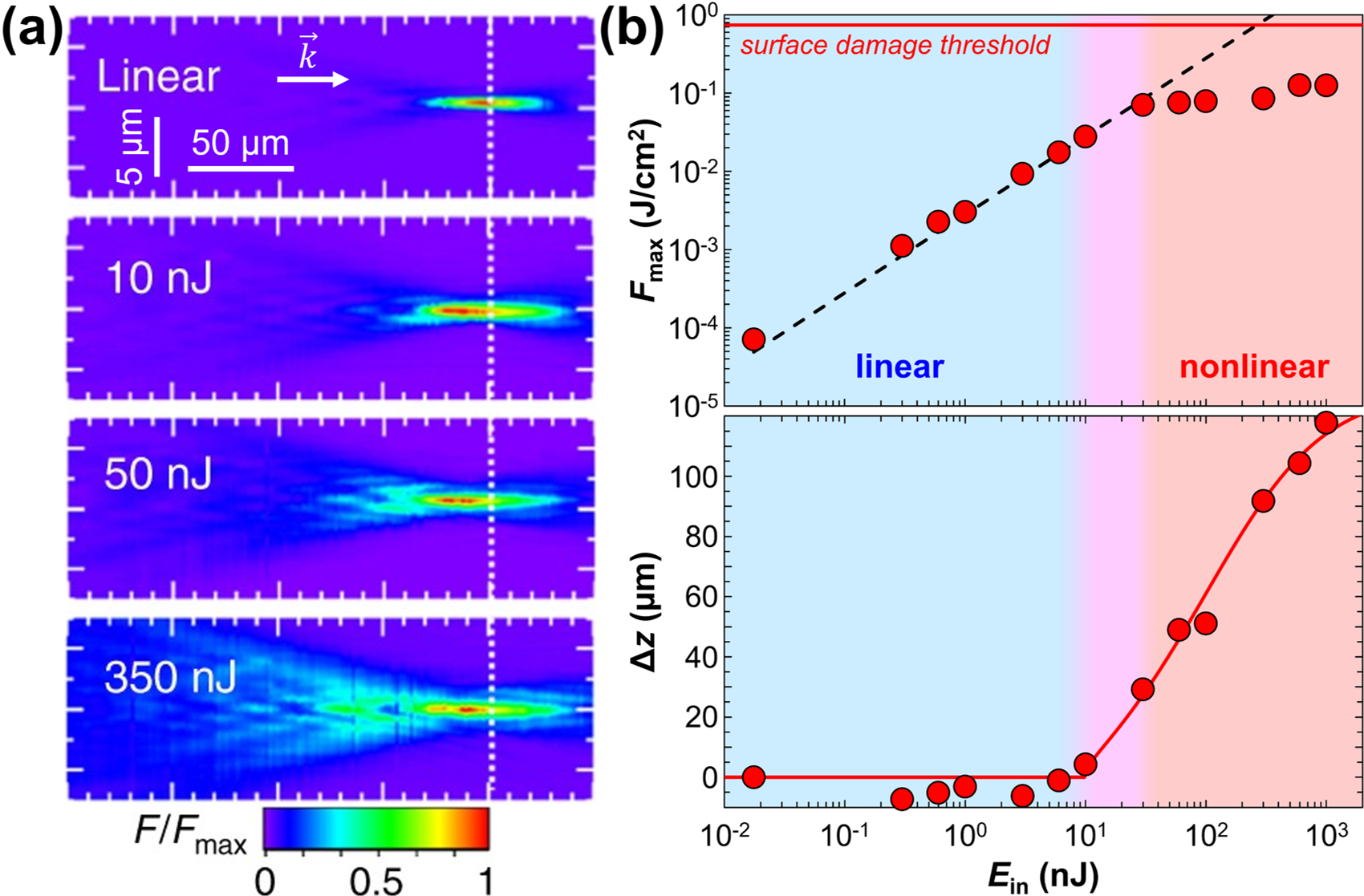}
\caption{(a) Experimentally measured fluence distributions in the bulk of silicon for various input pulse energies at $\lambda=1.3$~$\mu$m (pulse duration: 60~fs). The vector $\vec{k}$ indicates the direction of laser propagation. \emph{Adapted from M.~Chanal et al. ``Crossing the threshold of ultrafast laser writing in bulk silicon,'' Nature Commun. \textbf{8}, 773 (2017) \cite{Chanal2017}, Copyright 2017 under CC BY 4.0 (https://creativecommons.org/licenses/by/4.0/)}. (b) Evolution of the maximum fluence $F_{\text{max}}$ inside silicon and the nonlinear focal shift $\Delta z$ as a function of the input pulse energy $E_{ \text{in}}$ at $\lambda=1.55$~$\mu$m (pulse duration: 9.8~ps). \emph{Adapted from M.~Chambonneau et al. ``Ultrafast Laser Welding of Silicon,'' Adv. Photonics Res. \textbf{4}, 2200300 (2023) \cite{Chambonneau2023}, Copyright 2023 under CC BY 4.0 (https://creativecommons.org/licenses/by/4.0/)}}
\label{fig:FigFilamentation}    
\end{figure}

Narrow-gap materials intrinsically exhibit high linear and nonlinear refractive indices \cite{Sheik-Bahae1990}. As an example, silicon ($E_{g}=1.1$~eV) shows an extreme nonlinear refractive index on the order of $n_{2} \approx 10^{-18}$~m$^{2}$/W \cite{Bristow2007}, which is two orders of magnitude higher than for fused silica ($n_{2} \approx 10^{-20}$~m$^{2}$/W, $E_{g}=9$~eV) \cite{Ensley2019}. A direct consequence is that, for the same laser intensity $I$, nonlinear refraction governed by Kerr effect will be much more pronounced in silicon compared to fused silica. The corresponding instantaneous refractive index change during propagation reads $\Delta n(t) = n_{2}I(t)$. As the intensity is higher for small beams, this refractive index change implies that, above a certain power value called ``critical power'', the laser beam self-collapses during propagation. However, simultaneously, as the beam gets smaller---and thus, the intensity increases---plasma formation governed by nonlinear absorption inevitably occurs. The plasma will also contribute to refraction by defocusing the beam. The extremely nonlinear propagation regime as it is expected in the conditions shown in Fig.~\ref{fig:FigFilamentation}(a) in which Kerr effect and plasma defocusing compete is called ``filamentation'', and has been extensively studied in transparent media \cite{Couairon2007,Berge2007}.\\

Filamentation exhibits countless applications in gases \cite{Kasparian2003,Stelmaszczyk2004,Schimmel2018,Fedorov2020,Fu2022,Houard2023,Goffin2023} as well as in wide-gap materials \cite{Tzortzakis2001,Papazoglou2011,Butkus2014,Xie2015,Garejev2017,Rahnama2020}. However, two main features of this propagation regime represent a major hurdle for internal processing of narrow-gap materials. The first feature called ``intensity clamping'' consists of a saturation of the maximum intensity---and thus, of the energy deposition---with respect to the input pulse energy, as shown in Fig.~\ref{fig:FigFilamentation}(b) \cite{Zavedeev2016,Chanal2017,Mareev2020a,Chambonneau2021}. The second feature---a direct consequence of Kerr effect---is the nonlinear focal shift $\Delta z$ upstream the incoming laser with respect to the geometrical focus as displayed in Fig.~\ref{fig:FigFilamentation}(b)
\cite{Kononenko2016a,Lei2016,Chambonneau2021}. Combined together, these two features lead to the conclusion that there is strong delocalization of the energy deposition towards the prefocal region. The key difference between wide- and narrow-gap solids, is that, in narrow-gap materials, the maximum fluence $F_{\text{max}}$ that can be reached in the bulk of the material saturates below the modification threshold [see Fig.~\ref{fig:FigFilamentation}(b)], while in wide-gap materials, this fluence saturates above the modification threshold. In other words, the energy deposited in narrow-gap materials is not sufficient to cause any permanent modifications with simple irradiation strategies. As demonstrated with pump-probe experiments \cite{Kononenko2012,Mouskeftaras2014,Mouskeftaras2016,Ganguly2024}, only transient underdense plasmas can be produced, which can nevertheless be advantageously used for carrier manipulation in transistors \cite{McMorrow2002,Pellish2008,El-Mamouni2011,Khachatrian2014,Hales2015,Marca2016a,Chambonneau2017a,Hales2017,Hales2018a,Ildefonso2019,Chiquet2019,Hales2019,Hales2020,Ryder2020,Ryder2021,Hales2022,Hales2023}, as well as for spatial \cite{Faccio2008a,Pinhas2018} and spectral \cite{Kadan2018,Marcinkeviciute2019} beam shaping. As a final important remark, it must be noted that filamentation and the associated limitations for internal laser processing have been recently shown to be universal in narrow-gap semiconductors \cite{Chambonneau2026}. Therefore, whereas only studies on silicon (which represent the overwhelming majority of works on internal ultrafast laser--semiconductor interaction) are presented hereafter, one must keep in mind that the main conclusions of these studies can potentially be qualitatively transposed to any other narrow-gap semiconductor.

\subsection{Solutions to Enhance the Interaction}
\label{subsec:solutions}

Various solutions have been developed in the last decade to circumvent the limitations to the energy deposition in narrow-gap materials---mostly silicon---caused by filamentation (see Sect.~\ref{subsec:limitations}). These solutions detailed below have led to internal ultrafast laser processing.\\

\begin{figure}[!ht]
\centering
\includegraphics[scale=0.5]{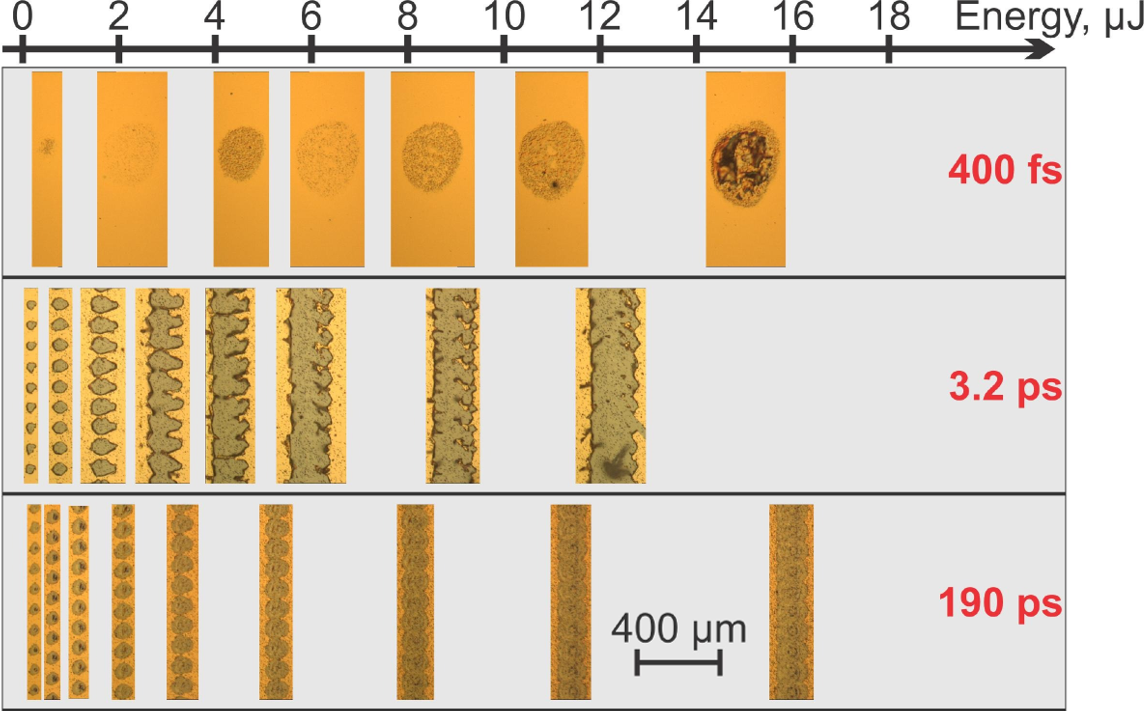}
\caption{Back-surface ablation of Au-coated silicon surface after through-substrate irradiation at $\lambda=2.09$~$\mu$m with various pulse durations and energies. The focusing lens is 6-mm focal length. \emph{Adapted from I.~Astrauskas et al. ``Influence of 2.09-$\mu$m pulse duration on through-silicon laser ablation of thin metal coatings,'' Opt. Laser Technol. \textbf{133}, 106535 (2021) \cite{Astrauskas2021}, Copyright 2020 under CC BY 4.0 (https://creativecommons.org/licenses/by/4.0/)}}
\label{fig:FigBackside}    
\end{figure}

\noindent \textbf{Surface seeds.} A first solution consists of a predesign of the entrance or the exit surface of the silicon sample by incorporating absorbing seeds of various physicochemical nature. Pioneering studies have shown that the deposition of a-SiO$_{2}$ \cite{Nejadmalayeri2005} or a-C \cite{Crawford2008} film at the entrance surface of silicon is beneficial for producing subsurface modifications. The modification depth is on the order of 1~$\mu$m when using 150-fs pulses at $\lambda=800$~nm \cite{Crawford2008}. This very short depth can be explained by high linear absorption of silicon at this wavelength, which is not in the transparency range of the material (cut-off wavelength of 1.1~$\mu$m). To solve this problem, 150-fs pulses at $\lambda=2.4$~$\mu$m  were employed, leading to continuous transverse inscription at depths on the order of 10~$\mu$m \cite{Nejadmalayeri2005}. On the flip side, attempts to modify silicon internally with ultrashort laser pulses starting from the exit surface is a considerable challenge. The nonlinear propagation phenomena in silicon described in Sect.~\ref{subsec:limitations} act as self-protection of the material. Randomly distributed defects may lead to unpredictable modifications as suggested in Ref.~\cite{Luong2020} using 900-fs pulses at $\lambda=1.55$~$\mu$m. As exemplified in Fig.~\ref{fig:FigBackside}, energy deposition on the exit surface of silicon was shown to be enhanced when interfaced with a metal \cite{Lei2016,Chambonneau2021,Astrauskas2021,Chambonneau2023}. This is a consequence of the presence of free electrons which linearly absorb light. To avoid introducing another material, enhanced absorption can also be obtained on the exit surface by rough polishing (e.g., grinding) \cite{Pavlov2017,Matthaus2018,Kammer2019,Alberucci2020}. In this configuration, topographical changes cause electric field enhancement.\\

\begin{figure}[!ht]
\centering
\includegraphics[scale=0.4]{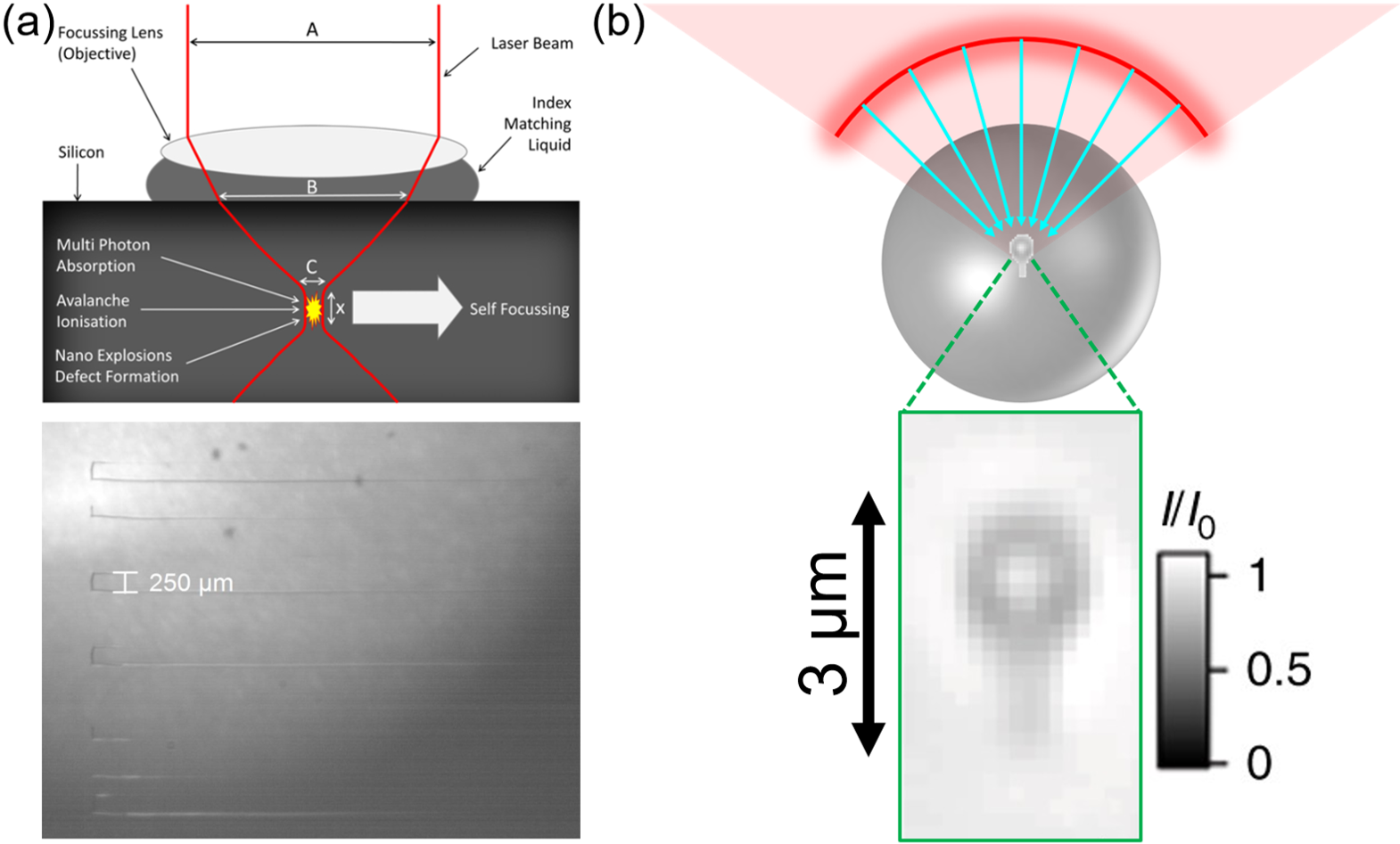}
\caption{(a) Liquid-immersion lens technique and corresponding buried lines transversely written in silicon with 800-fs pulses at $\lambda=1.55$~$\mu$m. The effective numerical aperture is $\text{NA}=1.25$. \emph{Adapted from V.~V.~Parsi~Sreenivas et al. ``Microsized subsurface modification of mono-crystalline silicon via non-linear absorption,'' J. Eur. Opt. Soc.: Rapid Publ. \textbf{7}, 12035 (2012) \cite{ParsiSreenivas2012}, Copyright 2012 under CC BY 4.0 (https://creativecommons.org/licenses/by/4.0/).} (b) Solid-immersion lens technique and corresponding transmission image of the micro-modification inside a silicon sphere after 100 consecutive 60-fs pulses at $\lambda=1.3$~$\mu$m. The effective numerical aperture is $\text{NA}_{\text{eff}}=2.97$. \emph{Adapted from M.~Chanal et al. ``Crossing the threshold of ultrafast laser writing in bulk silicon,'' Nature Commun. \textbf{8}, 773 (2017) \cite{Chanal2017}, Copyright 2017 under CC BY 4.0 (https://creativecommons.org/licenses/by/4.0/)}}
\label{fig:FigImmersionLens}    
\end{figure}

\noindent \textbf{Hyper-NA.} A second solution is to use hyper-NA focusing conditions. The idea is to increase the numerical aperture $\text{NA}$ with liquid- and solid-immersion lens techniques. The wide majority of objective lenses are designed for operation in air environment surrounding the exit pupil, which implies $\text{NA}<1$. In the case of oil-immersion lenses, the higher index of the immersion liquid causes an increase in $\text{NA}$, thus reducing the beam waist $w_{0}$ [Eq.~(\ref{eq:w0})], in turn increasing the intensity at the focus. By using oil-immersion lens with $\text{NA}=1.25$ for focusing 800-fs pulses at $\lambda=1.55$~$\mu$m, as illustrated in Fig.~\ref{fig:FigImmersionLens}(a), it was shown that tracks could be written transversely inside silicon \cite{ParsiSreenivas2012}. The maximum $\text{NA}$ which can be obtained with oil-immersion lens is mainly limited by the refractive index of the employed liquid. Numerical apertures of $\text{NA} \approx 1.6$ can be reached with high-magnification objectives combined with special immersion liquids. However, to obtain higher $\text{NA}$ values, a solid-immersion lens is required. This technique relies on focused pulses inside the solid, with a spherical air--material interface. This interface being perpendicular to all incident rays, these are not deviated due to refraction. This implies that the effective numerical aperture at the focus corresponds to the one of the objective lens multiplied by the refractive index of the material $n_{0}$, i.e., $\text{NA}_{\text{eff}}=n_{0} \times \text{NA}$. For once, the high refractive index of silicon ($n_{0}\approx 3.5$) represents a considerable advantage to reach extreme $\text{NA}_{\text{eff}}$ values. To capitalize on this, 60-fs pulses at $\lambda=1.3$~$\mu$m have been focused in silicon spheres with an effective numerical aperture $\text{NA}_{\text{eff}}=2.97$. As shown in Fig.~\ref{fig:FigImmersionLens}(b), these conditions have been used to produce modifications in this material \cite{Chanal2017}. For a sphere diameter of 1~mm, the effective pulse duration at the center of the sphere is estimated to be increased to 90~fs when accounting for group velocity dispersion (see Sect.~\ref{subsec:index}).\\

\begin{figure}[!ht]
\centering
\includegraphics[scale=0.18]{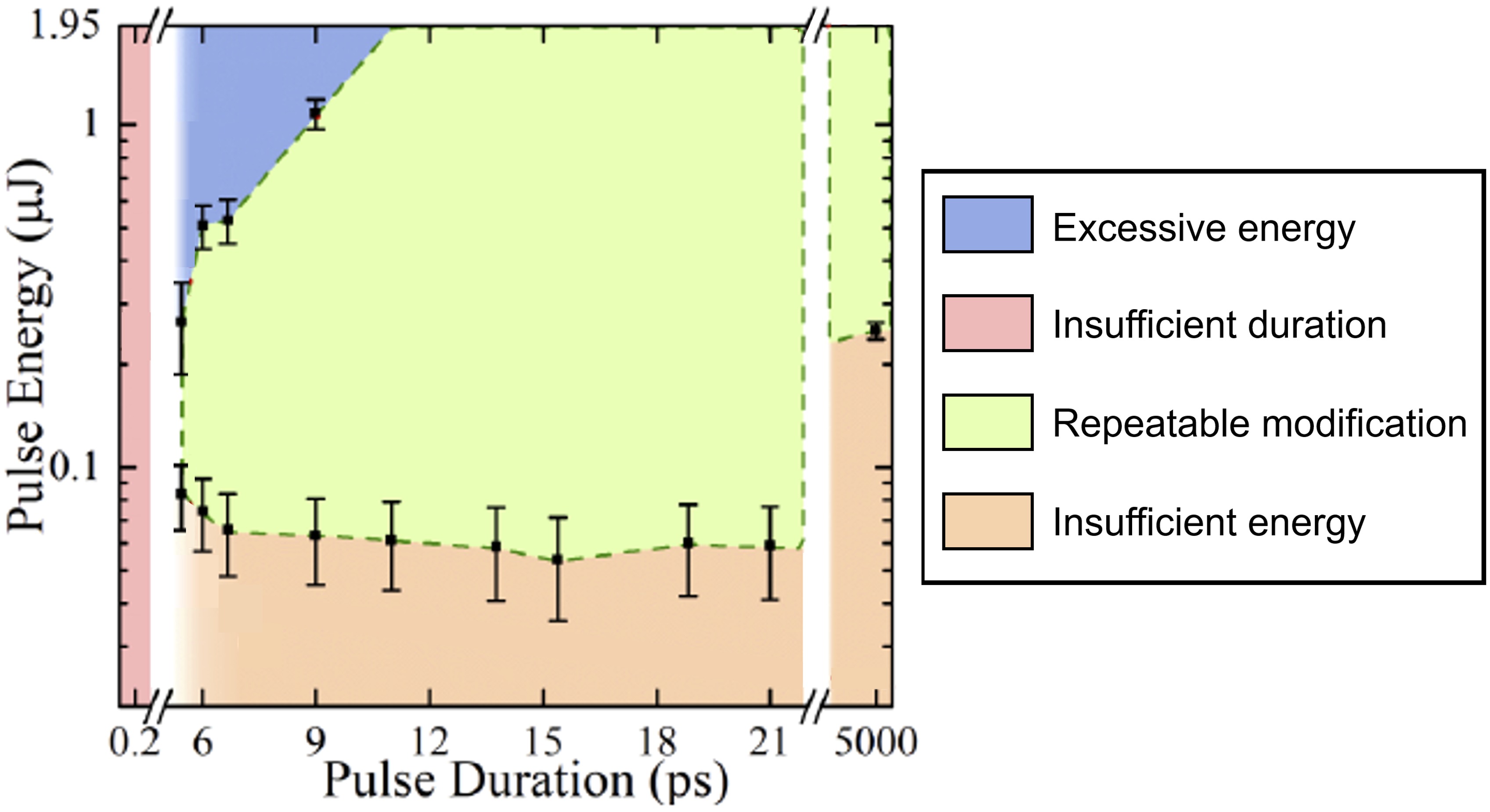}
\caption{Pulse energy and duration ranges to achieve repeatable internal modifications (in green) inside silicon at $\lambda=1.95$~$\mu$m. The blue, red, and orange areas correspond to conditions for which repeatable modifications cannot be produced. \emph{Adapted with permission from Ref.~\cite{Das2020}. © Optical Society of America}}
\label{fig:FigPsRegime}    
\end{figure}

\noindent \textbf{Picosecond regime.} A third solution consists of decreasing the laser intensity using longer pulse durations. In contrast to pulses with durations on the order of 100~fs, pulses in the picosecond regime lead to decreased propagation nonlinearities. Technically, this can be simply achieved by stretching bandwidth-limited femtosecond pulses, i.e., introducing a chirp, so that spectral components arrive at different times (see Sect.~\ref{subsec:spectralshaping}). Several works have examined the pulse duration dependence of internal laser-induced modifications in silicon \cite{Kammer2018,Chambonneau2019a,Das2020}. All concluded that employing longer pulses leads to higher---and, potentially 100$\%$---modification probability inside silicon. This result illustrated in Fig.~\ref{fig:FigPsRegime} is counter-intuitive as, traditionally in dielectrics, the modification threshold is decreased for shorter pulse durations \cite{Stuart1995,Stuart1996,Mero2005}. The surprising opposite result obtained in the bulk of silicon thus evidences the existence of nonlinear propagation effects which are detrimental for energy deposition inside the material (see Sect.~\ref{subsec:limitations}). Using pulses exhibiting durations $>10$~ps, it was shown that these effects could be circumvented, and that modifications could be produced both on the exit surface \cite{Blothe2022}, and also in the bulk \cite{Blothe2024} of silicon, at $\lambda=1.95$ and 1.55~$\mu$m, respectively. The solution based on picosecond pulses can be extended to even-longer pulses, in the nanosecond regime. While this regime is beyond the scope of this Chapter focused on ultrafast processing, one must note that repeatable modifications of diverse natures can be produced with nanosecond pulses, which has led to multiple proofs of concept and applications \cite{Verburg2014,Verburg2015,Chambonneau2016,Li2016a,Yu2016,Mingareev2016,Tokel2017a,Iwata2017,Iwata2018,Saka2018,Chambonneau2018b,Chambonneau2019,Kawaguchi2019,Turnali2019,Wang2019a,Kiyota2020,Wang2021,Trinh2024,Iwata2024,Saka2024,Saltik2024,Tauseef2024,Wang2024,Perez2024,Sopena2024,AsgariSabet2024}.\\

\begin{figure}[!ht]
\centering
\includegraphics[scale=0.29]{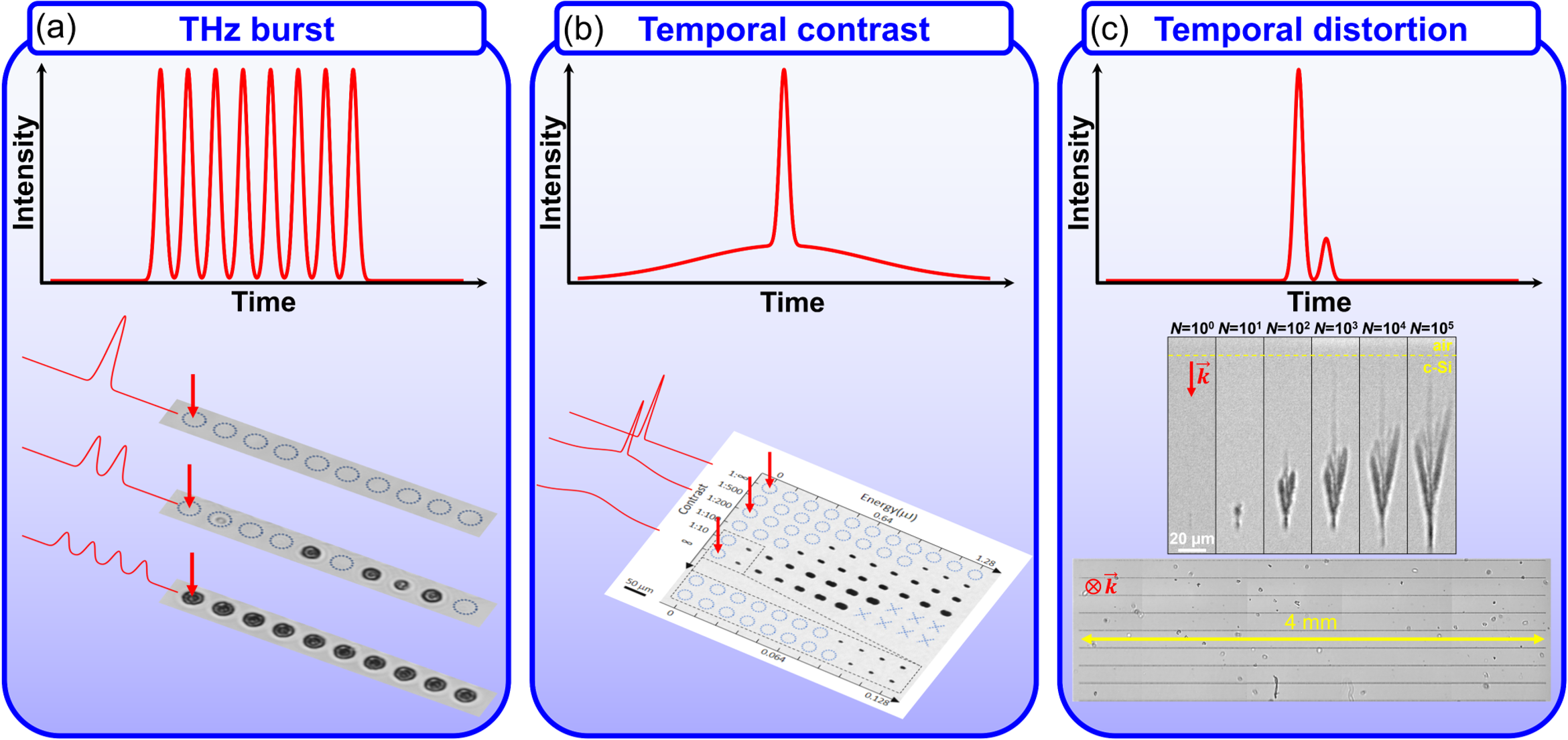}
\caption{Temporal shaping strategies for internal processing of silicon. (a) Transmission microscopy images of internal modifications produced inside silicon with a 0.37-THz burst of 5-ps pulses at $\lambda=1.55$~$\mu$m (pulse-to-pulse delay: 2.7~ps). \emph{Adapted from A.~Wang et al. ``Ultrafast Laser Writing Deep inside Silicon with THz-Repetition-Rate Trains of Pulses,'' Research \textbf{2020}, 8149764 (2020) \cite{Wang2020a}, Copyright 2020 under CC BY 4.0 (https://creativecommons.org/licenses/by/4.0/).} (b) Transmission microscopy images of modifications produced inside silicon with temporally contrasted irradiation consisting of 190-fs pulses combined with 11-ps pulses, both at $\lambda=1.55$~$\mu$m. \emph{Adapted from A.~Wang et al. ``Temporal-contrast imperfections as drivers for ultrafast laser modifications in bulk silicon,'' Phys. Rev. Research \textbf{2}, 033023 (2020) \cite{Wang2020}, Copyright 2020 under CC BY 4.0 (https://creativecommons.org/licenses/by/4.0/).} (c) Transmission microscopy images of the same modification in silicon obtained after applying $N$ repeated irradiation with temporally distorted 750-fs pulses at $\lambda=1.95$~$\mu$m (top), and 4-mm continuous tracks written transversely inside the material (bottom). \emph{Adapted from M.~Chambonneau et al. ``Transverse ultrafast laser inscription in bulk silicon,'' Phys. Rev. Research \textbf{3}, 043037 (2021) \cite{Chambonneau2021b}, Copyright 2021 under CC BY 4.0 (https://creativecommons.org/licenses/by/4.0/)}}
\label{fig:FigPulseTrains}    
\end{figure}

\noindent \textbf{Pulse trains.} With an analogous idea, a fourth solution consists of subdividing a high-energy femtosecond pulse into multiple low-energy pulses. The irradiation thus takes the form of a pulse train called ``burst''. For surface applications such as drilling, milling and cutting, the burst regime shows multiple advantages (see Chapter~18). The asset of this regime is that the hot material produced by one pulse is removed by the next pulse, so that the heat-affected zone is limited and the irradiated area is crack-free \cite{Kerse2016,Balage2023}. For in-volume laser processing where no ablation takes place, the burst regime can advantageously be utilized to cause heat- or carrier-accumulation effects. The prerequisite to obtain the latter is that heat or carriers produced by one pulse should not dissipate before the subsequent pulse arrives, so that the temperature increases on a pulse-to-pulse basis. To examine if cumulative effects occur, the diffusion length can be estimated as $L=\sqrt{2D/\Omega}$, where $D$ is the diffusivity related to the considered process, and $\Omega$ is the repetition rate of the pulse train. Considering $L=1$~$\mu$m, which corresponds to the order of magnitude for the radius of tightly focused beams, $\Omega$ must be $\ge 176$~MHz to obtain heat accumulation (heat diffusion coefficient $D_{\text{heat}}=8.8\times 10^{-5}$~m$^{2}$/s \cite{Shanks1963}), and $\ge 500$~MHz to obtain carrier accumulation (ambipolar diffusion coefficient $D_{\text{ambipolar}}=2.5\times 10^{-4}$~m$^{2}$/s \cite{Mouskeftaras2016}) in silicon. These high repetition rates compared to dielectrics \cite{Eaton2005} show that at least nanosecond pulse-to-pulse delays are required for causing cumulative effects in silicon. In the double-pulse configuration, it was shown that 110-fs pulses at $\lambda=1.24$~$\mu$m with 1--800~ps delays could be used to produce internal modifications in silicon \cite{Mori2015a}. A delay line was used to achieve double-pulse irradiation. However, it is technically challenging to implement and align delay lines to achieve much higher number of pulses ($N$ delay lines correspond to $2^{N}$ pulses), or much lower delays (1~ps corresponds to a movement of 300~$\mu$m on the delay line). Both these issues were solved by using a set of birefringent crystals which are capable of producing THz-bursts [Fig.~\ref{fig:FigPulseTrains}(a)]. It was shown that subdividing a pulse at $\lambda=1.55$~$\mu$m with a duration of 5~ps into closely delayed 4 pulses (equivalent to a less intense quasi-square longer irradiation) leads to repeatable modifications inside silicon \cite{Wang2020a}. An analogous multi-pulse approach consists of adjusting the temporal contrast of 190-fs pulses at $\lambda=1.55$~$\mu$m by a combination with either picosecond or nanosecond pulses \cite{Wang2020}. Compared to high-contrast femtosecond pulses, temporal-contrast imperfections act as seeds to reach the internal modification regime in silicon [Fig.~\ref{fig:FigPulseTrains}(b)]. Another way to obtain pulse trains is to rely on temporal distortion caused by higher-order dispersion due to some mismatch between the optical stretcher and compressor. The resulting distorted temporal profile exhibits satellite peaks. Compared to Gaussian pulses, 750-fs distorted pulses at $\lambda=1.95$~$\mu$m were shown to drastically improve energy deposition inside silicon \cite{Chambonneau2021b}. As a result, repeatable internal modifications could be produced, leading to transverse inscription of continuous tracks in the material [Fig.~\ref{fig:FigPulseTrains}(c)].\\

\begin{figure}[!ht]
\centering
\includegraphics[scale=0.3]{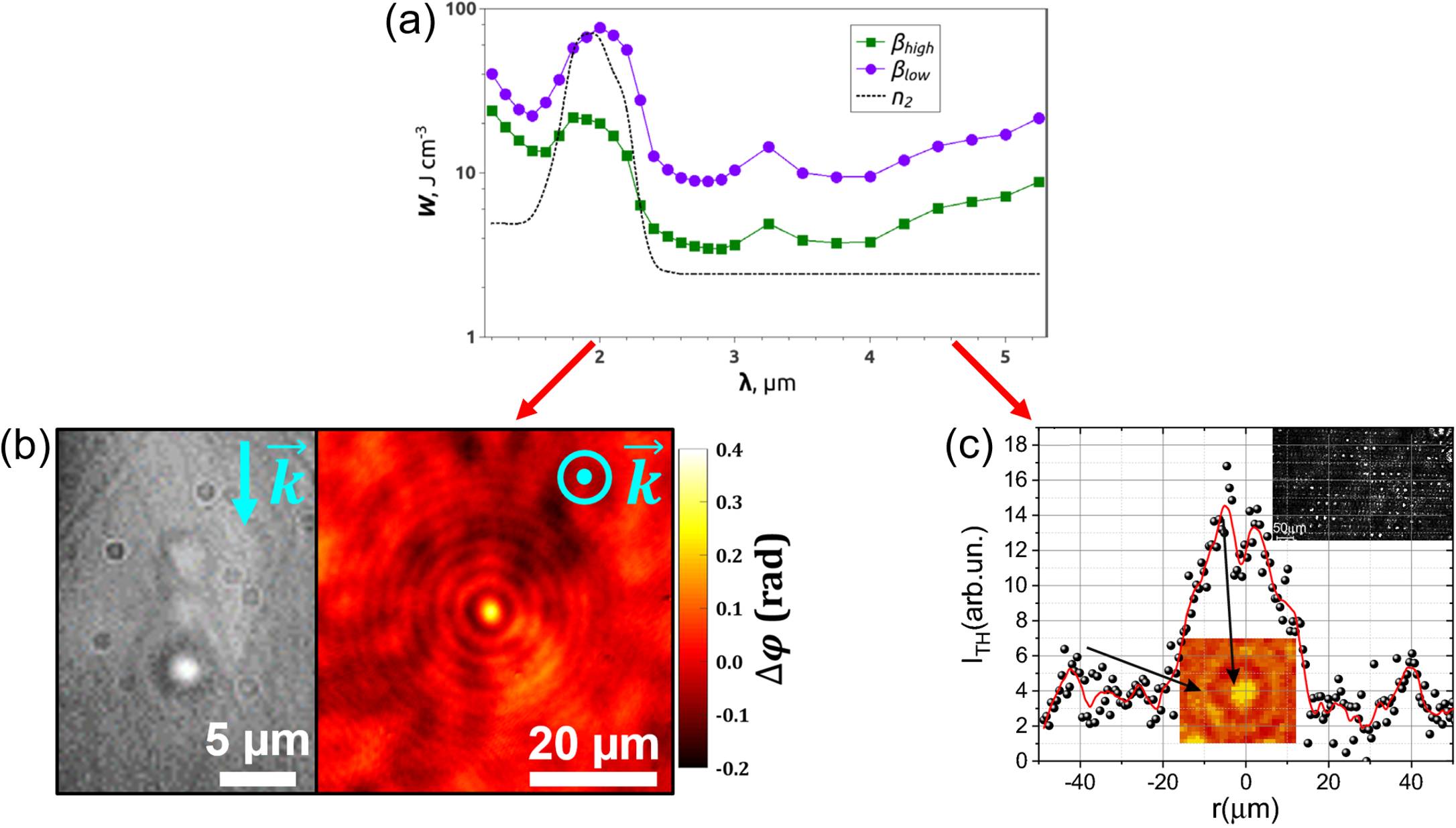}
\caption{(a) Theoretical evolution of the deposited laser energy $W$ as a function of the wavelength $\lambda$ (pulse duration: 250~fs). \emph{Adapted from E.~V.~Zavedeev et al. ``Delocalization of femtosecond laser radiation in crystalline Si in the mid-IR range,'' Laser Phys. \textbf{26}, 016101 (2016) \cite{Zavedeev2016}. ©Astro Ltd. Reproduced by permission of IOP Publishing Ltd. All rights reserved.} (b) Single-pulse modifications produced at $\lambda=1.95$~$\mu$m (pulse duration: 750~fs). The morphology is characterized in amplitude and in phase (left- and right-hand side, respectively). \emph{Adapted from M.~Chambonneau et al. ``Transverse ultrafast laser inscription in bulk silicon,'' Phys. Rev. Research \textbf{3}, 043037 (2021) \cite{Chambonneau2021b}, Copyright 2021 under CC BY 4.0 (https://creativecommons.org/licenses/by/4.0/).} (c) Radial profile of single-pulse modification produced at $\lambda=4.6$~$\mu$m (pulse duration: 160~fs) and observed with third-harmonic imaging. The insets show single micro-modification and a micro-modification array. \emph{Adapted from E.~Mareev et al. ``Single-shot femtosecond bulk micromachining of silicon with mid-IR tightly focused beams,'' Sci. Rep. \textbf{12}, 7517 (2022) \cite{Mareev2022}, Copyright 2022 under CC BY 4.0 (https://creativecommons.org/licenses/by/4.0/)}}
\label{fig:FigOptimizedLambda}    
\end{figure}

\noindent \textbf{Wavelength optimization.} Finally, a fifth solution consists of selecting a wavelength which is favorable for energy deposition inside the material. Theoretical calculations \cite{Zavedeev2016} based on the nonlinear Schr\"{o}dinger equation have predicted higher energy deposition in silicon around $\lambda = 2$~$\mu$m [Fig.~\ref{fig:FigOptimizedLambda}(a)], as a consequence of the existence of a peak in the nonlinear refractive index $n_{2}$ at this wavelength \cite{Bristow2007,Lin2007,Wang2013}. Although this result may seem contradictory to the discussion in Sect.~\ref{subsec:limitations}, one has to keep in mind that, when aiming at the production of internal modifications, sufficient intensity is required. Given the high $n_{2}$ value of silicon for all wavelengths \cite{Bristow2007}, the critical power for self-focusing is exceeded by a large amount in this intensity range. In other words, filamentation governs the interaction in the laser processing regime. Therefore, selecting the wavelength for which $n_{2}$ is the highest may help to partly counterbalance plasma defocusing---and thus, increase energy deposition.\\

This advantageous spectral range around $\lambda = 2$~$\mu$m for internal processing of silicon was verified experimentally by using 750-fs, 2-ps, 9-ps, and 400-ps pulses emitted by a Tm:doped fiber laser ($\lambda = 1.97$~$\mu$m), as illustrated in Fig.~\ref{fig:FigOptimizedLambda}(b) \cite{Chambonneau2019a,Chambonneau2021b}, and confirmed using 5-ps pulses emitted by a Ho:doped fiber laser ($\lambda = 2.09$~$\mu$m) \cite{Richter2020}. Modifications were also produced using 25-ps pulses at $\lambda = 1.95$~$\mu$m \cite{Blothe2022}. Besides the $2$~$\mu$m spectral region which is still in the 2-photon absorption regime (1.1--2.2~$\mu$m), the theoretical predictions in Fig.~\ref{fig:FigOptimizedLambda}(a) suggested that more energy deposition could be achieved inside silicon by using longer wavelengths \cite{Zavedeev2016}. This phenomenon was ascribed to free-carrier absorption, which is all the more efficient at longer wavelengths (see Sect.~\ref{subsec:absorption}). Despite some technical difficulties which inevitably arise in this spectral range where Si-, Ge-, and InGaAs-based sensors are inoperative, it was recently shown that 160-fs pulses at $\lambda = 4.6$~$\mu$m could be used to produce modifications inside silicon (5-photon absorption process), as revealed in Fig.~\ref{fig:FigOptimizedLambda}(c) under third-harmonic imaging \cite{Mareev2022}. While technological developments and experimental efforts are required to determine experimentally the wavelength dependence of in-volume laser-silicon interaction, this first observation of internal modifications with such ultrashort pulses holds a considerable potential. Moreover, the potential of multi-photon absorption orders $N>2$ for modifying semiconductors internally was recently demonstrated in gallium arsenide ($E_{g}=1.43$~eV) as well as indium phosphide ($E_{g}=1.34$~eV), where 3-ps pulses at $\lambda=1555$~nm (2-photon absorption) underperform pulses at $\lambda=1960$~nm (3-photon absorption) in terms of energy deposition \cite{Chambonneau2026}. The measured maximum fluence reached in the bulk of these materials was shown to be more than one order of magnitude higher for $\lambda=1960$~nm compared to $\lambda=1555$~nm.\\

Aiming at industrial applications, all above-detailed solutions show pros and cons, which are summarized in Table~\ref{tab:solutions}. One must note that these solutions are not mutually exclusive, i.e., multiple of these can be combined.

\begin{table}[!ht]
\caption{Overview on how the solutions to bypass nonlinear propagation effects in narrow-gap materials address specific desired processing properties.}
\label{tab:solutions}
\begin{tabular}{p{3.2cm}p{1.6cm}p{1.6cm}p{1.6cm}p{1.6cm}p{1.6cm}}

\hline\noalign{\smallskip}

    Desired processing property & Surface seeds & Hyper-NA & Picosecond pulses & Pulse trains & Wavelength optimization\\
    
\noalign{\smallskip}\svhline\noalign{\smallskip}

    Three-dimensional (3D) & no & no & yes & yes & yes\\
    Limited heat-affected zone & yes & yes & no & no & yes\\
    Contactless & no & no & yes & yes & yes\\
    Monolithic & no & yes & yes & yes & yes\\
    Technologically mature & yes & yes & yes & no & no\\
    
\noalign{\smallskip}\hline\noalign{\smallskip}

\end{tabular}
\end{table}

\subsection{From Fundamental Physics to Processing Applications}
\label{subsec:applications}

\begin{figure}[!ht]
\centering
\includegraphics[scale=0.3]{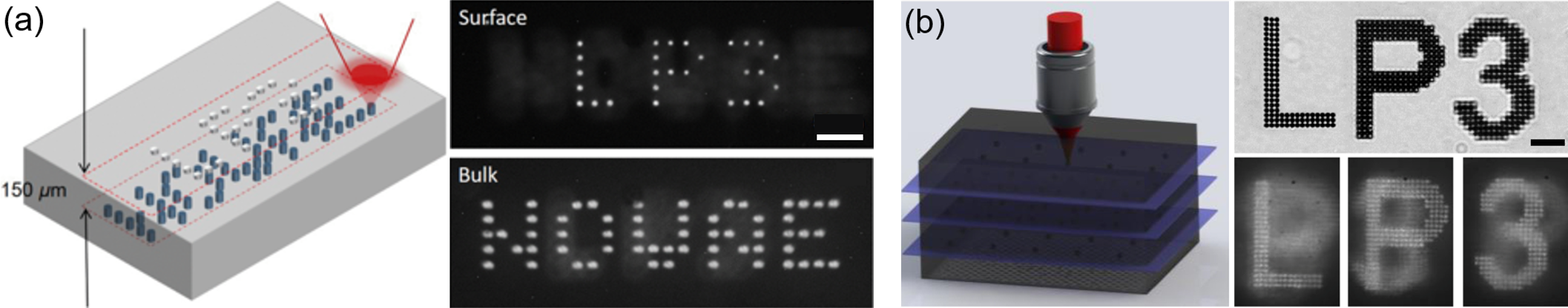}
\caption{Data storage inside silicon. (a) Infrared microscopy images of modifications written on the surface and in the bulk of silicon with 2-ps pulses at $\lambda=1.95$~$\mu$m. \emph{Adapted from M.~Chambonneau et al. ``Competing Nonlinear Delocalization of Light for Laser Inscription Inside Silicon with a 2-$\mu$m Picosecond Laser,'' Phys. Rev. Applied \textbf{12}, 024009 (2019) \cite{Chambonneau2019a}, Copyright 2019 under CC BY 4.0 (https://creativecommons.org/licenses/by/4.0/).} (b) Infrared microscopy images of modifications written at three distinct planes inside silicon (100~$\mu$m, 300~$\mu$m, and 500~$\mu$m under the front surface) with a burst of 4 pulses at $\lambda=1.55$~$\mu$m. In these experiments, the applied train is composed with 4.7-ps pulses separated by a pulse-to-pulse delay of 2.7~ps. The scale bar is 50~$\mu$m. \emph{Adapted from A.~Wang et al. ``Ultrafast Laser Writing Deep inside Silicon with THz-Repetition-Rate Trains of Pulses,'' Research \textbf{2020}, 8149764 (2020) \cite{Wang2020a}, Copyright 2020 under CC BY 4.0 (https://creativecommons.org/licenses/by/4.0/)}}
\label{fig:FigDataStorage}    
\end{figure}

\noindent \textbf{Data storage.} Employing the solutions presented in Sect.~\ref{subsec:solutions}, numerous in-volume and back-surface applications have been developed. A first in-volume application is data storage in silicon. This is directly inspired by achievements in glass \cite{Royon2010,Zhang2014} which have led to the advent of 5D optical storage of $>10$ terabytes of data in glass disks \cite{Wang2022}. The five dimensions of this technique include three in space, one in the size of the modifications in glass, and one as the birefringence. In a less sophisticated form, it was demonstrated that binary data could also be stored in a single plane in the bulk of silicon \cite{Chambonneau2019a}. This demonstration shown in Fig.~\ref{fig:FigDataStorage}(a) relying on picosecond laser pulses was also used to discriminate bulk and surface modifications. This paved the way to data storage in silicon in multiple planes separated by 200-$\mu$m with THz-burst irradiations, as shown in Fig.~\ref{fig:FigDataStorage}(b) \cite{Wang2020a}. More recently a double-pulse method to enhance even further the controllability of the data inscription process was introduced. Interestingly, with an appropriate ultrafast component to adjust the plasma state and a synchronized picosecond laser writing radiation, extremely confined energy deposition has been achieved to write structures with unprecedented resolution \cite{Wang2025}. With the approach, drastic improvement is also found on the material change controllability leading to the capacity to rely on a mechanism of erasure by thermal annealing with nanosecond laser irradiation. Exploiting this property, a set of QR codes was repeatedly written and erased, supporting the vision of rewritable data storage in silicon. Based on the same principle, rewritable and reconfigurable photonics components have been demonstrated.\\

\begin{figure}[!ht]
\centering
\includegraphics[scale=0.4]{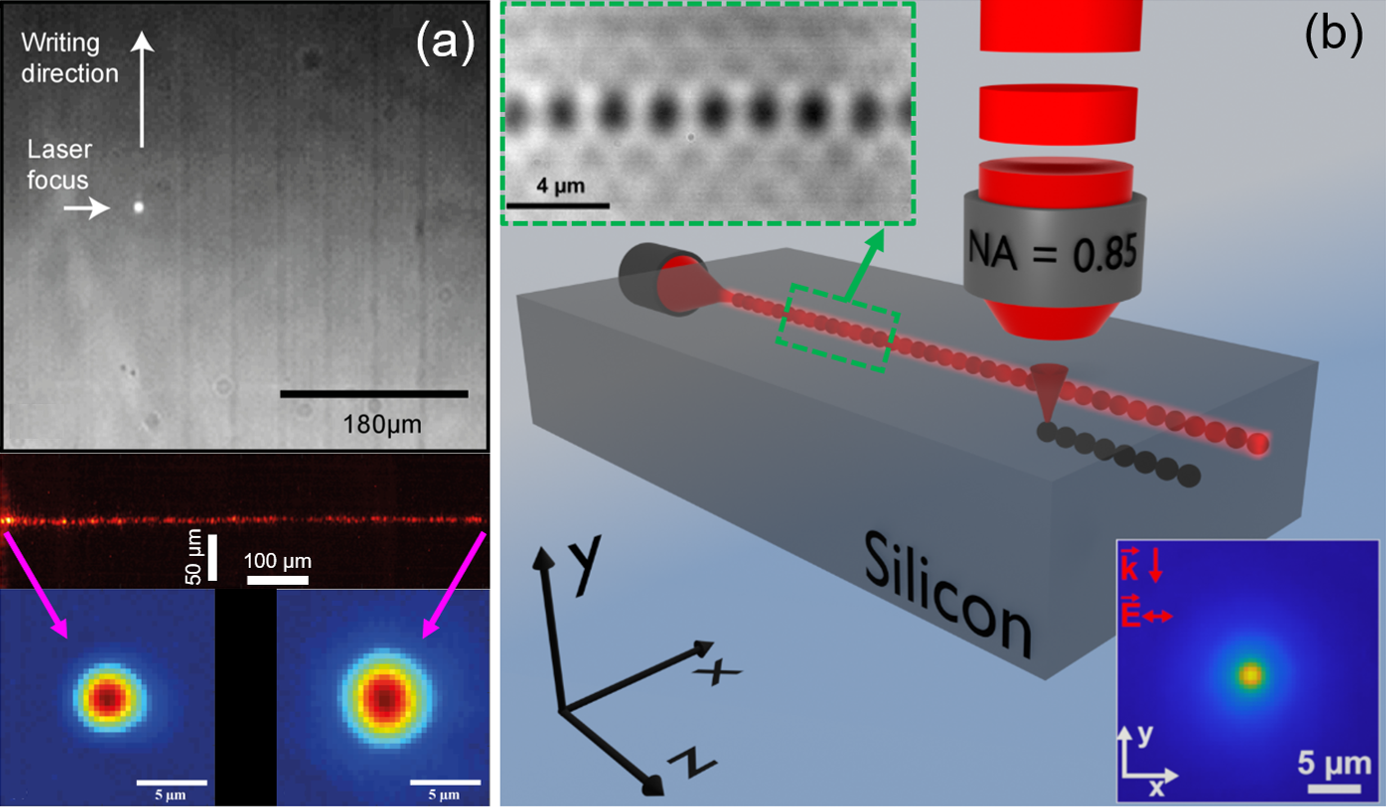}
\caption{Waveguide inscription in the bulk of silicon. (a) In-situ monitoring of exit-surface seeded waveguide longitudinally written inside silicon with 860-fs pulses at $\lambda=1.55$~$\mu$m. The bottom image shows scattering after injection of continuous light, and the insets are near-field images at the input and output of the waveguide. \emph{Adapted with permission from Ref.~\cite{Matthaus2018}. © Optical Society of America.} (b) Segmented waveguides written transversely inside silicon with 43-ps pulses at $\lambda=1.55$~$\mu$m. The insets show an infrared shadowgraphy image of the waveguide (top) and near-field image at the output of a 5-mm long waveguide (bottom). \emph{Adapted from M.~Blothe et al. ``Transverse Inscription of Silicon Waveguides by Picosecond Laser Pulses,'' Laser Photon. Rev. \textbf{18}, 2400535 (2024) \cite{Blothe2024}, Copyright 2024 under CC BY 4.0 (https://creativecommons.org/licenses/by/4.0/)}}
\label{fig:FigWaveguide}    
\end{figure}

\noindent \textbf{Waveguide writing.} A second in-volume application is optical functionalization, mainly for writing waveguide. One must distinguish two inscription techniques, which are longitudinal and transverse writing, along and perpendicular to the optical axis, respectively. Surface-seeded inscription of waveguides in silicon was demonstrated with 350-fs pulses \cite{Pavlov2017}, as well as 860-fs pulses \cite{Matthaus2018,Alberucci2020}, both at $\lambda=1.55$~$\mu$m [Fig.~\ref{fig:FigWaveguide}(a)]. These waveguides rely on a positive refractive index change $6 \times 10^{-4} \le \Delta n \le 4 \times 10^{-3}$ resulting from a disturbed crystal structure \cite{Kammer2019}. The main advantage of longitudinal writing with centrosymmetric beams is that it intrinsically leads to cylindrically symmetric waveguides. However, in this configuration, the maximum waveguide length is limited by the working distance of the focusing optics. Moreover, focusing deep inside the material may lead to prominent spherical aberration (see Sect.~\ref{subsec:focus}).\\

In contrast, waveguides inscribed transversely are not limited in length. In a pioneering study, subsurface waveguides were written in silicon through a 20-$\mu$m layer of amorphous silica \cite{Nejadmalayeri2005}. This achievement relies on 150-fs pulses at $\lambda=2.4$~$\mu$m. More recently, transverse waveguide inscription has been achieved deep inside monolithic silicon with 43-ps pulses at $\lambda=1.55$~$\mu$m \cite{Blothe2024}. This technique illustrated in Fig.~\ref{fig:FigWaveguide}(b) is based on the repetitive single-pulse production of individual spheres exhibiting a diameter of 2~$\mu$m. This discontinuous track of micro-modifications guides light when the effective average refractive index change along the track is sufficient (maximum effective refractive index change of $\Delta n=7\times10^{-3}$). One remarkable aspect of the waveguides shown in Fig.~\ref{fig:FigWaveguide}(b) is the mode field diameter of $\approx 2$~$\mu$m (FWHM). This is a direct consequence of repeated single-pulse approach instead of multiple pulses per point, and this is to date the smallest diameter among laser-written waveguides inside silicon \cite{Chambonneau2016,Pavlov2017,Tokel2017a,Matthaus2018,Kammer2019,Turnali2019,Wang2019a}. This small waveguide diameter represents an important step for envisioning inscription of high-compactness photonic integrated circuits inside silicon. While all the aforementioned waveguide writing techniques currently exhibit high losses (typically on the order of a few dB/mm), one must keep in mind that these achievements are relatively recent and efforts should be put to further reduce losses.\\

\begin{figure}[htbp]
\centering
\includegraphics[scale=0.35]{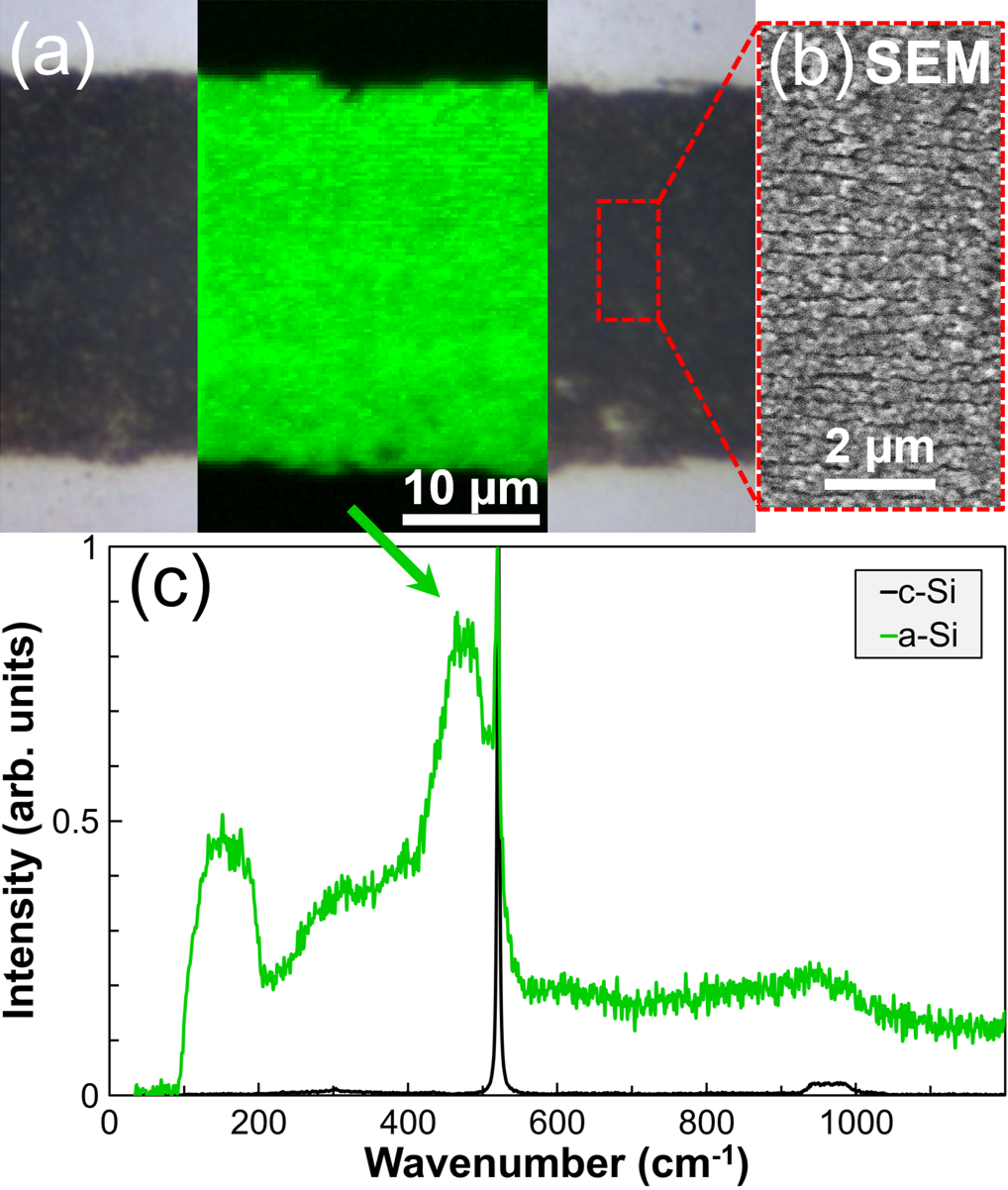}
\caption{Selective backside amorphization of silicon with 25-ps pulses at $\lambda=1.95$~$\mu$m. (a) Optical micrograph of a laser-modified track on the back surface of silicon appearing in dark. (b) Scanning electron microscopy (SEM) image of laser-induced periodic surface structures in the track. (c) Raman spectroscopy measurements of the green and black areas in (a), showing different crystalline states. \emph{Adapted from M.~Blothe et al. ``Nanostructured back surface amorphization of silicon with picosecond laser pulses,'' Appl. Phys. Lett. \textbf{121}, 101602 (2022) \cite{Blothe2022}, Copyright 2022 under CC BY 4.0 (https://creativecommons.org/licenses/by/4.0/)}}
\label{fig:FigBacksideAmorphization}    
\end{figure}

\noindent \textbf{Backside structuring.} A first example of back-surface application is backside structuring. By focusing 25-ps pulses at $\lambda=1.95$~$\mu$m, it was shown that modifications could be produced on the exit surface of silicon, as shown in Fig.~\ref{fig:FigBacksideAmorphization}. To ensure that the fluence at the exit surface exceeds the modification threshold, Bessel beam shaping was implemented instead of working with standard Gaussian beams which are more prone to exhibit nonlinear focal shift (see Sect.~\ref{subsec:limitations}). Raman spectroscopy analyses revealed a nearly full amorphization of silicon \cite{Blothe2022}. While amorphization on the front surface of silicon is now a standard technique (see Chapter~28), its extension to the back surface allows one to envision selective tailoring of the crystallinity of in-built devices. This aspect is especially relevant as silicon properties (e.g., refractive index, electrical conductivity) depend on the crystalline state. Interestingly, the amorphized areas exhibit nanostructuring in the form of laser-induced periodic surface structures (LIPSS), which are also traditionally produced in silicon exclusively on the front surface (see Chapter~10). The observed LIPSS on the back surface exhibit a period of $\Lambda \approx \lambda/2n_{0}$, in excellent agreement with typical values on the front surface \cite{Bonse2020}, as discussed in Sect.~\ref{subsec:functionalization}.\\

\begin{figure}[!ht]
\centering
\includegraphics[scale=0.5]{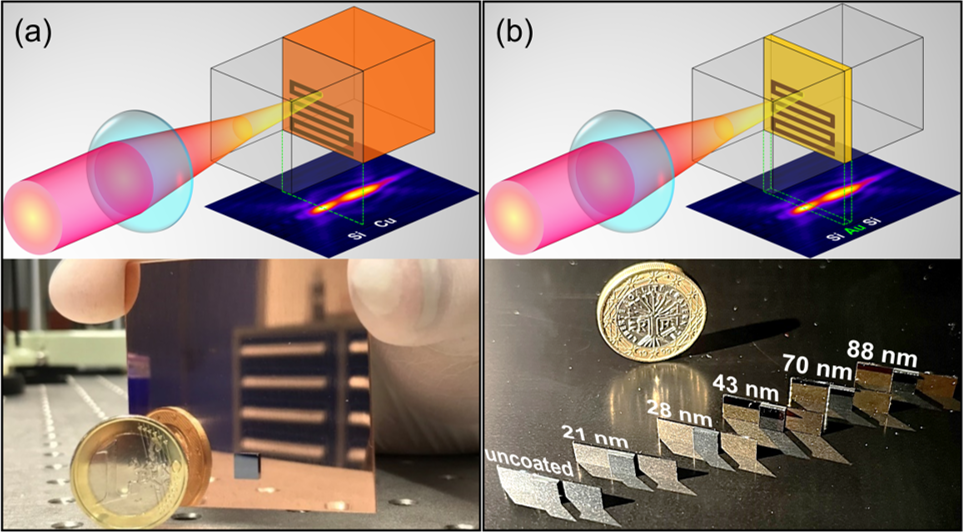}
\caption{Through-silicon ultrafast laser welding with 10-ps pulses at $\lambda=1.55$~$\mu$m. (a) Application in the Si--Cu configuration. \emph{Adapted from M.~Chambonneau et al. ``Taming Ultrafast Laser Filaments for Optimized Semiconductor--Metal Welding,'' Laser Photon. Rev. \textbf{15}, 2000433 (2021) \cite{Chambonneau2021}, Copyright 2020 under CC BY 4.0 (https://creativecommons.org/licenses/by/4.0/).} (b) Application in the Si--Si configuration. \emph{Adapted from M.~Chambonneau et al. ``Ultrafast Laser Welding of Silicon,'' Adv. Photonics Res. \textbf{4}, 2200300 (2023) \cite{Chambonneau2023}, Copyright 2023 under CC BY 4.0 (https://creativecommons.org/licenses/by/4.0/)}}
\label{fig:FigWelding}    
\end{figure}

\noindent \textbf{Laser welding.} Finally, a second back-surface application is ultrafast laser welding. This technique consists of energy deposition at the interface between two workpieces to join them. This laser-based method is clean, fast and contactless, and it represents an attractive alternative to traditional joining techniques (e.g., adhesive bonding, brazing, anodic bonding). When one aims at through-silicon ultrafast laser welding, a considerable bottleneck is the non-localized energy deposition related to filamentation effects during the propagation in silicon (see Sect.~\ref{subsec:limitations}). As illustrated in Fig.~\ref{fig:FigWelding}(a), this challenge was recently tackled by predetermining the nonlinear focal shift $\Delta z$ with nonlinear propagation imaging \cite{Chambonneau2021}. It was shown that precompensating for $\Delta z$ during silicon--copper welding with 10-ps pulses at $\lambda=1.55$~$\mu$m leads to MPa shear joining strength values. While, in principle, this strategy is applicable for various metals which efficiently absorb the laser pulses, joining silicon to silicon represents another level of difficulty. Indeed, an apparent paradox is that welding is much more efficient for silicon samples in optical contact \cite{Richter2015,Tan2017,Sopena2022}, while, by definition, these would behave optically as a single silicon sample which cannot be modified due to filamentation in the ultrafast regime. To sidestep this paradox, metallic nanolayers were deposited at the interface between both silicon samples for enhancing absorption during welding with 10-ps pulses at $\lambda=1.55$~$\mu$m \cite{Chambonneau2023}, as illustrated in Fig.~\ref{fig:FigWelding}(b). Counter-intuitively, it was found that the thinnest tested nanolayers (21~nm) lead to the highest shear joining strength ($>4$~MPa), as a consequence of the confinement of laser-produced hot electrons in a restricted volume \cite{Tsibidis2022}. The advantage of such thin nanolayers is that the optical transmission of silicon is only marginally impacted compared to thicker films.

\section{Conclusion and Outlook}
\label{sec:conclusion}

To summarize, the question of the wavelength-dependence of ultrafast laser--matter interactions in bandgap materials has been reviewed. This important control parameter changes the interaction through the spectral dependence of the apparent material properties (refractive index), and the 3D spatial distribution of the focal zone (waist and Rayleigh length). It is also a basis parameter to trigger different nonlinear absorption mechanisms governing the interaction (multi-photon, tunnel, and avalanche ionization). In principle, to optimize the processing performance, a compromise has to be found between the smallest achievable spot size defining the optical resolution and the order of nonlinearity that permit to access beneficial effects, including three-dimensional writing capabilities.\\

The impact of the wavelength in laser--dielectric interaction has then been presented, starting with recent counter-intuitive results showing the independence of surface ablation in dielectrics from the wavelength, from near-ultraviolet to mid-infrared. In contrast, the wavelength plays a key role to control material nanostructuring modalities. This includes the direct control of the periodicity in laser interference patterning approaches. The wavelength and polarization are also the main driving parameters in self-organized material responses exploited for instance for the formation of well-defined arrays on the surface and in the bulk of irradiated dielectrics. For improved efficiency of ultrafast laser processing in dielectrics, more degrees of freedom can be obtained by spectrally shaping pulses, which includes configurations such as multi-wavelength irradiation, chirped pulses, and simultaneous spatial and temporal focusing.\\

Ultimately, it has been shown that the growing development of near- and mid-infrared femtosecond laser systems allows one to envision ultrafast laser processing in the bulk and on the backside of narrow-gap materials (i.e., semiconductors, exemplified with silicon). In this spectral range, these materials are transparent, which is a prerequisite for internal energy deposition. Nevertheless, a major difference with dielectrics is that energy deposition inside silicon is drastically reduced due to nonlinear propagation effects. Multiple strategies have been devised to bypass these physical limitations, which include surface seeds, hyper-NA configurations, picosecond pulses, pulse trains, and wavelength optimization---all these solutions offering advantages and disadvantages. Various applications have been demonstrated, including data storage and optical functionalization for in-volume processing, as well as backside structuring and welding for backside processing. One can anticipate that the development of infraredder femtosecond laser sources will make possible ultrafast internal processing of materials exhibiting bandgaps even lower than silicon such as germanium and indium antimonide, establishing new processing capabilities.

\section*{Acknowledgments}

\noindent The German Federal Ministry of Education and Research (Bundesministerium für Bildung und Forschung, BMBF) through the RUBIN-UKPi\~no project (Grants no. 03RU2U033H and 03RU2U032F).\\

\noindent The German Research Foundation (Deutsche Forschungsgemeinschaft, DFG) through the Silabus project (Grant no. 530105422), and the Inseption project (Grant no. 545531713).\\

\noindent The French National Research Agency (Grants ANR-22-CE92-0057-01 and ANR-23-CE08-0029-03).\\

\noindent The PEPR LUMA, part of the France 2030 program through the PROSECO project (Grant ANR-24-EXLU-0012).\\

\sloppy \noindent Project HyperSpec (PID2023-148178OB-C22) funded by MICIU/AEI/10.13039/501100011033 and by ERDF/EU.\\

\noindent CSIC through the call for internationalization i-LINK 2022 (project ILINK22056)
 
\bibliographystyle{spphys_custom}
\bibliography{refs}

@article{grojo2010exciton,
  title={Exciton-seeded multiphoton ionization in bulk SiO$_2$},
  author={Grojo, D and Gertsvolf, M and Lei, S and Barillot, T and Rayner, DM and Corkum, PB},
  journal={Physical Review B},
  volume={81},
  number={21},
  pages={212301},
  year={2010},
  doi = {10.1103/PhysRevB.81.212301},
  url = {https://doi.org/10.1103/PhysRevB.81.212301},
  publisher={APS}
}

@article{shen1989laser,
  title={Laser heating of free electrons in wide-gap optical materials at 1064 nm},
  author={Shen, XA and Jones, Scott C and Braunlich, Peter},
  journal={Physical Review Letters},
  volume={62},
  number={23},
  pages={2711},
  year={1989},
  doi = {10.1103/PhysRevLett.62.2711},
  url = {https://doi.org/10.1103/PhysRevLett.62.2711},
  publisher={APS}
}

@article{peng2013control,
  title={Control of energy deposition in femtosecond laser dielectric interactions},
  author={Peng, Jiahui and Grojo, David and Rayner, David M and Corkum, Paul B},
  journal={Applied Physics Letters},
  volume={102},
  number={16},
  year={2013},
  doi = {10.1063/1.4802820},
  url = {https://doi.org/10.1063/1.4802820},
  publisher={AIP Publishing}
}

@article{leyder2013non,
  title={Non-linear absorption of focused femtosecond laser pulses at 1.3 $\mu$m inside silicon: Independence on doping concentration},
  author={Leyder, S and Grojo, D and Delaporte, P and Marine, W and Sentis, M and Ut{\'e}za, O},
  journal={Applied Surface Science},
  volume={278},
  pages={13--18},
  year={2013},
  doi = {10.1016/j.apsusc.2012.10.174},
  url = {https://doi.org/10.1016/j.apsusc.2012.10.174},
  publisher={Elsevier}
}

@article{gaudfrin2022fused,
  title={Fused silica ablation by double ultrashort laser pulses with dual wavelength and variable delays},
  author={Gaudfrin, K and Lopez, J and Gemini, L and Delaigue, M and H{\"o}nninger, C and Kling, R and Duchateau, G},
  journal={Optics Express},
  volume={30},
  number={22},
  pages={40120--40135},
  year={2022},
  doi = {10.1364/OE.461502},
  url = {https://doi.org/10.1364/OE.461502},
  publisher={Optica Publishing Group}
}

@article{yu2013near,
  title={Near-infrared femtosecond laser machining initiated by ultraviolet multiphoton ionization},
  author={Yu, Xiaoming and Bian, Qiumei and Zhao, B and Chang, Zenghu and Corkum, PB and Lei, Shuting},
  journal={Applied Physics Letters},
  volume={102},
  number={10},
  year={2013},
  doi = {10.1063/1.4794946},
  url = {https://doi.org/10.1063/1.4794946},
  publisher={AIP Publishing}
}

@article{yu2014fabricating,
  title={Fabricating nanostructures on fused silica using femtosecond infrared pulses combined with sub-nanojoule ultraviolet pulses},
  author={Yu, Xiaoming and Chang, Zenghu and Corkum, PB and Lei, Shuting},
  journal={Optics Letters},
  volume={39},
  number={19},
  pages={5638--5640},
  year={2014},
  doi = {10.1364/OL.39.005638},
  url = {https://doi.org/10.1364/OL.39.005638},
  publisher={Optica Publishing Group}
}

@article{zoppel2005enhancement,
  title={Enhancement of laser ablation yield by two color excitation},
  author={Zoppel, Sandra and Merz, Robert and Zehetner, Johann and Reider, Georg A},
  journal={Applied Physics A},
  volume={81},
  pages={847--850},
  year={2005},
  doi = {10.1007/s00339-005-3275-4},
  url = {https://doi.org/10.1007/s00339-005-3275-4},
  publisher={Springer}
}

@article{mouskeftaras2013mechanisms,
  title={Mechanisms of femtosecond laser ablation of dielectrics revealed by double pump--probe experiment},
  author={Mouskeftaras, Alexandros and Guizard, St{\'e}phane and Fedorov, Nikita and Klimentov, Sergey},
  journal={Applied Physics A},
  volume={110},
  pages={709--715},
  year={2013},
  doi = {10.1007/s00339-012-7217-7},
  url = {https://doi.org/10.1007/s00339-012-7217-7},
  publisher={Springer}
}

@article{bulgakov2024dual,
  title={Dual-wavelength femtosecond laser-induced single-shot damage and ablation of silicon},
  author={Bulgakov, Alexander V and Sl{\'a}dek, Juraj and Hrabovsk{\`y}, Jan and Mirza, Inam and Marine, Wladimir and Bulgakova, Nadezhda M},
  journal={Applied Surface Science},
  volume={643},
  pages={158626},
  year={2024},
  doi = {10.1016/j.apsusc.2023.158626},
  url = {https://doi.org/10.1016/j.apsusc.2023.158626},
  publisher={Elsevier}
}

@article{Rohloff2011,
author = {Rohloff, M. and Das, S. K. and H{\"{o}}hm, S. and Grunwald, R. and Rosenfeld, A. and Kr{\"{u}}ger, J. and Bonse, J.},
doi = {10.1063/1.3605513},
issn = {0021-8979},
journal = {Journal of Applied Physics},
month = {jul},
number = {1},
title = {{Formation of laser-induced periodic surface structures on fused silica upon multiple cross-polarized double-femtosecond-laser-pulse irradiation sequences}},
url = {https://doi.org/10.1063/1.3605513},
volume = {110},
year = {2011}
}

@article{Borowiec2005InP,
author = {Borowiec, A. and Tiedje, H. F. and Haugen, H. K.},
doi = {10.1016/j.apsusc.2004.09.105},
issn = {01694332},
journal = {Applied Surface Science},
number = {1-4},
pages = {129--137},
title = {{Wavelength dependence of the single pulse femtosecond laser ablation threshold of indium phosphide in the 400-2050 nm range}},
url = {https://doi.org/10.1016/j.apsusc.2004.09.105},
volume = {243},
year = {2005}
}

@article{Davis1996waveguide,
author = {Davis, K M and Miura, K and Sugimoto, N and Hirao, K},
doi = {10.1364/OL.21.001729},
issn = {0146-9592},
journal = {Optics Letters},
month = {nov},
number = {21},
pages = {1729},
publisher = {OSA},
title = {{Writing waveguides in glass with a femtosecond laser}},
url = {https://opg.optica.org/abstract.cfm?URI=ol-21-21-1729},
volume = {21},
year = {1996}
}

@article{Rudenko2017,
author = {Rudenko, Anton and Colombier, Jean-Philippe and H{\"{o}}hm, Sandra and Rosenfeld, Arkadi and Kr{\"{u}}ger, J{\"{o}}rg and Bonse, J{\"{o}}rn and Itina, Tatiana E.},
doi = {10.1038/s41598-017-12502-4},
issn = {2045-2322},
journal = {Scientific Reports},
month = {sep},
number = {1},
pages = {12306},
title = {{Spontaneous periodic ordering on the surface and in the bulk of dielectrics irradiated by ultrafast laser: a shared electromagnetic origin}},
url = {https://www.nature.com/articles/s41598-017-12502-4},
volume = {7},
year = {2017}
}

@phdthesis{Garcia-Lechuga2017thesis,
author = {Garcia-Lechuga, Mario},
school = {Universidad Autonoma de Madrid},
title = {{T{\'{e}}cnicas de imagen con resoluci{\'{o}}n temporal aplicadas al procesado de materiales con pulsos l{\'{a}}ser ultracortos}},
url = {https://repositorio.uam.es/handle/10486/679910},
year = {2017}
}

@article{Blumenstein2023DLIP,
author = {Blumenstein, Andreas and Simon, Peter and Ihlemann, J{\"{u}}rgen},
doi = {10.3390/nano13152240},
issn = {2079-4991},
journal = {Nanomaterials},
month = {aug},
number = {15},
pages = {2240},
title = {{High-Resolution Laser Interference Ablation and Amorphization of Silicon}},
url = {https://www.mdpi.com/2079-4991/13/15/2240},
volume = {13},
year = {2023}
}

@article{Oliveira2013DLIP,
author = {Oliveira, V. and Vilar, R. and Serra, R. and Oliveira, J.C. and Polushkin, N.I. and Conde, O.},
doi = {10.1016/j.optlastec.2013.06.031},
issn = {00303992},
journal = {Optics $\&$ Laser Technology},
month = {dec},
number = {1},
pages = {428--431},
publisher = {Elsevier},
title = {{Sub-micron structuring of silicon using femtosecond laser interferometry}},
url = {https://linkinghub.elsevier.com/retrieve/pii/S0030399213002478},
volume = {54},
year = {2013}
}

@article{Garcia-Lechuga2020,
author = {Garcia-Lechuga, M. and Ut{\'{e}}za, O. and Sanner, N. and Grojo, D.},
doi = {10.1364/OL.382610},
issn = {0146-9592},
journal = {Optics Letters},
month = {feb},
number = {4},
pages = {952},
title = {{Evidencing the nonlinearity independence of resolution in femtosecond laser ablation}},
url = {https://www.osapublishing.org/abstract.cfm?URI=ol-45-4-952},
volume = {45},
year = {2020}
}

@article{Korte2003,
author = {Korte, F. and Serbin, J. and Koch, J. and Egbert, A. and Fallnich, C. and Ostendorf, A. and Chichkov, B.N.},
doi = {10.1007/s00339-003-2110-z},
issn = {0947-8396},
journal = {Applied Physics A},
month = {jul},
number = {2},
pages = {229--235},
title = {{Towards nanostructuring with femtosecond laser pulses}},
url = {http://link.springer.com/10.1007/s00339-003-2110-z},
volume = {77},
year = {2003}
}

@article{Lenzner1999,
author = {Lenzner, M.},
doi = {10.1142/S0217979299001570},
issn = {0217-9792},
journal = {International Journal of Modern Physics B},
month = {may},
number = {13},
pages = {1559--1578},
title = {{Femtosecond laser-induced damage of dielectrics}},
url = {https://doi.org/10.1142/S0217979299001570},
volume = {13},
year = {1999}
}

@article{Sugioka2014,
author = {Sugioka, Koji and Cheng, Ya},
doi = {10.1063/1.4904320},
issn = {1931-9401},
journal = {Applied Physics Reviews},
month = {dec},
number = {4},
pages = {041303},
title = {{Femtosecond laser three-dimensional micro- and nanofabrication}},
url = {http://aip.scitation.org/doi/10.1063/1.4904320},
volume = {1},
year = {2014}
}

@article{Zipfel2003nonlinearmicroscopy,
author = {Zipfel, Warren R and Williams, Rebecca M and Webb, Watt W},
doi = {10.1038/nbt899},
issn = {1087-0156},
journal = {Nature Biotechnology},
month = {nov},
number = {11},
pages = {1369--1377},
pmid = {14595365},
title = {{Nonlinear magic: multiphoton microscopy in the biosciences}},
url = {https://www.nature.com/articles/nbt899},
volume = {21},
year = {2003}
}

@article{Richter2012NGs,
author = {Richter, S{\"{o}}ren and Heinrich, Matthias and D{\"{o}}ring, Sven and T{\"{u}}nnermann, Andreas and Nolte, Stefan and Peschel, Ulf},
doi = {10.2351/1.4718561},
issn = {1042-346X},
journal = {Journal of Laser Applications},
month = {sep},
number = {4},
pages = {1--8},
title = {{Nanogratings in fused silica: Formation, control, and applications}},
url = {https://doi.org/10.2351/1.4718561},
volume = {24},
year = {2012}
}

@article{Yang2006NGswavelength,
author = {Yang, Weijia and Bricchi, Erica and Kazansky, Peter G. and Bovatsek, James and Arai, Alan Y.},
doi = {10.1364/OE.14.010117},
issn = {1094-4087},
journal = {Optics Express},
number = {21},
pages = {10117},
title = {{Self-assembled periodic sub-wavelength structures by femtosecond laser direct writing}},
url = {https://opg.optica.org/oe/abstract.cfm?uri=oe-14-21-10117},
volume = {14},
year = {2006}
}

@article{Uteza2011fusedsilica,
author = {Ut{\'{e}}za, O. and Sanner, N. and Chimier, B. and Brocas, A. and Varkentina, N. and Sentis, M. and Lassonde, P. and L{\'{e}}gar{\'{e}}, F. and Kieffer, J. C.},
doi = {10.1007/s00339-011-6469-y},
issn = {0947-8396},
journal = {Applied Physics A},
month = {jun},
number = {1},
pages = {131--141},
title = {{Control of material removal of fused silica with single pulses of few optical cycles to sub-picosecond duration}},
url = {http://link.springer.com/10.1007/s00339-011-6469-y},
volume = {105},
year = {2011}
}

@article{Terasawa2020fusedsilica400nm,
author = {Terasawa, E. and Shibuya, T. and Satoh, D. and Moriai, Y. and Ogawa, H. and Tanaka, M. and Kuroda, R. and Kobayashi, Y. and Sakaue, K. and Washio, M.},
doi = {10.1007/s00339-020-03640-0},
isbn = {0033902003640},
issn = {0947-8396},
journal = {Applied Physics A},
month = {jun},
number = {6},
pages = {446},
publisher = {Springer Berlin Heidelberg},
title = {{Pulse duration dependence of ablation threshold for fused silica in the visible femtosecond regime}},
url = {https://link.springer.com/10.1007/s00339-020-03640-0},
volume = {126},
year = {2020}
}

@article{Christensen2009,
author = {Christensen, B and Balling, Peter},
doi = {10.1103/PhysRevB.79.155424},
issn = {1098-0121},
journal = {Physical Review B},
month = {apr},
number = {15},
pages = {155424},
title = {{Modeling ultrashort-pulse laser ablation of dielectric materials}},
url = {http://link.aps.org/doi/10.1103/PhysRevB.79.155424},
volume = {79},
year = {2009}
}

@article{Sipe1983,
archivePrefix = {arXiv},
arxivId = {cond-mat/0609178},
author = {Sipe, J. E. and Young, Jeff F. and Preston, J. S. and van Driel, H. M.},
doi = {10.1103/PhysRevB.27.1141},
eprint = {0609178},
isbn = {0163-1829},
issn = {0163-1829},
journal = {Physical Review B},
month = {jan},
number = {2},
pages = {1141--1154},
pmid = {21727564},
primaryClass = {cond-mat},
title = {{Laser-induced periodic surface structure. I. Theory}},
url = {https://link.aps.org/doi/10.1103/PhysRevB.27.1141},
volume = {27},
year = {1983}
}

@inproceedings{Buividas2010HSFL,
author = {Buividas, Ricardas and Kudrius, Tadas and Sliupas, Remigijus and Rosa, Lorenzo and Slekys, Gintas and Bagdonas, Saulius and Rotomskis, Ricardas and Juodkazis, Saulius},
booktitle = {Laser Applications in Life Sciences},
doi = {10.1117/12.871049},
editor = {Kinnunen, Matti and Myllyl{\"{a}}, Risto},
isbn = {9780819476524},
issn = {0277786X},
month = {jun},
pages = {737602},
title = {{Ripple-patterned substrates for light enhancement applications}},
url = {http://proceedings.spiedigitallibrary.org/proceeding.aspx?doi=10.1117/12.871049},
volume = {7376},
year = {2010}
}

@article{Buividas2014HSFLandNG,
author = {Buividas, Ri{\v{c}}ardas and Mikutis, Mindaugas and Juodkazis, Saulius},
doi = {10.1016/j.pquantelec.2014.03.002},
issn = {00796727},
journal = {Progress in Quantum Electronics},
month = {may},
number = {3},
pages = {119--156},
title = {{Surface and bulk structuring of materials by ripples with long and short laser pulses: Recent advances}},
url = {https://linkinghub.elsevier.com/retrieve/pii/S0079672714000044},
volume = {38},
year = {2014}
}

@article{Solana2024,
author = {Solana, Irene and Chacon-Sanchez, Fernando and Garcia-Lechuga, Mario and Siegel, Jan},
doi = {10.1016/j.optlastec.2024.111360},
issn = {00303992},
journal = {Optics $\&$ Laser Technology},
month = {dec},
number = {March},
pages = {111360},
publisher = {Elsevier Ltd},
title = {{Versatile femtosecond laser interference patterning applied to high-precision nanostructuring of silicon}},
url = {https://linkinghub.elsevier.com/retrieve/pii/S0030399224008181},
volume = {179},
year = {2024}
}

@article{MulkoSolderaLasagni2022,
author = {Mulko, Lucinda and Soldera, Marcos and Lasagni, Andr{\'{e}}s Fabi{\'{a}}n},
doi = {10.1515/nanoph-2021-0591},
issn = {2192-8614},
journal = {Nanophotonics},
month = {jan},
number = {2},
pages = {203--240},
title = {{Structuring and functionalization of non-metallic materials using direct laser interference patterning: a review}},
url = {https://www.degruyter.com/document/doi/10.1515/nanoph-2021-0591/html},
volume = {11},
year = {2022}
}

@incollection{Ihlemann2023chapterDLIP,
address = {Cham},
author = {Ihlemann, J{\"{u}}rgen and Blumenstein, Andreas and Klein-Wiele, Jan-Hendrik and Simon, Peter},
booktitle = {Ultrafast Laser Nanostructuring: The Pursuit of Extreme Scales},
doi = {10.1007/978-3-031-14752-4{\_}14},
editor = {Stoian, Razvan and Bonse, J{\"{o}}rn},
isbn = {978-3-031-14752-4},
pages = {495--532},
publisher = {Springer International Publishing},
title = {{Periodic Surface Structures by Laser Interference Ablation}},
url = {http://dx.doi.org/10.1007/978-3-031-14752-4{\_}14},
year = {2023}
}

@article{Bonse2012LIPPSreview,
author = {Bonse, J and Kr{\"{u}}ger, J. and H{\"{o}}hm, S. and Rosenfeld, A},
doi = {10.2351/1.4712658},
issn = {1042-346X},
journal = {Journal of Laser Applications},
month = {sep},
number = {4},
pages = {042006},
title = {{Femtosecond laser-induced periodic surface structures}},
url = {http://lia.scitation.org/doi/10.2351/1.4712658},
volume = {24},
year = {2012}
}

@article{Navickas2022LIPSSfusedsilica,
author = {Navickas, Marius and Grigutis, Robertas and Jukna, Vytautas and Tamo{\v{s}}auskas, Gintaras and Dubietis, Audrius},
doi = {10.1038/s41598-022-24771-9},
isbn = {0123456789},
issn = {2045-2322},
journal = {Scientific Reports},
month = {nov},
number = {1},
pages = {20231},
pmid = {36418435},
publisher = {Nature Publishing Group UK},
title = {{Low spatial frequency laser-induced periodic surface structures in fused silica inscribed by widely tunable femtosecond laser pulses}},
url = {https://www.nature.com/articles/s41598-022-24771-9},
volume = {12},
year = {2022}
}

@article{Puerto2016LIPSS,
author = {Puerto, Daniel and Garcia-Lechuga, Mario and Hernandez-Rueda, Javier and Garcia-Leis, Adianez and Sanchez-Cortes, Santiago and Solis, Javier and Siegel, Jan},
doi = {10.1088/0957-4484/27/26/265602},
issn = {0957-4484},
journal = {Nanotechnology},
month = {jul},
number = {26},
pages = {265602},
publisher = {IOP Publishing},
title = {{Femtosecond laser-controlled self-assembly of amorphous-crystalline nanogratings in silicon}},
url = {https://iopscience.iop.org/article/10.1088/0957-4484/27/26/265602},
volume = {27},
year = {2016}
}

@article{GarciaLechuga2021AdvOptMat,
author = {Garcia‐Lechuga, Mario and Casquero, Noemi and Wang, Andong and Grojo, David and Siegel, Jan},
doi = {10.1002/adom.202100400},
issn = {2195-1071},
journal = {Advanced Optical Materials},
month = {sep},
number = {17},
pages = {2100400},
title = {{Deep Silicon Amorphization Induced by Femtosecond Laser Pulses up to the Mid‐Infrared}},
url = {https://onlinelibrary.wiley.com/doi/10.1002/adom.202100400},
volume = {9},
year = {2021}
}

@article{GarciaLechuga2024LPR,
author = {Garcia‐Lechuga, Mario and Casquero, Noemi and Siegel, Jan and Solis, Javier and Clady, Raphael and Wang, Andong and Ut{\'{e}}za, Olivier and Grojo, David},
doi = {10.1002/lpor.202301327},
issn = {1863-8880},
journal = {Laser $\&$ Photonics Reviews},
month = {nov},
number = {11},
pages = {2301327},
title = {{Amorphization and Ablation of Crystalline Silicon Using Ultrafast Lasers: Dependencies on the Pulse Duration and Irradiation Wavelength}},
url = {https://onlinelibrary.wiley.com/doi/10.1002/lpor.202301327},
volume = {18},
year = {2024}
}

@article{Austin2018germanium,
author = {Austin, Drake R. and Kafka, Kyle R. P. and Lai, Yu Hang and Wang, Zhou and Blaga, Cosmin I. and Chowdhury, Enam A.},
doi = {10.1364/OL.43.003702},
issn = {0146-9592},
journal = {Optics Letters},
month = {aug},
number = {15},
pages = {3702},
title = {{Femtosecond laser damage of germanium from near- to mid-infrared wavelengths}},
url = {https://opg.optica.org/abstract.cfm?URI=ol-43-15-3702},
volume = {43},
year = {2018}
}

@article{Migal2020plasmathrehold,
author = {Migal, Ekaterina and Mareev, Evgenii and Smetanina, Evgeniya and Duchateau, Guillaume and Potemkin, Fedor},
doi = {10.1038/s41598-020-70862-w},
isbn = {0123456789},
issn = {2045-2322},
journal = {Scientific Reports},
month = {aug},
number = {1},
pages = {14007},
publisher = {Nature Publishing Group UK},
title = {{Role of wavelength in photocarrier absorption and plasma formation threshold under excitation of dielectrics by high-intensity laser field tunable from visible to mid-IR}},
url = {https://www.nature.com/articles/s41598-020-70862-w},
volume = {10},
year = {2020}
}

@article{Skliutas2023xphoton,
author = {Skliutas, Edvinas and Samsonas, Danielius and {\v{C}}iburys, Arūnas and Kontenis, Lukas and Gailevi{\v{c}}ius, Darius and Berzin{\v{s}}, Jonas and Narbutas, Donatas and Jukna, Vytautas and Vengris, Mikas and Juodkazis, Saulius and Malinauskas, Mangirdas},
doi = {10.1080/17452759.2023.2228324},
issn = {17452767},
journal = {Virtual and Physical Prototyping},
number = {1},
title = {{X-photon laser direct write 3D nanolithography}},
url = {http://doi.org/10.1080/17452759.2023.2228324},
volume = {18},
year = {2023}
}

@article{Kawata2001,
author = {Kawata, Satoshi and Sun, Hong-Bo and Tanaka, Tomokazu and Takada, Kenji},
doi = {10.1038/35089130},
issn = {0028-0836},
journal = {Nature},
month = {aug},
number = {6848},
pages = {697--698},
title = {{Finer features for functional microdevices}},
url = {https://www.nature.com/articles/35089130},
volume = {412},
year = {2001}
}

@article{GordonANDFarsari2024,
author = {Zyla, Gordon and Farsari, Maria},
title = {Frontiers of Laser-Based 3D Printing: A Perspective on Multi-Photon Lithography},
journal = {Laser \& Photonics Reviews},
volume = {18},
number = {7},
pages = {2301312},
doi = {https://doi.org/10.1002/lpor.202301312},
url = {https://onlinelibrary.wiley.com/doi/abs/10.1002/lpor.202301312},
year = {2024}
}

@article{Bhardwaj2006periodNG,
author = {Bhardwaj, V. R. and Simova, E. and Rajeev, P. P. and Hnatovsky, C. and Taylor, R. S. and Rayner, D. M. and Corkum, P. B.},
doi = {10.1103/PhysRevLett.96.057404},
issn = {0031-9007},
journal = {Physical Review Letters},
month = {feb},
number = {5},
pages = {057404},
title = {{Optically Produced Arrays of Planar Nanostructures inside Fused Silica}},
url = {https://link.aps.org/doi/10.1103/PhysRevLett.96.057404},
volume = {96},
year = {2006}
}

@article{Lancry2024reviewNGs,
author = {Yao, Heng and Pugliese, Diego and Lancry, Matthieu and Dai, Ye},
title = {Ultrafast Laser Direct Writing Nanogratings and their Engineering in Transparent Materials},
journal = {Laser \& Photonics Reviews},
volume = {18},
number = {9},
pages = {2300891},
doi = {https://doi.org/10.1002/lpor.202300891},
url = {https://onlinelibrary.wiley.com/doi/abs/10.1002/lpor.202300891},
eprint = {https://onlinelibrary.wiley.com/doi/pdf/10.1002/lpor.202300891},
year = {2024}
}

@article{Beresna2011PolSensitive,
author = {Beresna, Martynas and Gecevi{\v{c}}ius, Mindaugas and Kazansky, Peter G.},
doi = {10.1364/OME.1.000783},
issn = {2159-3930},
journal = {Optical Materials Express},
month = {aug},
number = {4},
pages = {783},
title = {{Polarization sensitive elements fabricated by femtosecond laser nanostructuring of glass [Invited]}},
url = {https://opg.optica.org/ome/abstract.cfm?uri=ome-1-4-783},
volume = {1},
year = {2011}
}

@article{Shimotsuma2003nanograting,
author = {Shimotsuma, Yasuhiko and Kazansky, Peter G. and Qiu, Jiarong and Hirao, Kazuoki},
doi = {10.1103/PhysRevLett.91.247405},
issn = {0031-9007},
journal = {Physical Review Letters},
month = {dec},
number = {24},
pages = {247405},
title = {{Self-Organized Nanogratings in Glass Irradiated by Ultrashort Light Pulses}},
url = {https://link.aps.org/doi/10.1103/PhysRevLett.91.247405},
volume = {91},
year = {2003}
}

@article{michele2019near,
  title={Near-IR-and UV-femtosecond laser waveguide inscription in silica glasses},
  author={Michele, Vincenzo De and Royon, Maxime and Marin, Emmanuel and Alessi, Antonino and Morana, Adriana and Boukenter, Aziz and Cannas, Marco and Girard, Sylvain and Ouerdane, Youcef},
  journal={Optical Materials Express},
  volume={9},
  number={12},
  pages={4624--4633},
  year={2019},
  publisher={Optical Society of America},
  doi = {10.1364/OME.9.004624},
  url = {https://doi.org/10.1364/OME.9.004624}
}

@article{Hernandez-Rueda2017waveguideLambda,
author = {Hernandez-Rueda, Javier and Clarijs, Jasper and van Oosten, Dries and Krol, Denise M.},
doi = {10.1063/1.4981124},
isbn = {9781628419757},
issn = {0003-6951},
journal = {Applied Physics Letters},
month = {apr},
number = {16},
pages = {161109},
title = {{The influence of femtosecond laser wavelength on waveguide fabrication inside fused silica}},
url = {http://aip.scitation.org/doi/10.1063/1.4981124},
volume = {110},
year = {2017}
}

@article{Fernandez2015ionmigration,
author = {Fernandez, Toney Teddy and Siegel, Jan and Hoyo, Jesus and Sotillo, Belen and Fernandez, Paloma and Solis, Javier},
doi = {10.1088/0022-3727/48/15/155101},
issn = {0022-3727},
journal = {Journal of Physics D: Applied Physics},
month = {apr},
number = {15},
pages = {155101},
publisher = {IOP Publishing},
title = {{Controlling plasma distributions as driving forces for ion migration during fs laser writing}},
url = {https://iopscience.iop.org/article/10.1088/0022-3727/48/15/155101},
volume = {48},
year = {2015}
}

@article{Rodenas2009RIchange,
author = {R{\'{o}}denas, A and Torchia, G A and Lifante, G and Cantelar, E and Lamela, J and Jaque, F. and Roso, L and Jaque, D},
doi = {10.1007/s00340-008-3353-3},
issn = {0946-2171},
journal = {Applied Physics B},
month = {apr},
number = {1},
pages = {85--96},
title = {{Refractive index change mechanisms in femtosecond laser written ceramic Nd:YAG waveguides: micro-spectroscopy experiments and beam propagation calculations}},
url = {http://link.springer.com/10.1007/s00340-008-3353-3},
volume = {95},
year = {2009}
}

@article{chen2014waveguide,
author = {Chen, Feng and de Aldana, J. R. Vázquez},
title = {Optical waveguides in crystalline dielectric materials produced by femtosecond-laser micromachining},
journal = {Laser \& Photonics Reviews},
volume = {8},
number = {2},
pages = {251-275},
doi = {https://doi.org/10.1002/lpor.201300025},
url = {https://onlinelibrary.wiley.com/doi/abs/10.1002/lpor.201300025},
eprint = {https://onlinelibrary.wiley.com/doi/pdf/10.1002/lpor.201300025},
year = {2014}
}

@article{Chung2009,
author = {Chung, Samuel H. and Mazur, Eric},
doi = {10.1007/s00339-009-5201-7},
issn = {09478396},
journal = {Applied Physics A},
number = {2},
pages = {335--341},
title = {{Femtosecond laser ablation of neurons in C. elegans for behavioral studies}},
url = {https://doi.org/10.1007/s00339-009-5201-7},
volume = {96},
year = {2009}
}

@article{Simanovskii2003,
author = {Simanovskii, D. M. and Schwettman, H. A. and Lee, H. and Welch, A. J.},
doi = {10.1103/PhysRevLett.91.107601},
issn = {0031-9007},
journal = {Physical Review Letters},
month = {sep},
number = {10},
pages = {107601},
pmid = {14525504},
title = {{Midinfrared Optical Breakdown in Transparent Dielectrics}},
url = {https://link.aps.org/doi/10.1103/PhysRevLett.91.107601},
volume = {91},
year = {2003}
}

@article{Olivie2008,
author = {Olivi{\'{e}}, G and Gigu{\`{e}}re, D and Vidal, F and Ozaki, T and Kieffer, J-C and Nada, O and Brunette, I},
doi = {10.1364/OE.16.004121},
isbn = {1094-4087},
issn = {1094-4087},
journal = {Optics Express},
number = {6},
pages = {4121},
pmid = {18542509},
title = {{Wavelength dependence of femtosecond laser ablation threshold of corneal stroma}},
url = {https://www.osapublishing.org/oe/abstract.cfm?uri=oe-16-6-4121},
volume = {16},
year = {2008}
}

@article{Gallais2015a,
author = {Gallais, L. and Douti, D.-B. and Commandr{\'{e}}, M. and Batavi{\v{c}}iūtė, G. and Pupka, E. and {\v{S}}{\v{c}}iuka, M. and Smalakys, L. and Sirutkaitis, V. and Melninkaitis, A.},
doi = {10.1063/1.4922353},
issn = {0021-8979},
journal = {Journal of Applied Physics},
month = {jun},
number = {22},
pages = {223103},
title = {{Wavelength dependence of femtosecond laser-induced damage threshold of optical materials}},
url = {http://aip.scitation.org/doi/10.1063/1.4922353},
volume = {117},
year = {2015}
}

@article{Jia2006,
author = {Jia, T. Q. and Chen, H. X. and Huang, M. and Zhao, F. L. and Li, X. X. and Xu, S. Z. and Sun, H. Y. and Feng, D. H. and Li, C. B. and Wang, X. F. and Li, R. X. and Xu, Z. Z. and He, X. K. and Kuroda, H.},
doi = {10.1103/PhysRevB.73.054105},
isbn = {1098-0121 1550-235X},
issn = {1098-0121},
journal = {Physical Review B},
month = {feb},
number = {5},
pages = {054105},
title = {{Ultraviolet-infrared femtosecond laser-induced damage in fused silica and CaF$_{2}$ crystals}},
url = {https://link.aps.org/doi/10.1103/PhysRevB.73.054105},
volume = {73},
year = {2006}
}

@article{Garcia-Lechuga2023,
archivePrefix = {arXiv},
arxivId = {2212.10326},
author = {Garcia-Lechuga, Mario and Ut{\'{e}}za, Olivier and Sanner, Nicolas and Grojo, David},
doi = {10.1103/PhysRevApplied.19.044047},
eprint = {2212.10326},
issn = {2331-7019},
journal = {Physical Review Applied},
month = {apr},
number = {4},
pages = {044047},
publisher = {American Physical Society},
title = {{Wavelength-Independent Performance of Femtosecond Laser Dielectric Ablation Spanning Over Three Octaves}},
url = {https://link.aps.org/doi/10.1103/PhysRevApplied.19.044047},
volume = {19},
year = {2023}
}

@article{Chambonneau2026,
   author = {Maxime Chambonneau and Markus Blothe and Vladimir Yu. Fedorov and Isaure de Kernier and Stelios Tzortzakis and Stefan Nolte},
   doi = {10.1038/s41467-026-69530-w},
   issn = {2041-1723},
   issue = {1},
   journal = {Nature Communications},
   month = {2},
   pages = {1701},
   publisher = {Nature Research},
   title = {Extreme optical nonlinearities unveiled by ultrafast laser filamentation in semiconductors},
   volume = {17},
   url = {https://www.nature.com/articles/s41467-026-69530-w},
   year = {2026}
}

@article{Wang2023,
author = {Wang, Ziyao and Heuermann, Tobias and Gebhardt, Martin and Lenski, Mathias and Gierschke, Philipp and Klas, Robert and Rothhardt, Jan and Jauregui, Cesar and Limpert, Jens},
doi = {10.1364/OL.487587},
issn = {0146-9592},
journal = {Optics Letters},
month = {may},
number = {10},
pages = {2647},
pmid = {37186730},
title = {{Nonlinear pulse compression to sub-two-cycle, 1.3 mJ pulses at 1.9 $\mu$m wavelength with 132 W average power}},
url = {https://opg.optica.org/abstract.cfm?URI=ol-48-10-2647},
volume = {48},
year = {2023}
}

@article{Nuter2025,
author = {Nuter, R. and {Karimbana Kandy}, A. and Gallais, L. and Grosjean, S. and Lamaign{\`{e}}re, L. and Wagner, F. and Natoli, J.-Y.},
doi = {10.1063/5.0242463},
issn = {0021-8979},
journal = {Journal of Applied Physics},
month = {jan},
number = {2},
publisher = {AIP Publishing LLC},
title = {{Influence of laser polarization on the laser-induced damage threshold: Comparison with photoionization theory}},
url = {https://doi.org/10.1063/5.0242463},
volume = {137},
year = {2025}
}

@article{Nuter2023,
author = {Nuter, Rachel},
doi = {10.1103/PhysRevA.108.033101},
issn = {2469-9926},
journal = {Physical Review A},
month = {sep},
number = {3},
pages = {033101},
publisher = {American Physical Society},
title = {{Theory of ionization in a crystal with a linear and a circular laser polarization: Influence of the laser wavelength}},
url = {https://link.aps.org/doi/10.1103/PhysRevA.108.033101},
volume = {108},
year = {2023}
}

@article{Prada-Rodrigo2024,
author = {Prada-Rodrigo, Javier and Maio, Yoan Di and Faure, Nicola and Kachan, E. and Colombier, J. P. and Sedao, Xxx},
doi = {10.2961/jlmn.2024.03.2002},
issn = {18800688},
journal = {Journal of Laser Micro/Nanoengineering},
month = {dec},
number = {3},
pages = {179--184},
title = {{Influence of the Wavelength on Femtosecond Laser Ablation Thresholds and Incubation Coefficients of Silicon and Germanium}},
url = {http://www.jlps.gr.jp/jlmn/uploads/24-031z.pdf},
volume = {19},
year = {2024}
}

@article{He2010,
author = {He, Fei and Xu, Han and Cheng, Ya and Ni, Jielei and Xiong, Hui and Xu, Zhizhan and Sugioka, Koji and Midorikawa, Katsumi},
doi = {10.1364/OL.35.001106},
issn = {0146-9592},
journal = {Optics Letters},
month = {apr},
number = {7},
pages = {1106},
title = {{Fabrication of microfluidic channels with a circular cross section using spatiotemporally focused femtosecond laser pulses}},
url = {https://opg.optica.org/abstract.cfm?URI=ol-35-7-1106},
volume = {35},
year = {2010}
}

@article{Kammel2014,
author = {Kammel, Robert and Ackermann, Roland and Thomas, Jens and G{\"{o}}tte, J{\"{o}}rg and Skupin, Stefan and T{\"{u}}nnermann, Andreas and Nolte, Stefan},
doi = {10.1038/lsa.2014.50},
issn = {2047-7538},
journal = {Light: Science $\&$ Applications},
month = {may},
number = {5},
pages = {e169--e169},
title = {{Enhancing precision in fs-laser material processing by simultaneous spatial and temporal focusing}},
url = {https://www.nature.com/articles/lsa201450},
volume = {3},
year = {2014}
}

@article{sanner2010toward,
author = {Sanner, N. and Ut{\'{e}}za, O. and Chimier, B. and Sentis, M. and Lassonde, P. and L{\'{e}}gar{\'{e}}, F. and Kieffer, J. C.},
doi = {10.1063/1.3309700},
issn = {0003-6951},
journal = {Applied Physics Letters},
month = {feb},
number = {7},
title = {{Toward determinism in surface damaging of dielectrics using few-cycle laser pulses}},
url = {https://doi.org/10.1063/1.3309700},
volume = {96},
year = {2010}
}

@article{Wang2022,
author = {Wang, Huijun and Lei, Yuhao and Wang, Lei and Sakakura, Masaaki and Yu, Yanhao and Shayeganrad, Gholamreza and Kazansky, Peter G.},
doi = {10.1002/lpor.202100563},
issn = {1863-8880},
journal = {Laser $\&$ Photonics Reviews},
month = {apr},
number = {4},
pages = {1--7},
title = {{100‐Layer Error‐Free 5D Optical Data Storage by Ultrafast Laser Nanostructuring in Glass}},
url = {https://onlinelibrary.wiley.com/doi/10.1002/lpor.202100563},
volume = {16},
year = {2022}
}

@article{Royon2010,
author = {Royon, Arnaud and Bourhis, Kevin and Bellec, Matthieu and Papon, Gautier and Bousquet, Bruno and Deshayes, Yannick and Cardinal, Thierry and Canioni, Lionel},
doi = {10.1002/adma.201002413},
issn = {0935-9648},
journal = {Advanced Materials},
month = {dec},
number = {46},
pages = {5282--5286},
title = {{Silver Clusters Embedded in Glass as a Perennial High Capacity Optical Recording Medium}},
url = {https://onlinelibrary.wiley.com/doi/10.1002/adma.201002413},
volume = {22},
year = {2010}
}

@article{AsgariSabet2024,
author = {{Asgari Sabet}, Rana and Ishraq, Aqiq and Saltik, Alperen and B{\"{u}}t{\"{u}}n, Mehmet and Tokel, Onur},
doi = {10.1038/s41467-024-49303-z},
issn = {2041-1723},
journal = {Nature Communications},
month = {jul},
number = {1},
pages = {5786},
pmid = {39013851},
publisher = {Springer US},
title = {{Laser nanofabrication inside silicon with spatial beam modulation and anisotropic seeding}},
url = {https://www.nature.com/articles/s41467-024-49303-z},
volume = {15},
year = {2024}
}

@article{Kiyota2020,
author = {Kiyota, H. and Hara, K. and Jankowski, M. and Fejer, M. M.},
doi = {10.1063/1.5130701},
issn = {10897550},
journal = {Journal of Applied Physics},
number = {8},
publisher = {AIP Publishing LLC},
title = {{Numerical simulation and validation of subsurface modification and crack formation induced by nanosecond-pulsed laser processing in monocrystalline silicon}},
url = {https://doi.org/10.1063/1.5130701},
volume = {127},
year = {2020}
}

@article{Trinh2024,
author = {Trinh, Lanh and Wang, Xinya and Zhang, Xiang and Hosseini-Zavareh, Sajed and Mao, Aofei and Yu, Xiaoming and Lei, Shuting and Cui, Bai},
doi = {10.1002/adem.202301377},
issn = {15272648},
journal = {Advanced Engineering Materials},
number = {2},
pages = {1--7},
title = {{Transmission Electron Microscopy Characterizations of Local Amorphization of Single Crystal Silicon by Nanosecond Pulsed Laser Direct Writing}},
url = {https://doi.org/10.1002/adem.202301377},
volume = {26},
year = {2024}
}

@article{Iwata2024,
author = {Iwata, Hiroyuki and Saka, Hiroyasu},
doi = {10.2320/matertrans.MT-M2024022},
issn = {13459678},
journal = {Materials Transactions},
number = {7},
pages = {711--722},
title = {{Electron Microscopy on Mechanism of Voidage and Cracking in Si by Injection of a Permeable Infra-Red Laser}},
url = {https://doi.org/10.2320/matertrans.MT-M2024022},
volume = {65},
year = {2024}
}

@article{Wang2024,
author = {Wang, Xinya and Trinh, Lanh and Yu, Xiaoming and Berg, Matthew J. and Hosseini-Zavareh, Sajed and Lacroix, Brice and Chen, Pingping and Chen, Ruqi and Cui, Bai and Lei, Shuting},
doi = {10.2351/7.0001305},
issn = {1042-346X},
journal = {Journal of Laser Applications},
number = {2},
publisher = {Laser Institute of America},
title = {{Direct observation and quantification of nanosecond laser induced amorphization inside silicon}},
url = {https://doi.org/10.2351/7.0001305},
volume = {36},
year = {2024}
}

@article{Tauseef2024,
author = {Tauseef, Muhammad Ahsan and {Asgari Sabet}, Rana and Tokel, Onur},
doi = {10.1002/admt.202301617},
issn = {2365709X},
journal = {Advanced Materials Technologies},
number = {9},
pages = {1--8},
title = {{Laser Lithography of Monolithically-Integrated Multi-Level Microchannels in Silicon}},
url = {https://doi.org/10.1002/admt.202301617},
volume = {9},
year = {2024}
}

@article{Sopena2024,
author = {Sope{\~{n}}a, Pol and Ganguly, Niladri and Sp{\"{u}}hler, Gabriel and Selivanau, Andrei and Grojo, David},
doi = {10.1016/j.optlastec.2024.111419},
issn = {00303992},
journal = {Optics and Laser Technology},
title = {{Multiphoton-initiated laser writing of semiconductors using nanosecond mid-infrared pulses}},
url = {https://doi.org/10.1016/j.optlastec.2024.111419},
volume = {179},
year = {2024}
}

@article{Saka2018,
author = {Saka, Hiroyasu and Iwata, Hiroyuki and Kawaguchi, Daisuke},
doi = {10.1093/jmicro/dfy008},
issn = {20505701},
journal = {Microscopy},
number = {2},
pages = {112--120},
pmid = {29474725},
title = {{Thermal stability of laser-induced modified volumes in Si as studied by in situ and ex situ heating experiments}},
url = {https://doi.org/10.1093/jmicro/dfy008},
volume = {67},
year = {2018}
}

@article{Perez2024,
author = {Perez, Asher and Kotzero, Tomer and Perez, Eytan and Juwiler, Irit and Noach, Salman},
doi = {10.1016/j.optlastec.2024.111316},
issn = {00303992},
journal = {Optics and Laser Technology},
pages = {111316},
publisher = {Elsevier Ltd},
title = {{Enhancing silicon wafer ablation using passive Q-switch 2$\mu$m lasers with third-order nonlinear effects}},
url = {https://doi.org/10.1016/j.optlastec.2024.111316},
volume = {179},
year = {2024}
}

@article{Saltik2024,
author = {Saltik, Alperen and Tokel, Onur},
doi = {10.1364/ol.504600},
issn = {0146-9592},
journal = {Optics Letters},
number = {1},
pages = {49},
pmid = {38134149},
publisher = {Optica Publishing Group},
title = {{Laser-written wave plates inside the silicon enabled by stress-induced birefringence}},
url = {https://doi.org/10.1364/ol.504600},
volume = {49},
year = {2024}
}

@article{Saka2024,
author = {Saka, Hiroyasu and Iwata, Hiroyuki},
doi = {10.1080/14786435.2023.2283115},
issn = {14786443},
journal = {Philosophical Magazine},
number = {1},
pages = {55--67},
title = {{Whereabouts of missing atoms in a laser-injected Si (part IV): interaction of plasma Si with dislocations in a laser-injected Si}},
url = {https://doi.org/10.1080/14786435.2023.2283115},
volume = {104},
year = {2024}
}

@article{Blothe2024,
archivePrefix = {arXiv},
arxivId = {2404.11187},
author = {Blothe, Markus and Alberucci, Alessandro and Alasgarzade, Namig and Chambonneau, Maxime and Nolte, Stefan},
doi = {10.1002/lpor.202400535},
eprint = {2404.11187},
issn = {1863-8880},
journal = {Laser $\&$ Photonics Reviews},
month = {jul},
number = {11},
pages = {2400535},
title = {{Transverse Inscription of Silicon Waveguides by Picosecond Laser Pulses}},
url = {https://onlinelibrary.wiley.com/doi/10.1002/lpor.202400535},
volume = {18},
year = {2024}
}

@book{Ashcroft1976,
author = {Ashcroft, Neil W. and Mermin, N. David},
isbn = {0030839939},
publisher = {Harcourt College Publishers},
title = {{Solid State Physics}},
year = {1976}
}

@article{Boroumand2022,
author = {Boroumand, N. and Thorpe, A. and Parks, A. M. and Brabec, T.},
doi = {10.1088/1361-6455/ac9205},
issn = {13616455},
journal = {Journal of Physics B: Atomic, Molecular and Optical Physics},
number = {21},
publisher = {IOP Publishing},
title = {{Keldysh ionization theory of atoms: Mathematical details}},
url = {http://doi.org/10.1088/1361-6455/ac9205},
volume = {55},
year = {2022}
}

@article{Butkus2014,
author = {Butkus, S. and Gai{\v{z}}auskas, E. and Paipulas, D. and Viburys, {\v{Z}}. and Ka{\v{s}}kelyė, D. and Barkauskas, M. and Alesenkov, A. and Sirutkaitis, V.},
doi = {10.1007/s00339-013-8108-2},
issn = {0947-8396},
journal = {Applied Physics A},
month = {jan},
number = {1},
pages = {81--90},
title = {{Rapid microfabrication of transparent materials using filamented femtosecond laser pulses}},
url = {http://link.springer.com/10.1007/s00339-013-8108-2},
volume = {114},
year = {2014}
}

@article{Ganguly2024,
author = {Ganguly, Niladri and Sope{\~{n}}a, Pol and Grojo, David},
doi = {10.37188/LAM.2024.022},
issn = {28314093},
journal = {Light: Advanced Manufacturing},
title = {{Ultra-high-aspect-ratio structures through silicon using infrared laser pulses focused with axicon-lens doublets}},
url = {https://doi.org/10.37188/lam.2024.022},
volume = {5},
year = {2024}
}

@article{Sellmeier1871,
author = {Sellmeier, W},
doi = {10.1002/andp.18712190612},
issn = {0003-3804},
journal = {Annalen der Physik},
month = {jan},
number = {6},
pages = {272--282},
title = {{Zur Erkl{\"{a}}rung der abnormen Farbenfolge im Spectrum einiger Substanzen}},
url = {https://onlinelibrary.wiley.com/doi/10.1002/andp.18712190612},
volume = {219},
year = {1871}
}

@article{Marcinkeviciute2019,
author = {Marcinkevi{\v{c}}iūtė, A. and Jukna, V. and {\v{S}}uminas, R. and Garejev, N. and Tamo{\v{s}}auskas, G. and Dubietis, A.},
doi = {10.1364/OL.44.001343},
issn = {0146-9592},
journal = {Optics Letters},
month = {mar},
number = {6},
pages = {1343},
pmid = {30874646},
title = {{Femtosecond filamentation and supercontinuum generation in bulk silicon}},
url = {https://opg.optica.org/abstract.cfm?URI=ol-44-6-1343},
volume = {44},
year = {2019}
}

@article{Ryder2021,
author = {Ryder, Landen D. and Ryder, Kaitlyn L. and Sternberg, Andrew L. and Kozub, John A. and Khachatrian, Ani and Buchner, Steven P. and McMorrow, Dale and Hales, Joel M. and Zhao, Yuanfu and Wang, Liang and Wang, Chuanmin and Weller, Robert A. and Schrimpf, Ronald D. and Weiss, Sharon M. and Reed, Robert A.},
doi = {10.1109/TNS.2021.3111864},
issn = {15581578},
journal = {IEEE Transactions on Nuclear Science},
number = {10},
pages = {2496--2507},
publisher = {IEEE},
title = {{Simulation of Pulsed-Laser-Induced Testing in Microelectronic Devices}},
url = {https://doi.org/10.1109/TNS.2021.3111864},
volume = {68},
year = {2021}
}

@article{Hales2019,
author = {Hales, Joel M. and Khachatrian, Ani and Warner, Jeffrey and Buchner, Stephen and Ildefonso, Adrian and Tzintzarov, George N. and Nergui, Delgermaa and Monahan, Daniele M. and LaLumondiere, Stephen D. and Cressler, John D. and McMorrow, Dale},
doi = {10.1364/OE.27.037652},
issn = {1094-4087},
journal = {Optics Express},
month = {dec},
number = {26},
pages = {37652},
pmid = {31878543},
title = {{Using Bessel beams and two-photon absorption to predict radiation effects in microelectronics}},
url = {https://opg.optica.org/abstract.cfm?URI=oe-27-26-37652},
volume = {27},
year = {2019}
}

@article{Pinhas2018,
author = {Pinhas, Hadar and Wagner, Omer and Danan, Yossef and Danino, Meir and Zalevsky, Zeev and Sinvani, Moshe},
doi = {10.1364/OE.26.025370},
issn = {1094-4087},
journal = {Optics Express},
month = {sep},
number = {19},
pages = {25370},
pmid = {30469640},
title = {{Plasma dispersion effect based super-resolved imaging in silicon}},
url = {https://opg.optica.org/abstract.cfm?URI=oe-26-19-25370},
volume = {26},
year = {2018}
}

@article{Hales2023,
author = {Hales, Joel M. and Ildefonso, Adrian and Buchner, Stephen P. and Khachatrian, Ani and Allen, Greg and McMorrow, Dale},
doi = {10.1109/TNS.2022.3232724},
issn = {0018-9499},
journal = {IEEE Transactions on Nuclear Science},
month = {apr},
number = {4},
pages = {354--362},
publisher = {IEEE},
title = {{Quantitative Prediction of Ion-Induced Single-Event Transients in an Operational Amplifier Using a Quasi-Bessel Beam Pulsed-Laser Approach}},
url = {https://ieeexplore.ieee.org/document/9999719/},
volume = {70},
year = {2023}
}

@article{Hales2022,
author = {Hales, Joel M. and Khachatrian, Ani and Ildefonso, Adrian and Buchner, Stephen and Adams, Dennis and Wheeler, David and Messenger, Scott and Mishler, Codie and Budzinski, Nicholas and Jordan, Scott and {Van Art}, Roger and Mcmorrow, Dale},
doi = {10.1109/TNS.2021.3129416},
issn = {0018-9499},
journal = {IEEE Transactions on Nuclear Science},
month = {mar},
number = {3},
pages = {429--435},
publisher = {IEEE},
title = {{Pulsed-Laser Testing to Quantitatively Evaluate Latchup Sensitivity in Mixed-Signal ASICs}},
url = {https://ieeexplore.ieee.org/document/9620067/},
volume = {69},
year = {2022}
}

@article{Tzortzakis2001,
author = {Tzortzakis, S. and Sudrie, L. and Franco, M. and Prade, B. and Mysyrowicz, A. and Couairon, A. and Berg{\'{e}}, L.},
doi = {10.1103/PhysRevLett.87.213902},
issn = {0031-9007},
journal = {Physical Review Letters},
month = {nov},
number = {21},
pages = {213902},
title = {{Self-Guided Propagation of Ultrashort IR Laser Pulses in Fused Silica}},
url = {https://link.aps.org/doi/10.1103/PhysRevLett.87.213902},
volume = {87},
year = {2001}
}

@article{Garejev2017,
author = {Garejev, Nail and Tamo{\v{s}}auskas, Gintaras and Dubietis, Audrius},
doi = {10.1364/JOSAB.34.000088},
issn = {0740-3224},
journal = {Journal of the Optical Society of America B},
month = {jan},
number = {1},
pages = {88},
title = {{Comparative study of multioctave supercontinuum generation in fused silica, YAG, and LiF in the range of anomalous group velocity dispersion}},
url = {https://opg.optica.org/abstract.cfm?URI=josab-34-1-88},
volume = {34},
year = {2017}
}

@article{Papazoglou2011,
author = {Papazoglou, D. G. and Tzortzakis, S.},
doi = {10.1364/OME.1.000625},
issn = {2159-3930},
journal = {Optical Materials Express},
month = {aug},
number = {4},
pages = {625},
title = {{Physical mechanisms of fused silica restructuring and densification after femtosecond laser excitation [Invited]}},
url = {https://opg.optica.org/ome/abstract.cfm?uri=ome-1-4-625},
volume = {1},
year = {2011}
}

@article{Shanks1963,
author = {Shanks, H. R. and Maycock, P. D. and Sidles, P. H. and Danielson, G. C.},
doi = {10.1103/PhysRev.130.1743},
issn = {0031-899X},
journal = {Physical Review},
month = {jun},
number = {5},
pages = {1743--1748},
title = {{Thermal Conductivity of Silicon from 300 to 1400°K}},
url = {https://link.aps.org/doi/10.1103/PhysRev.130.1743},
volume = {130},
year = {1963}
}

@article{El-Mamouni2011,
author = {El-Mamouni, F. and Zhang, E. X. and Pate, N. D. and Hooten, N. and Schrimpf, R. D. and Reed, R. A. and Galloway, K. F. and McMorrow, D. and Warner, J. and Simoen, E. and Claeys, C. and Griffoni, A. and Linten, D. and Vizkelethy, G.},
doi = {10.1109/TNS.2011.2171994},
issn = {0018-9499},
journal = {IEEE Transactions on Nuclear Science},
month = {dec},
number = {6},
pages = {2563--2569},
publisher = {IEEE},
title = {{Laser- and Heavy Ion-Induced Charge Collection in Bulk FinFETs}},
url = {http://ieeexplore.ieee.org/document/6084846/},
volume = {58},
year = {2011}
}

@article{Tolstik2013,
author = {Tolstik, Nikolai and Sorokin, Evgeni and Sorokina, Irina T.},
doi = {10.1364/OL.38.000299},
isbn = {978-1-55752-933-6},
issn = {0146-9592},
journal = {Optics Letters},
month = {feb},
number = {3},
pages = {299},
publisher = {OSA},
title = {{Kerr-lens mode-locked Cr:ZnS laser}},
url = {https://opg.optica.org/abstract.cfm?URI=ol-38-3-299},
volume = {38},
year = {2013}
}

@article{Chudoba2001,
author = {Chudoba, C. and Fujimoto, J. G. and Ippen, E. P. and Haus, H. A. and Morgner, U. and K{\"{a}}rtner, F. X. and Scheuer, V. and Angelow, G. and Tschudi, T.},
doi = {10.1364/OL.26.000292},
isbn = {1557526346},
issn = {0146-9592},
journal = {Optics Letters},
month = {mar},
number = {5},
pages = {292},
title = {{All-solid-state Cr:forsterite laser generating 14-fs pulses at 13 $\mu$m}},
url = {https://opg.optica.org/abstract.cfm?URI=ol-26-5-292},
volume = {26},
year = {2001}
}

@article{Duval2016,
author = {Duval, Simon and Gauthier, Jean-Christophe and Robichaud, Louis-Rafa{\"{e}}l and Paradis, Pascal and Olivier, Michel and Fortin, Vincent and Bernier, Martin and Pich{\'{e}}, Michel and Vall{\'{e}}e, R{\'{e}}al},
doi = {10.1364/OL.41.005294},
issn = {0146-9592},
journal = {Optics Letters},
month = {nov},
number = {22},
pages = {5294},
title = {{Watt-level fiber-based femtosecond laser source tunable from 2.8 to 3.6 $\mu$m}},
url = {https://opg.optica.org/abstract.cfm?URI=ol-41-22-5294},
volume = {41},
year = {2016}
}

@article{Nakazawa1990,
author = {Nakazawa, Masataka and Kurokawa, Kenji and Kubota, Hirokazu and Suzuki, Kazunori and Kimura, Yasuo},
doi = {10.1063/1.104102},
issn = {0003-6951},
journal = {Applied Physics Letters},
number = {7},
pages = {653--655},
title = {{Femtosecond erbium-doped optical fiber amplifier}},
url = {https://doi.org/10.1063/1.104102},
volume = {57},
year = {1990}
}

@article{Duchateau2014,
author = {Duchateau, Guillaume and Bourgeade, Antoine},
doi = {10.1103/PhysRevA.89.053837},
issn = {1050-2947},
journal = {Physical Review A},
month = {may},
number = {5},
pages = {053837},
title = {{Influence of the time-dependent pulse spectrum on ionization and laser propagation in nonlinear optical materials}},
url = {https://link.aps.org/doi/10.1103/PhysRevA.89.053837},
volume = {89},
year = {2014}
}

@article{Louzon2005,
author = {Louzon, E. and Henis, Z. and Pecker, S. and Ehrlich, Y. and Fisher, D. and Fraenkel, M. and Zigler, A.},
doi = {10.1063/1.2140476},
issn = {0003-6951},
journal = {Applied Physics Letters},
month = {dec},
number = {24},
pages = {1--3},
title = {{Reduction of damage threshold in dielectric materials induced by negatively chirped laser pulses}},
url = {https://doi.org/10.1063/1.2140476},
volume = {87},
year = {2005}
}

@article{Sun2022,
author = {Sun, Ying and Yin, Weiyi and Yao, Qian and Ren, Xiangyu and Song, Juan and Dai, Ye},
doi = {10.1364/OL.471276},
issn = {0146-9592},
journal = {Optics Letters},
month = {dec},
number = {23},
pages = {6045},
pmid = {37219168},
publisher = {Optica Publishing Group},
title = {{Temporal modulation toward femtosecond laser-induced nonlinear ionization process}},
url = {https://opg.optica.org/abstract.cfm?URI=ol-47-23-6045},
volume = {47},
year = {2022}
}

@article{Xie2015,
author = {Xie, Chen and Jukna, Vytautas and Mili{\'{a}}n, Carles and Giust, Remo and Ouadghiri-Idrissi, Ismail and Itina, Tatiana and Dudley, John M. and Couairon, Arnaud and Courvoisier, Francois},
doi = {10.1038/srep08914},
issn = {2045-2322},
journal = {Scientific Reports},
month = {mar},
number = {1},
pages = {8914},
title = {{Tubular filamentation for laser material processing}},
url = {https://www.nature.com/articles/srep08914},
volume = {5},
year = {2015}
}

@article{Chen2008,
author = {Chen, L. M. and Kando, M. and Xu, M. H. and Li, Y. T. and Koga, J. and Chen, M. and Xu, H. and Yuan, X. H. and Dong, Q. L. and Sheng, Z. M. and Bulanov, S. V. and Kato, Y. and Zhang, J. and Tajima, T.},
doi = {10.1103/PhysRevLett.100.045004},
issn = {0031-9007},
journal = {Physical Review Letters},
month = {jan},
number = {4},
pages = {045004},
title = {{Study of X-Ray Emission Enhancement via a High-Contrast Femtosecond Laser Interacting with a Solid Foil}},
url = {https://link.aps.org/doi/10.1103/PhysRevLett.100.045004},
volume = {100},
year = {2008}
}

@article{Mareev2020a,
author = {Mareev, E. I. and Lvov, K. V. and Rumiantsev, B. V. and Migal, E. A. and Novikov, I. D. and Stremoukhov, S. Yu and Potemkin, F. V.},
doi = {10.1088/1612-202X/ab5d23},
issn = {1612-2011},
journal = {Laser Physics Letters},
month = {jan},
number = {1},
pages = {015402},
publisher = {IOP Publishing},
title = {{Effect of pulse duration on the energy delivery under nonlinear propagation of tightly focused Cr:forsterite laser radiation in bulk silicon}},
url = {https://iopscience.iop.org/article/10.1088/1612-202X/ab5d23},
volume = {17},
year = {2020}
}

@article{Stelmaszczyk2004,
author = {Stelmaszczyk, Kamil and Rohwetter, Philipp and M{\'{e}}jean, Guillaume and Yu, Jin and Salmon, Estelle and Kasparian, J{\'{e}}r{\^{o}}me and Ackermann, Roland and Wolf, Jean-Pierre and W{\"{o}}ste, Ludger},
doi = {10.1063/1.1812843},
issn = {0003-6951},
journal = {Applied Physics Letters},
number = {18},
pages = {3977--3979},
title = {{Long-distance remote laser-induced breakdown spectroscopy using filamentation in air}},
url = {https://pubs.aip.org/aip/apl/article/85/18/3977-3979/939897},
volume = {85},
year = {2004}
}

@incollection{Grojo2023,
author = {Grojo, David and Chambonneau, Maxime and Lei, Shuting and Mouskeftaras, Alexandros and Ut{\'{e}}za, Olivier and Wang, Andong},
booktitle = {Ultrafast Laser Nanostructuring},
doi = {10.1007/978-3-031-14752-4{\_}27},
pages = {979--1018},
publisher = {Springer},
title = {{Internal Structuring of Semiconductors with Ultrafast Lasers: Opening a Route to Three-Dimensional Silicon Photonics}},
url = {https://link.springer.com/10.1007/978-3-031-14752-4{\_}27},
year = {2023}
}

@article{Houard2023,
archivePrefix = {arXiv},
arxivId = {2207.03769},
author = {Houard, Aur{\'{e}}lien and Walch, Pierre and Produit, Thomas and Moreno, Victor and Mahieu, Benoit and Sunjerga, Antonio and Herkommer, Clemens and Mostajabi, Amirhossein and Andral, Ugo and Andr{\'{e}}, Yves-Bernard and Lozano, Magali and Bizet, Laurent and Schroeder, Malte C. and Schimmel, Guillaume and Moret, Michel and Stanley, Mark and Rison, W. A. and Maurice, Oliver and Esmiller, Bruno and Michel, Knut and Haas, Walter and Metzger, Thomas and Rubinstein, Marcos and Rachidi, Farhad and Cooray, Vernon and Mysyrowicz, Andr{\'{e}} and Kasparian, J{\'{e}}r{\^{o}}me and Wolf, Jean-Pierre},
doi = {10.1038/s41566-022-01139-z},
eprint = {2207.03769},
issn = {1749-4885},
journal = {Nature Photonics},
number = {3},
pages = {231--235},
publisher = {Springer US},
title = {{Laser-guided lightning}},
url = {https://www.nature.com/articles/s41566-022-01139-z},
volume = {17},
year = {2023}
}

@article{Fu2022,
author = {Fu, Silin and Mahieu, Benoit and Mysyrowicz, Andr{\'{e}} and Houard, Aurelien},
doi = {10.1364/OL.472143},
issn = {0146-9592},
journal = {Optics Letters},
month = {oct},
number = {19},
pages = {5228},
pmid = {36181228},
publisher = {Optica Publishing Group},
title = {{Femtosecond filamentation of optical vortices for the generation of optical air waveguides}},
url = {https://opg.optica.org/abstract.cfm?URI=ol-47-19-5228},
volume = {47},
year = {2022}
}

@article{Fedorov2020,
author = {Fedorov, Vladimir Yu and Tzortzakis, Stelios},
doi = {10.1038/s41377-020-00423-3},
issn = {2047-7538},
journal = {Light: Science $\&$ Applications},
month = {nov},
number = {1},
pages = {186},
publisher = {Springer US},
title = {{Powerful terahertz waves from long-wavelength infrared laser filaments}},
url = {https://www.nature.com/articles/s41377-020-00423-3},
volume = {9},
year = {2020}
}

@article{Schimmel2018,
author = {Schimmel, Guillaume and Produit, Thomas and Mongin, Denis and Kasparian, J{\'{e}}r{\^{o}}me and Wolf, Jean-Pierre},
doi = {10.1364/OPTICA.5.001338},
issn = {2334-2536},
journal = {Optica},
month = {oct},
number = {10},
pages = {1338},
title = {{Free space laser telecommunication through fog}},
url = {https://opg.optica.org/abstract.cfm?URI=optica-5-10-1338},
volume = {5},
year = {2018}
}

@article{Goffin2023,
author = {Goffin, A and Tartaro, A and Milchberg, H M},
doi = {10.1364/OPTICA.487292},
issn = {2334-2536},
journal = {Optica},
month = {apr},
number = {4},
pages = {505},
publisher = {Optica Publishing Group},
title = {{Quasi-steady-state air waveguide}},
url = {https://opg.optica.org/abstract.cfm?URI=optica-10-4-505},
volume = {10},
year = {2023}
}

@article{Rahnama2020,
author = {Rahnama, Abdullah and {Mahmoud Aghdami}, Keivan and Kim, Young Hwan and Herman, Peter R.},
doi = {10.1002/adpr.202000026},
issn = {2699-9293},
journal = {Advanced Photonics Research},
month = {dec},
number = {2},
pages = {2000026},
title = {{Ultracompact Lens‐Less “Spectrometer in Fiber” Based on Chirped Filament‐Array Gratings}},
url = {https://onlinelibrary.wiley.com/doi/10.1002/adpr.202000026},
volume = {1},
year = {2020}
}

@article{Ensley2019,
author = {Ensley, Trenton R. and Bambha, Neal K.},
doi = {10.1364/OE.380702},
issn = {1094-4087},
journal = {Optics Express},
month = {dec},
number = {26},
pages = {37940},
pmid = {31878566},
title = {{Ultrafast nonlinear refraction measurements of infrared transmitting materials in the mid-wave infrared}},
url = {https://opg.optica.org/abstract.cfm?URI=oe-27-26-37940},
volume = {27},
year = {2019}
}

@article{Malitson1965,
author = {Malitson, I. H.},
doi = {10.1364/JOSA.55.001205},
issn = {0030-3941},
journal = {Journal of the Optical Society of America},
month = {oct},
number = {10},
pages = {1205},
title = {{Interspecimen Comparison of the Refractive Index of Fused Silica}},
url = {https://opg.optica.org/abstract.cfm?URI=josa-55-10-1205},
volume = {55},
year = {1965}
}

@article{Chambonneau2023,
archivePrefix = {arXiv},
arxivId = {2211.03518},
author = {Chambonneau, Maxime and Li, Qingfeng and Blothe, Markus and Arumugam, Stree Vithya and Nolte, Stefan},
doi = {10.1002/adpr.202200300},
eprint = {2211.03518},
issn = {2699-9293},
journal = {Advanced Photonics Research},
month = {may},
number = {5},
pages = {2200300},
title = {{Ultrafast Laser Welding of Silicon}},
url = {https://onlinelibrary.wiley.com/doi/10.1002/adpr.202200300},
volume = {4},
year = {2023}
}

@article{Bonse2020,
author = {Bonse, J{\"{o}}rn and Gr{\"{a}}f, Stephan},
doi = {10.1002/lpor.202000215},
issn = {1863-8880},
journal = {Laser $\&$ Photonics Reviews},
month = {oct},
number = {10},
pages = {2000215},
title = {{Maxwell Meets Marangoni—A Review of Theories on Laser‐Induced Periodic Surface Structures}},
url = {https://onlinelibrary.wiley.com/doi/10.1002/lpor.202000215},
volume = {14},
year = {2020}
}

@article{Blothe2022,
author = {Blothe, Markus and Chambonneau, Maxime and Nolte, Stefan},
doi = {10.1063/5.0103276},
issn = {0003-6951},
journal = {Applied Physics Letters},
month = {sep},
number = {10},
pages = {101602},
publisher = {AIP Publishing LLC},
title = {{Nanostructured back surface amorphization of silicon with picosecond laser pulses}},
url = {https://aip.scitation.org/doi/10.1063/5.0103276},
volume = {121},
year = {2022}
}

@article{Alberucci2020,
author = {Alberucci, Alessandro and Alasgarzade, Namig and Chambonneau, Maxime and Blothe, Markus and K{\"{a}}mmer, Helena and Matth{\"{a}}us, Gabor and Jisha, Chandroth P. and Nolte, Stefan},
doi = {10.1103/PhysRevApplied.14.024078},
issn = {2331-7019},
journal = {Physical Review Applied},
month = {aug},
number = {2},
pages = {024078},
publisher = {American Physical Society},
title = {{In-Depth Optical Characterization of Femtosecond-Written Waveguides in Silicon}},
url = {https://doi.org/10.1103/PhysRevApplied.14.024078},
volume = {14},
year = {2020}
}

@article{Mareev2022,
author = {Mareev, Evgenii and Pushkin, Andrey and Migal, Ekaterina and Lvov, Kirill and Stremoukhov, Sergey and Potemkin, Fedor},
doi = {10.1038/s41598-022-11501-4},
isbn = {0123456789},
issn = {2045-2322},
journal = {Scientific Reports},
month = {may},
number = {1},
pages = {7517},
pmid = {35525844},
publisher = {Nature Publishing Group UK},
title = {{Single-shot femtosecond bulk micromachining of silicon with mid-IR tightly focused beams}},
url = {https://www.nature.com/articles/s41598-022-11501-4},
volume = {12},
year = {2022}
}

@article{Richter2015,
author = {Richter, S. and Zimmermann, F. and Eberhardt, R. and T{\"{u}}nnermann, A. and Nolte, S.},
doi = {10.1007/s00339-015-9377-8},
issn = {0947-8396},
journal = {Applied Physics A},
month = {oct},
number = {1},
pages = {1--9},
title = {{Toward laser welding of glasses without optical contacting}},
url = {http://link.springer.com/10.1007/s00339-015-9377-8},
volume = {121},
year = {2015}
}

@article{Chambonneau2021a,
author = {Chambonneau, Maxime and Grojo, David and Tokel, Onur and Ilday, Fatih {\"{O}}mer and Tzortzakis, Stelios and Nolte, Stefan},
doi = {10.1002/lpor.202100140},
issn = {1863-8880},
journal = {Laser $\&$ Photonics Reviews},
month = {nov},
number = {11},
pages = {2100140},
title = {{In‐Volume Laser Direct Writing of Silicon—Challenges and Opportunities}},
url = {https://onlinelibrary.wiley.com/doi/10.1002/lpor.202100140},
volume = {15},
year = {2021}
}

@article{Couairon2007,
author = {Couairon, A and Mysyrowicz, A},
doi = {10.1016/j.physrep.2006.12.005},
issn = {03701573},
journal = {Physics Reports},
month = {mar},
number = {2-4},
pages = {47--189},
title = {{Femtosecond filamentation in transparent media}},
url = {https://linkinghub.elsevier.com/retrieve/pii/S037015730700021X},
volume = {441},
year = {2007}
}

@article{Tan2017,
author = {Tan, Hua and Duan, Ji'an},
doi = {10.1007/s00339-017-1079-y},
isbn = {1378609883},
issn = {0947-8396},
journal = {Applied Physics A},
month = {jul},
number = {7},
pages = {481},
publisher = {Springer Berlin Heidelberg},
title = {{Welding of glasses in optical and partial-optical contact via focal position adjustment of femtosecond-laser pulses at moderately high repetition rate}},
url = {http://link.springer.com/10.1007/s00339-017-1079-y},
volume = {123},
year = {2017}
}

@article{Tsibidis2022,
archivePrefix = {arXiv},
arxivId = {2205.05342},
author = {Tsibidis, G. D. and Mansour, Dimitris and Stratakis, Emmanuel},
doi = {10.1016/j.optlastec.2022.108484},
eprint = {2205.05342},
issn = {00303992},
journal = {Optics and Laser Technology},
month = {dec},
pages = {108484},
title = {{Damage threshold evaluation of thin metallic films exposed to femtosecond laser pulses: The role of material thickness}},
url = {https://linkinghub.elsevier.com/retrieve/pii/S0030399222006375},
volume = {156},
year = {2022}
}

@article{Chambonneau2021b,
archivePrefix = {arXiv},
arxivId = {2104.12084},
author = {Chambonneau, M. and Blothe, M. and Li, Q. and Fedorov, V. Yu. and Heuermann, T. and Gebhardt, M. and Gaida, C. and Tertelmann, S. and Sotier, F. and Limpert, J. and Tzortzakis, S. and Nolte, S.},
doi = {10.1103/PhysRevResearch.3.043037},
eprint = {2104.12084},
issn = {2643-1564},
journal = {Physical Review Research},
month = {oct},
number = {4},
pages = {043037},
title = {{Transverse ultrafast laser inscription in bulk silicon}},
url = {https://link.aps.org/doi/10.1103/PhysRevResearch.3.043037},
volume = {3},
year = {2021}
}

@article{Berner2018,
author = {Berner, Andrea and Kasperkiewicz, Ewa and Zhang, Yueqian and Gross, Herbert},
doi = {10.1364/JOSAA.35.001368},
issn = {1084-7529},
journal = {Journal of the Optical Society of America A},
month = {aug},
number = {8},
pages = {1368},
pmid = {30110298},
title = {3rd-order spherochromatism surface contribution and its intrinsic and induced aberration parts},
url = {https://www.osapublishing.org/abstract.cfm?URI=josaa-35-8-1368},
volume = {35},
year = {2018}
}

@article{Berge2007,
author = {Berg{\'{e}}, L. and Skupin, S. and Nuter, R. and Kasparian, J. and Wolf, J-P},
doi = {10.1088/0034-4885/70/10/R03},
issn = {0034-4885},
journal = {Reports on Progress in Physics},
month = {oct},
number = {10},
pages = {1633--1713},
title = {{Ultrashort filaments of light in weakly ionized, optically transparent media}},
url = {https://iopscience.iop.org/article/10.1088/0034-4885/70/10/R03},
volume = {70},
year = {2007}
}

@article{Wang2021,
author = {Wang, Xinya and Yu, Xiaoming and Berg, Matthew J. and Chen, Pingping and Lacroix, Brice and Fathpour, Sasan and Lei, Shuting},
doi = {10.1364/OE.419074},
issn = {1094-4087},
journal = {Optics Express},
month = {may},
number = {10},
pages = {14201},
title = {{Curved waveguides in silicon written by a shaped laser beam}},
url = {https://www.osapublishing.org/abstract.cfm?URI=oe-29-10-14201},
volume = {29},
year = {2021}
}

@article{Lin2007,
author = {Lin, Q. and Zhang, J. and Piredda, G. and Boyd, R. W. and Fauchet, P. M. and Agrawal, G. P.},
doi = {10.1063/1.2750523},
issn = {0003-6951},
journal = {Applied Physics Letters},
month = {jul},
number = {2},
pages = {021111},
title = {{Dispersion of silicon nonlinearities in the near infrared region}},
url = {http://aip.scitation.org/doi/10.1063/1.2750523},
volume = {91},
year = {2007}
}

@article{Chambonneau2021,
author = {Chambonneau, Maxime and Li, Qingfeng and Fedorov, Vladimir Yu and Blothe, Markus and Schaarschmidt, Kay and Lorenz, Martin and Tzortzakis, Stelios and Nolte, Stefan},
doi = {10.1002/lpor.202000433},
issn = {1863-8880},
journal = {Laser $\&$ Photonics Reviews},
month = {feb},
number = {2},
pages = {2000433},
title = {{Taming Ultrafast Laser Filaments for Optimized Semiconductor–Metal Welding}},
url = {https://onlinelibrary.wiley.com/doi/10.1002/lpor.202000433},
volume = {15},
year = {2021}
}

@article{Turnali2019,
author = {Turnali, Ahmet and Han, Mertcan and Tokel, Onur},
doi = {10.1364/JOSAB.36.000966},
issn = {0740-3224},
journal = {Journal of the Optical Society of America B},
month = {apr},
number = {4},
pages = {966},
title = {{Laser-written depressed-cladding waveguides deep inside bulk silicon}},
url = {https://opg.optica.org/abstract.cfm?URI=josab-36-4-966},
volume = {36},
year = {2019}
}

@article{Pavlov2017,
author = {Pavlov, Ihor and Tokel, Onur and Pavlova, Svitlana and Kadan, Viktor and Makey, Ghaith and Turnali, Ahmet and Yavuz, {\"{O}}. and Ilday, Fatih {\"{O}}.},
doi = {10.1364/OL.42.003028},
isbn = {9781557528209},
issn = {0146-9592},
journal = {Optics Letters},
month = {aug},
number = {15},
pages = {3028},
title = {{Femtosecond laser written waveguides deep inside silicon}},
url = {https://opg.optica.org/abstract.cfm?URI=ol-42-15-3028},
volume = {42},
year = {2017}
}

@article{Kammer2019,
author = {K{\"{a}}mmer, Helena and Matth{\"{a}}us, Gabor and Lammers, Kim A. and Vetter, Christian and Chambonneau, Maxime and Nolte, Stefan},
doi = {10.1002/lpor.201800268},
issn = {1863-8880},
journal = {Laser $\&$ Photonics Reviews},
month = {feb},
number = {2},
pages = {1800268},
title = {{Origin of Waveguiding in Ultrashort Pulse Structured Silicon}},
url = {https://onlinelibrary.wiley.com/doi/10.1002/lpor.201800268},
volume = {13},
year = {2019}
}

@article{Matthaus2018,
author = {Matth{\"{a}}us, G. and K{\"{a}}mmer, H. and Lammers, K. A. and Vetter, C. and Watanabe, W. and Nolte, S.},
doi = {10.1364/OE.26.024089},
issn = {1094-4087},
journal = {Optics Express},
month = {sep},
number = {18},
pages = {24089},
pmid = {30184901},
title = {{Inscription of silicon waveguides using picosecond pulses}},
url = {https://opg.optica.org/abstract.cfm?URI=oe-26-18-24089},
volume = {26},
year = {2018}
}

@article{Astrauskas2021,
author = {Astrauskas, I. and Pova{\v{z}}ay, B. and Baltu{\v{s}}ka, A. and Pug{\v{z}}lys, A.},
doi = {10.1016/j.optlastec.2020.106535},
issn = {00303992},
journal = {Optics $\&$ Laser Technology},
month = {jan},
pages = {106535},
publisher = {Elsevier Ltd},
title = {{Influence of 2.09-$\mu$m pulse duration on through-silicon laser ablation of thin metal coatings}},
url = {https://linkinghub.elsevier.com/retrieve/pii/S0030399220311683},
volume = {133},
year = {2021}
}

@article{Chambonneau2018b,
author = {Chambonneau, M. and Richter, D. and Nolte, S. and Grojo, D.},
doi = {10.1364/OL.43.006069},
issn = {0146-9592},
journal = {Optics Letters},
month = {dec},
number = {24},
pages = {6069},
pmid = {30548006},
title = {{Inscribing diffraction gratings in bulk silicon with nanosecond laser pulses}},
url = {https://opg.optica.org/abstract.cfm?URI=ol-43-24-6069},
volume = {43},
year = {2018}
}

@article{Chambonneau2019,
author = {Chambonneau, M. and Wang, X. and Yu, X. and Li, Q. and Chaudanson, D. and Lei, S. and Grojo, D.},
doi = {10.1364/OL.44.001619},
issn = {0146-9592},
journal = {Optics Letters},
month = {apr},
number = {7},
pages = {1619},
pmid = {30933105},
title = {{Positive- and negative-tone structuring of crystalline silicon by laser-assisted chemical etching}},
url = {https://www.osapublishing.org/abstract.cfm?URI=ol-44-7-1619},
volume = {44},
year = {2019}
}

@article{Das2020,
author = {Das, Amlan and Wang, Andong and Uteza, Olivier and Grojo, David},
doi = {10.1364/OE.398984},
issn = {1094-4087},
journal = {Optics Express},
month = {aug},
number = {18},
pages = {26623},
pmid = {32906932},
title = {{Pulse-duration dependence of laser-induced modifications inside silicon}},
url = {https://opg.optica.org/abstract.cfm?URI=oe-28-18-26623},
volume = {28},
year = {2020}
}

@article{Wang2020,
author = {Wang, Andong and Das, Amlan and Grojo, David},
doi = {10.1103/PhysRevResearch.2.033023},
issn = {2643-1564},
journal = {Physical Review Research},
month = {jul},
number = {3},
pages = {033023},
publisher = {American Physical Society},
title = {{Temporal-contrast imperfections as drivers for ultrafast laser modifications in bulk silicon}},
url = {https://link.aps.org/doi/10.1103/PhysRevResearch.2.033023},
volume = {2},
year = {2020}
}

@article{Wang2020a,
author = {Wang, Andong and Das, Amlan and Grojo, David},
doi = {10.34133/2020/8149764},
issn = {2639-5274},
journal = {Research},
month = {jan},
pages = {1--11},
title = {{Ultrafast Laser Writing Deep inside Silicon with THz-Repetition-Rate Trains of Pulses}},
url = {https://spj.sciencemag.org/journals/research/2020/8149764/},
volume = {2020},
year = {2020}
}

@article{Sundaram2002c,
author = {Sundaram, S. K. and Mazur, E.},
doi = {10.1038/nmat767},
issn = {1476-1122},
journal = {Nature Materials},
month = {dec},
number = {4},
pages = {217--224},
pmid = {12618781},
title = {{Inducing and probing non-thermal transitions in semiconductors using femtosecond laser pulses}},
url = {https://www.nature.com/articles/nmat767},
volume = {1},
year = {2002}
}

@article{Richter2020,
author = {Richter, Roland A. and Tolstik, Nikolai and Rigaud, Sebastien and Valle, Paul Dalla and Erbe, Andreas and Ebbinghaus, Petra and Astrauskas, Ignas and Kalashnikov, Vladimir and Sorokin, Evgeni and Sorokina, Irina T.},
doi = {10.1364/JOSAB.396074},
issn = {0740-3224},
journal = {Journal of the Optical Society of America B},
month = {sep},
number = {9},
pages = {2543},
title = {{Sub-surface modifications in silicon with ultra-short pulsed lasers above 2 $\mu$m}},
url = {https://opg.optica.org/abstract.cfm?URI=josab-37-9-2543},
volume = {37},
year = {2020}
}

@article{Ryder2020,
author = {Ryder, Landen D. and Ryder, Kaitlyn L. and Sternberg, Andrew L. and Kozub, John A. and Gong, Huiqi and Zhang, En Xia and Linten, Dimitri and Mitard, Jerome and Weller, Robert A. and Schrimpf, Ronald D. and Weiss, Sharon M. and Reed, Robert A.},
doi = {10.1109/TNS.2019.2956911},
issn = {0018-9499},
journal = {IEEE Transactions on Nuclear Science},
month = {jan},
number = {1},
pages = {38--43},
title = {{Polarization Dependence of Pulsed Laser-Induced SEEs in SOI FinFETs}},
url = {https://ieeexplore.ieee.org/document/8918115/},
volume = {67},
year = {2020}
}

@article{Kononenko2016a,
author = {Kononenko, V. V. and Zavedeev, E. V. and Gololobov, V. M.},
doi = {10.1007/s00339-016-9844-x},
issn = {0947-8396},
journal = {Applied Physics A},
month = {apr},
number = {4},
pages = {293},
publisher = {Springer Berlin Heidelberg},
title = {{The effect of light-induced plasma on propagation of intense fs laser radiation in c-Si}},
url = {http://link.springer.com/10.1007/s00339-016-9844-x},
volume = {122},
year = {2016}
}

@article{Salzberg1957,
author = {Salzberg, Calvin D. and Villa, John J.},
doi = {10.1364/JOSA.47.000244},
issn = {0030-3941},
journal = {Journal of the Optical Society of America},
month = {mar},
number = {3},
pages = {244},
title = {{Infrared Refractive Indexes of Silicon Germanium and Modified Selenium Glass}},
url = {https://opg.optica.org/abstract.cfm?URI=josa-47-3-244},
volume = {47},
year = {1957}
}

@article{Ildefonso2019,
author = {Ildefonso, Adrian and Fleetwood, Zachary E. and Tzintzarov, George N. and Hales, Joel M. and Nergui, Delgermaa and Frounchi, Milad and Khachatrian, Ani and Buchner, Stephen P. and Mcmorrow, Dale and Warner, Jeffrey H. and Harms, Joseph and Erickson, Anna and Voss, Kay and Ferlet-Cavrois, Veronique and Cressler, John D.},
doi = {10.1109/TNS.2018.2882821},
issn = {0018-9499},
journal = {IEEE Transactions on Nuclear Science},
month = {jan},
number = {1},
pages = {359--367},
title = {{Optimizing Optical Parameters to Facilitate Correlation of Laser- and Heavy-Ion-Induced Single-Event Transients in SiGe HBTs}},
url = {https://ieeexplore.ieee.org/document/8543177/},
volume = {66},
year = {2019}
}

@article{Hales2018a,
author = {Hales, Joel M. and Khachatrian, Ani and Buchner, Stephen and Roche, Nicolas J.H. and Warner, Jeffrey and Fleetwood, Zachary E. and Ildefonso, Adrian and Cressler, John D. and Ferlet-Cavrois, Veronique and Mcmorrow, Dale},
doi = {10.1109/TNS.2018.2828332},
issn = {0018-9499},
journal = {IEEE Transactions on Nuclear Science},
month = {aug},
number = {8},
pages = {1724--1733},
publisher = {IEEE},
title = {{Experimental Validation of an Equivalent LET Approach for Correlating Heavy-Ion and Laser-Induced Charge Deposition}},
url = {https://ieeexplore.ieee.org/document/8340873/},
volume = {65},
year = {2018}
}

@article{Hales2017,
author = {Hales, Joel M. and Roche, Nicolas J.H. and Khachatrian, Ani and Mcmorrow, Dale and Buchner, Stephen and Warner, Jeffrey and Turowski, Marek and Lilja, Klas and Hooten, Nicholas C. and Zhang, En Xia and Reed, Robert A. and Schrimpf, Ronald D.},
doi = {10.1109/TNS.2017.2686010},
issn = {0018-9499},
journal = {IEEE Transactions on Nuclear Science},
month = {may},
number = {5},
pages = {1133--1136},
publisher = {IEEE},
title = {{Strong Correlation Between Experiment and Simulation for Two-Photon Absorption Induced Carrier Generation}},
url = {http://ieeexplore.ieee.org/document/7886295/},
volume = {64},
year = {2017}
}

@article{Hales2020,
author = {Hales, Joel M. and Cressler, John D. and McMorrow, Dale and Khachatrian, Ani and Buchner, Stephen and Warner, Jeffrey and Ildefonso, Adrian and Tzintzarov, George N. and Nergui, Delgermaa and Monahan, Daniele M. and LaLumondiere, Stephen D.},
doi = {10.1109/TNS.2019.2950431},
issn = {0018-9499},
journal = {IEEE Transactions on Nuclear Science},
month = {jan},
number = {1},
pages = {81--90},
title = {{New Approach for Pulsed-Laser Testing That Mimics Heavy-Ion Charge Deposition Profiles}},
url = {https://ieeexplore.ieee.org/document/8887522/},
volume = {67},
year = {2020}
}

@article{McMorrow2002,
author = {McMorrow, Dale and Lotshaw, W.T. and Melinger, J.S. and Buchner, Stephen and Pease, R.L.},
doi = {10.1109/TNS.2002.805337},
issn = {0018-9499},
journal = {IEEE Transactions on Nuclear Science},
month = {dec},
number = {6},
pages = {3002--3008},
title = {{Subbandgap laser-induced single event effects: carrier generation via two-photon absorption}},
url = {https://ieeexplore.ieee.org/document/1134253/},
volume = {49},
year = {2002}
}

@article{Khachatrian2014,
author = {Khachatrian, Ani and Roche, Nicolas J.H. and McMorrow, Dale and Warner, Jeffrey H. and Buchner, Stephen P. and Melinger, Joseph S.},
doi = {10.1109/TNS.2014.2369006},
issn = {0018-9499},
journal = {IEEE Transactions on Nuclear Science},
month = {dec},
number = {6},
pages = {3416--3423},
publisher = {IEEE},
title = {{A Dosimetry Methodology for Two-Photon Absorption Induced Single-Event Effects Measurements}},
url = {http://ieeexplore.ieee.org/document/6976039/},
volume = {61},
year = {2014}
}

@article{Hales2015,
author = {Hales, Joel M. and Khachatrian, Ani and Roche, Nicolas J.H. and Warner, Jeffrey and Buchner, Stephen P. and McMorrow, Dale},
doi = {10.1109/TNS.2015.2422793},
issn = {0018-9499},
journal = {IEEE Transactions on Nuclear Science},
month = {aug},
number = {4},
pages = {1550--1557},
publisher = {IEEE},
title = {{Simulation of Laser-Based Two-Photon Absorption Induced Charge Carrier Generation in Silicon}},
url = {http://ieeexplore.ieee.org/document/7132797/},
volume = {62},
year = {2015}
}

@article{Lei2016,
author = {Lei, Shuting and Grojo, David and Ma, Jianfeng and Yu, Xiaoming and Wu, Han},
doi = {10.1016/j.promfg.2016.08.049},
issn = {23519789},
journal = {Procedia Manufacturing},
pages = {594--608},
publisher = {The Author(s)},
title = {{Femtosecond Laser Backside Ablation of Gold Film on Silicon Substrate}},
url = {https://linkinghub.elsevier.com/retrieve/pii/S2351978916300543},
volume = {5},
year = {2016}
}

@article{Luong2020,
author = {Luong, Khanh Phu and Tanabe-Yamagishi, Rie and Yamada, Noboru and Ito, Yoshiro},
doi = {10.2351/1.5123309},
issn = {1042-346X},
journal = {Journal of Laser Applications},
month = {feb},
number = {1},
pages = {012017},
publisher = {Laser Institute of America},
title = {{Micromachining of the back surface and interior of Si using infrared femtosecond laser pulses}},
url = {http://lia.scitation.org/doi/10.2351/1.5123309},
volume = {32},
year = {2020}
}

@article{Rousse2001,
author = {Rousse, A. and Rischel, C. and Fourmaux, S. and Uschmann, I. and Sebban, S. and Grillon, G. and Balcou, Ph and F{\"{o}}rster, E. and Geindre, J.P. and Audebert, P. and Gauthier, J.C. and Hulin, D.},
doi = {10.1038/35065045},
issn = {0028-0836},
journal = {Nature},
month = {mar},
number = {6824},
pages = {65--68},
title = {{Non-thermal melting in semiconductors measured at femtosecond resolution}},
url = {https://www.nature.com/articles/35065045},
volume = {410},
year = {2001}
}

@article{Chambonneau2019a,
author = {Chambonneau, M. and Lavoute, L. and Gaponov, D. and Fedorov, V.Y. and Hideur, A. and F{\'{e}}vrier, S. and Tzortzakis, S. and Ut{\'{e}}za, O. and Grojo, D.},
doi = {10.1103/PhysRevApplied.12.024009},
issn = {2331-7019},
journal = {Physical Review Applied},
month = {aug},
number = {2},
pages = {024009},
publisher = {American Physical Society},
title = {{Competing Nonlinear Delocalization of Light for Laser Inscription Inside Silicon with a 2-$\mu$m Picosecond Laser}},
url = {https://doi.org/10.1103/PhysRevApplied.12.024009},
volume = {12},
year = {2019}
}

@article{Wang2013,
author = {Wang, Ting and Venkatram, Nalla and Gosciniak, Jacek and Cui, Yuanjing and Qian, Guodong and Ji, Wei and Tan, Dawn T. H.},
doi = {10.1364/OE.21.032192},
issn = {1094-4087},
journal = {Optics Express},
month = {dec},
number = {26},
pages = {32192},
title = {{Multi-photon absorption and third-order nonlinearity in silicon at mid-infrared wavelengths}},
url = {https://opg.optica.org/oe/abstract.cfm?uri=oe-21-26-32192},
volume = {21},
year = {2013}
}

@article{Chambonneau2017a,
author = {Chambonneau, Maxime and Souiki-Figuigui, Sarra and Chiquet, Philippe and {Della Marca}, Vincenzo and Postel-Pellerin, J{\'{e}}r{\'{e}}my and Canet, Pierre and Portal, Jean-Michel and Grojo, David},
doi = {10.1063/1.4982032},
issn = {0003-6951},
journal = {Applied Physics Letters},
month = {apr},
number = {16},
pages = {6},
title = {{Suppressing the memory state of floating gate transistors with repeated femtosecond laser backside irradiations}},
url = {https://doi.org/10.1063/1.4982032},
volume = {110},
year = {2017}
}

@article{ParsiSreenivas2012,
author = {{Parsi Sreenivas}, V. V. and B{\"{u}}lters, M. and Bergmann, R. B.},
doi = {10.2971/jeos.2012.12035},
issn = {1990-2573},
journal = {Journal of the European Optical Society-Rapid Publications},
month = {sep},
pages = {12035},
title = {{Microsized subsurface modification of mono-crystalline silicon via non-linear absorption}},
url = {https://jeos.edpsciences.org/10.2971/jeos.2012.12035},
volume = {7},
year = {2012}
}

@article{Tokel2017a,
archivePrefix = {arXiv},
arxivId = {1409.2827},
author = {Tokel, Onur and Turnall, Ahmet and Makey, Ghaith and Elahi, Parviz and {\c{C}}olakoǧlu, Tahir and Erge{\c{c}}en, Emre and Yavuz, {\"{O}}zg{\"{u}}n and H{\"{u}}bner, Ren{\'{e}} and {Zolfaghari Borra}, Mona and Pavlov, Ihor and Bek, Alpan and Turan, Raşit and Kesim, Denizhan Koray and Tozburun, Serhat and Ilday, Serim and Ilday, F. {\"{O}}mer},
doi = {10.1038/s41566-017-0004-4},
eprint = {1409.2827},
issn = {17494893},
journal = {Nature Photonics},
number = {10},
pages = {639--645},
title = {{In-chip microstructures and photonic devices fabricated by nonlinear laser lithography deep inside silicon}},
url = {https://doi.org/10.1038/s41566-017-0004-4},
volume = {11},
year = {2017}
}

@article{Yu2016,
author = {Yu, Xiaoming and Wang, Xinya and Chanal, Margaux and Trallero-Herrero, Carlos A. and Grojo, David and Lei, Shuting},
doi = {10.1007/s00339-016-0540-7},
issn = {14320630},
journal = {Applied Physics A},
number = {12},
pages = {1--7},
publisher = {Springer Berlin Heidelberg},
title = {{Internal modification of intrinsic and doped silicon using infrared nanosecond laser}},
url = {https://doi.org/10.1007/s00339-016-0540-7},
volume = {122},
year = {2016}
}

@article{Mingareev2016,
author = {Mingareev, Ilya and Gehlich, Nils and Bonhoff, Tobias and Abdulfattah, Ali and Sincore, Alex M. and Kadwani, Pankaj and Shah, Lawrence and Richardson, Martin},
doi = {10.1007/s00170-015-7870-z},
issn = {0268-3768},
journal = {The International Journal of Advanced Manufacturing Technology},
month = {jun},
number = {9-12},
pages = {2567--2578},
publisher = {The International Journal of Advanced Manufacturing Technology},
title = {{Principles and applications of trans-wafer processing using a 2-$\mu$m thulium fiber laser}},
url = {http://link.springer.com/10.1007/s00170-015-7870-z},
volume = {84},
year = {2016}
}

@article{Kawaguchi2019,
author = {Kawaguchi, Daisuke and Iwata, Hiroyuki and Saka, Hiroyasu},
doi = {10.1080/14786435.2019.1605216},
issn = {14786443},
journal = {Philosophical Magazine},
number = {15},
pages = {1849--1865},
publisher = {Taylor & Francis},
title = {{Whereabouts of missing atoms in a laser-injected Si: Part 1}},
url = {https://doi.org/10.1080/14786435.2019.1605216},
volume = {99},
year = {2019}
}

@article{Kadan2018,
author = {Kadan, Viktor and Pavlova, Svitlana and Pavlov, Ihor and Rezaei, Hossein and Ilday, {\"{O}}mer and Blonskyi, Ivan},
doi = {10.1007/s00339-018-1986-6},
isbn = {0123456789},
issn = {0947-8396},
journal = {Applied Physics A},
month = {aug},
number = {8},
pages = {560},
publisher = {Springer Berlin Heidelberg},
title = {{Spatio-temporal dynamics of femtosecond laser pulses at 1550 nm wavelength in crystal silicon}},
url = {http://link.springer.com/10.1007/s00339-018-1986-6},
volume = {124},
year = {2018}
}

@article{Chiquet2019,
author = {Chiquet, Philippe and Chambonneau, Maxime and {Della Marca}, Vincenzo and Postel-Pellerin, J{\'{e}}r{\'{e}}my and Canet, Pierre and Souiki-Figuigui, Sarra and Idda, Guillaume and Portal, Jean-Michel and Grojo, David},
doi = {10.1038/s41598-019-43344-x},
isbn = {4159801943},
issn = {2045-2322},
journal = {Scientific Reports},
month = {may},
number = {1},
pages = {7392},
title = {{Phenomenological modelling of non-volatile memory threshold voltage shift induced by nonlinear ionization with a femtosecond laser}},
url = {https://www.nature.com/articles/s41598-019-43344-x},
volume = {9},
year = {2019}
}

@article{Wang2019a,
author = {Wang, Xinya and Yu, Xiaoming and Shi, Hongyu and Tian, Xianhua and Chambonneau, Maxime and Grojo, David and DePaola, Brett and Berg, Matthew and Lei, Shuting},
doi = {10.2351/1.5096086},
issn = {1042-346X},
journal = {Journal of Laser Applications},
month = {may},
number = {2},
pages = {022601},
title = {{Characterization and control of laser induced modification inside silicon}},
url = {http://lia.scitation.org/doi/10.2351/1.5096086},
volume = {31},
year = {2019}
}

@article{Iwata2018,
author = {Iwata, Hiroyuki and Kawaguchi, Daisuke and Saka, Hiroyasu},
doi = {10.1093/jmicro/dfx128},
issn = {20505701},
journal = {Microscopy},
number = {1},
pages = {30--36},
title = {{Crystal structures of high-pressure phases formed in Si by laser irradiation}},
url = {https://doi.org/10.1093/jmicro/dfx128},
volume = {67},
year = {2018}
}

@article{Iwata2017,
author = {Iwata, Hiroyuki and Kawaguchi, Daisuke and Saka, Hiroyasu},
doi = {10.1093/jmicro/dfx024},
issn = {20505701},
journal = {Microscopy},
number = {5},
pages = {328--336},
title = {{Electron microscopy of voids in Si formed by permeable pulse laser irradiation}},
url = {https://doi.org/10.1093/jmicro/dfx024},
volume = {66},
year = {2017}
}

@article{Kammer2018,
author = {K{\"{a}}mmer, H. and Matth{\"{a}}us, G. and Nolte, S. and Chanal, M. and Ut{\'{e}}za, O. and Grojo, D.},
doi = {10.1007/s00339-018-1715-1},
isbn = {0123456789},
issn = {0947-8396},
journal = {Applied Physics A},
month = {apr},
number = {4},
pages = {302},
publisher = {Springer Berlin Heidelberg},
title = {{In-volume structuring of silicon using picosecond laser pulses}},
url = {http://link.springer.com/10.1007/s00339-018-1715-1},
volume = {124},
year = {2018}
}

@article{Zhang2014,
author = {Zhang, Jingyu and Gecevi{\v{c}}ius, Mindaugas and Beresna, Martynas and Kazansky, Peter G.},
doi = {10.1103/PhysRevLett.112.033901},
issn = {0031-9007},
journal = {Physical Review Letters},
month = {jan},
number = {3},
pages = {033901},
pmid = {24484138},
title = {{Seemingly Unlimited Lifetime Data Storage in Nanostructured Glass}},
url = {https://link.aps.org/doi/10.1103/PhysRevLett.112.033901},
volume = {112},
year = {2014}
}

@inproceedings{Marca2016a,
author = {{Della Marca}, V. and Chambonneau, M. and Souiki-Figuigui, S. and Postel-Pellerin, J. and Canet, P. and Chiquet, P. and Kussener, E. and Yengui, F. and Wacquez, R. and Grojo, D. and Portal, J.-M. and Lisart, M.},
booktitle = {2016 IEEE International Reliability Physics Symposium (IRPS)},
doi = {10.1109/IRPS.2016.7574580},
isbn = {978-1-4673-9137-5},
issn = {15417026},
month = {apr},
pages = {7B--4--1--7B--4--7},
publisher = {IEEE},
title = {{NVM cell degradation induced by femtosecond laser backside irradiation for reliability tests}},
url = {http://ieeexplore.ieee.org/document/7574580/},
volume = {2016},
year = {2016}
}

@article{Li2016a,
author = {Li, Q and Chambonneau, M and Chanal, M and Grojo, D.},
doi = {10.1364/AO.55.009577},
issn = {0003-6935},
journal = {Applied Optics},
month = {nov},
number = {33},
pages = {9577},
title = {{Quantitative-phase microscopy of nanosecond laser-induced micro-modifications inside silicon}},
url = {https://www.osapublishing.org/abstract.cfm?URI=ao-55-33-9577},
volume = {55},
year = {2016}
}

@article{Chambonneau2016,
author = {Chambonneau, M. and Li, Q. and Chanal, M. and Sanner, N. and Grojo, D.},
doi = {10.1364/OL.41.004875},
isbn = {0146-9592},
issn = {0146-9592},
journal = {Optics Letters},
number = {21},
pages = {4875},
title = {{Writing waveguides inside monolithic crystalline silicon with nanosecond laser pulses}},
url = {https://www.osapublishing.org/abstract.cfm?URI=ol-41-21-4875},
volume = {41},
year = {2016}
}

@article{Chanal2017,
author = {Chanal, Margaux and Fedorov, Vladimir Yu and Chambonneau, Maxime and Clady, Rapha{\"{e}}l and Tzortzakis, Stelios and Grojo, David},
doi = {10.1038/s41467-017-00907-8},
issn = {2041-1723},
journal = {Nature Communications},
month = {oct},
number = {1},
pages = {773},
publisher = {Springer US},
title = {{Crossing the threshold of ultrafast laser writing in bulk silicon}},
url = {https://www.nature.com/articles/s41467-017-00907-8},
volume = {8},
year = {2017}
}

@article{Kerse2016,
archivePrefix = {arXiv},
arxivId = {NIHMS150003},
author = {Kerse, Can and Kalaycıoğlu, Hamit and Elahi, Parviz and {\c{C}}etin, Barbaros and Kesim, Denizhan K. and Ak{\c{c}}aalan, {\"{O}}nder and Yavaş, Seydi and Aşık, Mehmet D. and {\"{O}}ktem, B{\"{u}}lent and Hoogland, Heinar and Holzwarth, Ronald and Ilday, Fatih {\"{O}}mer},
doi = {10.1038/nature18619},
eprint = {NIHMS150003},
isbn = {0028-0836},
issn = {0028-0836},
journal = {Nature},
month = {sep},
number = {7618},
pages = {84--88},
pmid = {27409814},
publisher = {Nature Publishing Group},
title = {{Ablation-cooled material removal with ultrafast bursts of pulses}},
url = {https://www.nature.com/articles/nature18619},
volume = {537},
year = {2016}
}

@article{Kononenko2012,
author = {Kononenko, Vitali V. and Konov, Vitali V. and Dianov, Evgeny M.},
doi = {10.1364/OL.37.003369},
isbn = {0146-9592},
issn = {0146-9592},
journal = {Optics Letters},
month = {aug},
number = {16},
pages = {3369},
pmid = {23381260},
title = {{Delocalization of femtosecond radiation in silicon}},
url = {https://opg.optica.org/abstract.cfm?URI=ol-37-16-3369},
volume = {37},
year = {2012}
}

@article{Sokolowski-Tinten2003,
author = {Sokolowski-Tinten, Klaus and Blome, Christian and Blums, Juris and Cavalleri, Andrea and Dietrich, Clemens and Tarasevitch, Alexander and Uschmann, Ingo and F{\"{o}}rster, Eckhard and Kammler, Martin and Horn-von-Hoegen, Michael and von der Linde, Dietrich},
doi = {10.1038/nature01490},
isbn = {0028-0836},
issn = {0028-0836},
journal = {Nature},
month = {mar},
number = {6929},
pages = {287--289},
pmid = {12646915},
title = {{Femtosecond X-ray measurement of coherent lattice vibrations near the Lindemann stability limit}},
url = {https://www.nature.com/articles/nature01490},
volume = {422},
year = {2003}
}

@article{Mouskeftaras2016,
author = {Mouskeftaras, Alexandros and Chanal, Margaux and Chambonneau, Maxime and Clady, Rapha{\"{e}}l and Ut{\'{e}}za, Olivier and Grojo, David},
doi = {10.1063/1.4941031},
issn = {0003-6951},
journal = {Applied Physics Letters},
month = {jan},
number = {4},
pages = {041107},
title = {{Direct measurement of ambipolar diffusion in bulk silicon by ultrafast infrared imaging of laser-induced microplasmas}},
url = {http://aip.scitation.org/doi/10.1063/1.4941031},
volume = {108},
year = {2016}
}

@article{Nejadmalayeri2005,
author = {Nejadmalayeri, Amir H and Herman, Peter R and Burghoff, Jonas and Will, Matthias and Nolte, Stefan and T{\"{u}}nnermann, Andreas},
doi = {10.1364/OL.30.000964},
issn = {0146-9592},
journal = {Optics Letters},
month = {may},
number = {9},
pages = {964},
title = {{Inscription of optical waveguides in crystalline silicon by mid-infrared femtosecond laser pulses}},
url = {https://opg.optica.org/abstract.cfm?URI=ol-30-9-964},
volume = {30},
year = {2005}
}

@article{Crawford2008,
author = {Crawford, T. H. R. and Yamanaka, J and Botton, G A and Haugen, H K},
doi = {10.1063/1.2885111},
issn = {0021-8979},
journal = {Journal of Applied Physics},
month = {mar},
number = {5},
pages = {1--8},
title = {{High-resolution observations of an amorphous layer and subsurface damage formed by femtosecond laser irradiation of silicon}},
url = {https://doi.org/10.1063/1.2885111},
volume = {103},
year = {2008}
}

@article{Eaton2005,
author = {Eaton, Shane M and Zhang, Haibin and Herman, Peter R and Yoshino, Fumiyo and Shah, Lawrence and Bovatsek, James and Arai, Alan Y},
doi = {10.1364/OPEX.13.004708},
isbn = {1094-4087},
issn = {1094-4087},
journal = {Optics Express},
number = {12},
pages = {4708},
pmid = {19495387},
title = {{Heat accumulation effects in femtosecond laser-written waveguides with variable repetition rate}},
url = {https://opg.optica.org/oe/abstract.cfm?uri=oe-13-12-4708},
volume = {13},
year = {2005}
}

@article{Stuart1996,
author = {Stuart, B C and Feit, M D and Herman, S and Rubenchik, A M and Shore, B W and Perry, M D},
doi = {10.1103/PhysRevB.53.1749},
issn = {0163-1829},
journal = {Physical Review B},
month = {jan},
number = {4},
pages = {1749--1761},
title = {{Nanosecond-to-femtosecond laser-induced breakdown in dielectrics}},
url = {https://link.aps.org/doi/10.1103/PhysRevB.53.1749},
volume = {53},
year = {1996}
}

@article{Temnov2006,
author = {Temnov, V V and Sokolowski-Tinten, K. and Zhou, P and El-Khamhawy, A. and von der Linde, D.},
doi = {10.1103/PhysRevLett.97.237403},
issn = {0031-9007},
journal = {Physical Review Letters},
month = {dec},
number = {23},
pages = {237403},
title = {{Multiphoton Ionization in Dielectrics: Comparison of Circular and Linear Polarization}},
url = {https://link.aps.org/doi/10.1103/PhysRevLett.97.237403},
volume = {97},
year = {2006}
}

@article{Gattass2008,
author = {Gattass, Rafael R. and Mazur, Eric},
doi = {10.1038/nphoton.2008.47},
isbn = {1749-4885},
issn = {1749-4885},
journal = {Nature Photonics},
month = {apr},
number = {4},
pages = {219--225},
pmid = {226},
title = {{Femtosecond laser micromachining in transparent materials}},
url = {https://www.nature.com/articles/nphoton.2008.47},
volume = {2},
year = {2008}
}

@article{grojo2013long,
author = {Grojo, D. and Leyder, S. and Delaporte, P. and Marine, W. and Sentis, M. and Ut{\'{e}}za, O.},
doi = {10.1103/PhysRevB.88.195135},
issn = {1098-0121},
journal = {Physical Review B},
month = {nov},
number = {19},
pages = {195135},
title = {{Long-wavelength multiphoton ionization inside band-gap solids}},
url = {https://link.aps.org/doi/10.1103/PhysRevB.88.195135},
volume = {88},
year = {2013}
}

@article{Mori2015a,
author = {Mori, Masahiro and Shimotsuma, Yasuhiko and Sei, Tomoaki and Sakakura, Masaaki and Miura, Kiyotaka and Udono, Haruhiko},
doi = {10.1002/pssa.201431777},
issn = {1862-6300},
journal = {physica status solidi (a)},
month = {apr},
number = {4},
pages = {715--721},
title = {{Tailoring thermoelectric properties of nanostructured crystal silicon fabricated by infrared femtosecond laser direct writing}},
url = {https://onlinelibrary.wiley.com/doi/10.1002/pssa.201431777},
volume = {212},
year = {2015}
}

@article{Mouskeftaras2014,
author = {Mouskeftaras, Alexandros and Rode, Andrei V. and Clady, Rapha{\"{e}}l and Sentis, Marc and Ut{\'{e}}za, Olivier and Grojo, David},
doi = {10.1063/1.4901528},
issn = {0003-6951},
journal = {Applied Physics Letters},
month = {nov},
number = {19},
pages = {191103},
title = {{Self-limited underdense microplasmas in bulk silicon induced by ultrashort laser pulses}},
url = {http://scitation.aip.org/content/aip/journal/apl/105/19/10.1063/1.4901528},
volume = {105},
year = {2014}
}

@article{Stuart1995,
author = {Stuart, B. C and Feit, M. D. and Rubenchik, A. M. and Shore, B. W. and Perry, M. D},
doi = {10.1103/PhysRevLett.74.2248},
issn = {0031-9007},
journal = {Physical Review Letters},
month = {mar},
number = {12},
pages = {2248--2251},
title = {{Laser-Induced Damage in Dielectrics with Nanosecond to Subpicosecond Pulses}},
url = {https://link.aps.org/doi/10.1103/PhysRevLett.74.2248},
volume = {74},
year = {1995}
}

@article{Mero2005,
author = {Mero, M. and Liu, J. and Rudolph, W. and Ristau, D. and Starke, K.},
doi = {10.1103/PhysRevB.71.115109},
issn = {1098-0121},
journal = {Physical Review B},
month = {mar},
number = {11},
pages = {115109},
title = {{Scaling laws of femtosecond laser pulse induced breakdown in oxide films}},
url = {https://link.aps.org/doi/10.1103/PhysRevB.71.115109},
volume = {71},
year = {2005}
}

@article{Verburg2014,
author = {Verburg, P. C. and R{\"{o}}mer, G. R. B. E. and {Huis in 't Veld}, A. J.},
doi = {10.1364/OE.22.021958},
issn = {1094-4087},
journal = {Optics Express},
month = {sep},
number = {18},
pages = {21958},
pmid = {25321571},
title = {{Two-photon–induced internal modification of silicon by erbium-doped fiber laser}},
url = {https://www.osapublishing.org/oe/abstract.cfm?uri=oe-22-18-21958},
volume = {22},
year = {2014}
}

@article{Faccio2008a,
author = {Faccio, D and Clerici, M and Averchi, A and Jedrkiewicz, O and Tzortzakis, S and Papazoglou, D. and Bragheri, F and Tartara, L and Trita, A and Henin, S and Cristiani, I and Couairon, A and {Di Trapani}, P},
doi = {10.1364/OE.16.008213},
isbn = {1094-4087; 1094-4087},
issn = {1094-4087},
journal = {Optics Express},
month = {may},
number = {11},
pages = {8213},
pmid = {18545532},
title = {{Kerr-induced spontaneous Bessel beam formation in the regime of strong two-photon absorption}},
url = {https://opg.optica.org/oe/abstract.cfm?uri=oe-16-11-8213},
volume = {16},
year = {2008}
}

@article{Tan2023,
author = {Tan, Yuanxin and Lv, Haotian and Xu, Jian and Zhang, Aodong and Song, Yunpeng and Yu, Jianping and Chen, Wei and Wan, Yuexin and Liu, Zhaoxiang and Liu, Zhaohui and Qi, Jia and Cai, Yangjian and Cheng, Ya},
doi = {10.29026/oea.2023.230066},
issn = {2096-4579},
journal = {Opto-Electronic Advances},
number = {10},
pages = {230066--230066},
title = {{Three-dimensional isotropic microfabrication in glass using spatiotemporal focusing of high-repetition-rate femtosecond laser pulses}},
url = {http://www.oejournal.org//article/doi/10.29026/oea.2023.230066},
volume = {6},
year = {2023}
}

@article{He2011,
author = {He, Fei and Cheng, Ya and Lin, Jintian and Ni, Jielei and Xu, Zhizhan and Sugioka, Koji and Midorikawa, Katsumi},
doi = {10.1088/1367-2630/13/8/083014},
issn = {1367-2630},
journal = {New Journal of Physics},
month = {aug},
number = {8},
pages = {083014},
title = {{Independent control of aspect ratios in the axial and lateral cross sections of a focal spot for three-dimensional femtosecond laser micromachining}},
url = {https://iopscience.iop.org/article/10.1088/1367-2630/13/8/083014},
volume = {13},
year = {2011}
}

@article{Suzuki2024,
author = {Suzuki, Anna and Wang, Yicheng and Tomilov, Sergei and Pan, Zhongben and Saraceno, Clara J.},
doi = {10.35848/1882-0786/ad346b},
isbn = {9781957171395},
issn = {1882-0778},
journal = {Applied Physics Express},
month = {apr},
number = {4},
pages = {042002},
publisher = {IOP Publishing},
title = {{Diode-pumped 88 fs SESAM mode-locked Tm,Ho:CLNGG laser at 2090 nm}},
url = {https://iopscience.iop.org/article/10.35848/1882-0786/ad346b},
volume = {17},
year = {2024}
}

@article{Sopena2022,
archivePrefix = {arXiv},
arxivId = {2203.16313},
author = {Sope{\~{n}}a, Pol and Wang, Andong and Mouskeftaras, Alexandros and Grojo, David},
doi = {10.1002/lpor.202200208},
eprint = {2203.16313},
issn = {18638899},
journal = {Laser $\&$ Photonics Reviews},
number = {11},
pages = {1--8},
title = {{Transmission Laser Welding of Similar and Dissimilar Semiconductor Materials}},
url = {https://doi.org/10.1002/lpor.202200208},
volume = {16},
year = {2022}
}

@article{Wang2025,
   author = {Andong Wang and Amlan Das and Vladimir Yu Fedorov and Pol Sopeña and Stelios Tzortzakis and David Grojo},
   doi = {10.1038/s41467-025-61983-9},
   issn = {2041-1723},
   issue = {1},
   journal = {Nature Communications},
   month = {7},
   pages = {6733},
   pmid = {40695822},
   publisher = {Nature Research},
   title = {In-chip critical plasma seeds for laser writing of reconfigurable silicon photonics systems},
   volume = {16},
   url = {https://www.nature.com/articles/s41467-025-61983-9},
   year = {2025}
}

@article{Chavilkkadan2025,
  doi = {10.1063/5.0255787},
  title={Purely geometric spin--orbit Laguerre--Gauss waveplates from 3D laser-nanostructured silica glass},
  author={Chavilkkadan, Mufeeduzaman and Shevtsov, Sergei and Kazansky, Peter and Brasselet, Etienne},
  journal={APL Photonics},
  url = {https://doi.org/10.1063/5.0255787},
  volume={10},
  number={4},
  year={2025},
  publisher={AIP Publishing}
}

@article{Lapointe2024,
author = {Lapointe, Jerome and Pouliot, Samuel and Dupont, Albert and Gu{\'{e}}rineau, Th{\'{e}}o and Gaulin, Joseph and Pelletier, {\'{E}}tienne and Delarosbil, Jean-Luc and Lafreni{\`{e}}re-Greig, Jonathan and Olivier, Lo{\"{i}}c and Gagnon, St{\'{e}}phane and Messaddeq, Younes and Vall{\'{e}}e, R{\'{e}}al},
doi = {10.1364/OL.542171},
issn = {0146-9592},
journal = {Optics Letters},
month = {dec},
number = {24},
pages = {7048},
title = {{Optimization of fs-laser-induced voxels in nonlinear materials via over-correction of spherical aberration}},
url = {https://opg.optica.org/abstract.cfm?URI=ol-49-24-7048},
volume = {49},
year = {2024}
}

@article{Kolner1989,
author = {Kolner, Brian H. and Nazarathy, Moshe},
doi = {10.1364/OL.14.000630},
issn = {0146-9592},
journal = {Optics Letters},
month = {jun},
number = {12},
pages = {630},
title = {{Temporal imaging with a time lens}},
url = {https://opg.optica.org/abstract.cfm?URI=ol-14-12-630},
volume = {14},
year = {1989}
}

@article{Zhu2005,
author = {Zhu, Guanghao and Howe, Jamesvan and Durst, Michael and Zipfel, Warren and Xu, Chris},
doi = {10.1364/opex.13.002153},
isbn = {1557527709},
issn = {21622701},
journal = {Optics Express},
number = {6},
pages = {2153--2159},
pmid = {19495103},
title = {{Simultaneous spatial and temporal focusing of femtosecond pulses}},
url = {https://doi.org/10.1364/OPEX.13.002153},
volume = {13},
year = {2005}
}

@article{joglekar2004optics,
author = {Joglekar, Ajit P. and Liu, Hsiao-hua and Meyh{\"{o}}fer, Edgar and Mourou, Gerard and Hunt, Alan J.},
doi = {10.1073/pnas.0307470101},
issn = {0027-8424},
journal = {Proceedings of the National Academy of Sciences},
month = {apr},
number = {16},
pages = {5856--5861},
title = {{Optics at critical intensity: Applications to nanomorphing}},
url = {https://pnas.org/doi/full/10.1073/pnas.0307470101},
volume = {101},
year = {2004}
}

@article{joglekar2003study,
author = {Joglekar, A.P. and Liu, H. and Spooner, G.J. and Meyh{\"{o}}fer, E. and Mourou, G. and Hunt, A.J.},
doi = {10.1007/s00340-003-1246-z},
issn = {0946-2171},
journal = {Applied Physics B},
month = {aug},
number = {1},
pages = {25--30},
title = {{A study of the deterministic character of optical damage by femtosecond laser pulses and applications to nanomachining}},
url = {http://link.springer.com/10.1007/s00340-003-1246-z},
volume = {77},
year = {2003}
}

@article{Zhou2021,
author = {Zhou, Boyang and Kar, Aravinda and Soileau, M. J. and Yu, Xiaoming},
doi = {10.1364/OE.418288},
issn = {1094-4087},
journal = {Optics Express},
month = {feb},
number = {4},
pages = {5635},
pmid = {33726097},
title = {{Invariance of the r$^{2}$-$\ln(F)$ relationship and attainable precision in ultrafast laser ablation experiments}},
url = {https://opg.optica.org/abstract.cfm?URI=oe-29-4-5635},
volume = {29},
year = {2021}
}

@article{Balage2023,
author = {Balage, Pierre and Lopez, John and Bonamis, Guillaume and H{\"{o}}nninger, Clemens and Manek-H{\"{o}}nninger, Inka},
doi = {10.1088/2631-7990/acaa14},
issn = {2631-8644},
journal = {International Journal of Extreme Manufacturing},
month = {mar},
number = {1},
pages = {015002},
title = {{Crack-free high-aspect ratio holes in glasses by top–down percussion drilling with infrared femtosecond laser GHz-bursts}},
url = {https://iopscience.iop.org/article/10.1088/2631-7990/acaa14},
volume = {5},
year = {2023}
}

@article{Keldysh1965,
author = {Keldysh, L. V.},
doi = {10.1134/S1063776116030043},
isbn = {9780335226375},
issn = {0038-5646},
journal = {Journal of Experimental and Theoretical Physics},
number = {5},
pages = {1307--1314},
pmid = {14199369},
title = {{Ionization in the field of a strong electromagnetic wave}},
url = {https://doi.org/10.1134/S1063776116030043},
volume = {20},
year = {1965}
}

@book{Kingslake2010,
author = {Kingslake, R},
booktitle = {Optica Acta: International Journal of Optics},
doi = {10.1016/C2009-0-22069-1},
isbn = {9780123743015},
issn = {0030-3909},
publisher = {Elsevier},
title = {{Lens Design Fundamentals}},
url = {https://linkinghub.elsevier.com/retrieve/pii/C20090220691},
year = {2010}
}

@article{Pellish2008,
author = {Pellish, Jonathan A and Reed, Robert A and McMorrow, Dale and Melinger, Joseph S and Jenkins, Phillip and Sutton, Akil K and Diestelhorst, Ryan M and Phillips, Stanley D and Cressler, John D and Pouget, Vincent and Pate, Nicholas D and Kozub, John A and Mendenhall, Marcus H and Weller, Robert A and Schrimpf, Ronald D and Marshall, Paul W and Tipton, Alan D and Niu, Guofu},
doi = {10.1109/TNS.2008.2007954},
issn = {0018-9499},
journal = {IEEE Transactions on Nuclear Science},
month = {dec},
number = {6},
pages = {2936--2942},
title = {{Laser-Induced Current Transients in Silicon-Germanium HBTs}},
url = {http://ieeexplore.ieee.org/document/4723759/},
volume = {55},
year = {2008}
}

@article{Kasparian2003,
author = {Kasparian, J. and Rodriguez, M. and M{\'{e}}jean, G. and Yu, J. and Salmon, E. and Wille, H. and Bourayou, R. and Frey, S. and Andr{\'{e}}, Y.-B. and Mysyrowicz, A. and Sauerbrey, R. and Wolf, J.-P. and W{\"{o}ste, L.}},
doi = {10.1126/science.1085020},
issn = {0036-8075},
journal = {Science},
month = {jul},
number = {5629},
pages = {61--64},
pmid = {12843384},
title = {{White-Light Filaments for Atmospheric Analysis}},
url = {https://www.science.org/doi/10.1126/science.1085020},
volume = {301},
year = {2003}
}

@article{Bristow2007,
author = {Bristow, Alan D. and Rotenberg, Nir and van Driel, Henry M.},
doi = {10.1063/1.2737359},
issn = {0003-6951},
journal = {Applied Physics Letters},
month = {may},
number = {19},
pages = {191104},
title = {{Two-photon absorption and Kerr coefficients of silicon for 850–2200nm}},
url = {http://aip.scitation.org/doi/10.1063/1.2737359},
volume = {90},
year = {2007}
}

@article{Sheik-Bahae1990,
author = {Sheik-Bahae, M. and Hagan, D. J. and {Van Stryland}, E. W.},
doi = {10.1103/PhysRevLett.65.96},
issn = {0031-9007},
journal = {Physical Review Letters},
month = {jul},
number = {1},
pages = {96--99},
title = {{Dispersion and band-gap scaling of the electronic Kerr effect in solids associated with two-photon absorption}},
url = {https://link.aps.org/doi/10.1103/PhysRevLett.65.96},
volume = {65},
year = {1990}
}

@article{Zavedeev2016,
author = {Zavedeev, E V and Kononenko, V V and Konov, V I},
doi = {10.1088/1054-660X/26/1/016101},
issn = {1054-660X},
journal = {Laser Physics},
month = {jan},
number = {1},
pages = {016101},
publisher = {IOP Publishing},
title = {{Delocalization of femtosecond laser radiation in crystalline Si in the mid-IR range}},
url = {https://iopscience.iop.org/article/10.1088/1054-660X/26/1/016101},
volume = {26},
year = {2016}
}

@article{Verburg2015,
author = {Verburg, P. C. and Smillie, L. A. and R{\"{o}}mer, G. R. B. E. and Haberl, B. and Bradby, J. E. and Williams, J. S. and {Huis in 't Veld}, A. J.},
doi = {10.1007/s00339-015-9238-5},
issn = {0947-8396},
journal = {Applied Physics A},
month = {aug},
number = {2},
pages = {683--691},
title = {{Crystal structure of laser-induced subsurface modifications in Si}},
url = {http://link.springer.com/10.1007/s00339-015-9238-5},
volume = {120},
year = {2015}
}

\end{document}